%% file: 3dEMD_ptesta.tex
\documentclass[apj,numberedappendix]{emulateapj}
\usepackage{graphicx}

\shorttitle{Testing Coronal Temperature Diagnostics}
\shortauthors{Testa et al.}
\usepackage{color} 
\usepackage{amsmath}           
\usepackage{amsfonts}

\def \ll   {$\lambda$}

\def \emt {$EM(T)$}
\def \dens {$n_{\rm e}$}

\def \hinode   {{\em Hinode}}

\def \eis  {{\sc EIS}}

\def \ov     {O\,{\sc v}}
\def \mgvi   {Mg\,{\sc vi}}
\def \mgvii  {Mg\,{\sc vii}}

\def \feviii  {Fe\,{\sc viii}}
\def \feix    {Fe\,{\sc ix}}
\def \fex     {Fe\,{\sc x}}
\def \fexi    {Fe\,{\sc xi}}
\def \fexii   {Fe\,{\sc xii}}

\def \fexiv   {Fe\,{\sc xiv}}
\def \fexv    {Fe\,{\sc xv}}
\def \fexvi   {Fe\,{\sc xvi}}
\def \fexvii  {Fe\,{\sc xvii}}

\def \s9     {S\,{\sc ix}}

\def \sivii   {Si\,{\sc vii}}

\def \caxvi   {Ca\,{\sc xvi}}
\def \caxvii  {Ca\,{\sc xvii}}

\def\ion[#1 #2]{#1\,{\sc #2}}
\def\densr[#1 #2]{10$^{#1}$\hskip 1pt{--}\hskip .5pt{10$^{#2}$}\hskip 1.5pt{cm$^{-3}$}}
\def\fl[#1 #2]{{#1}$\pm${#2}}
\def\orb[#1 #2]{{$#1^{#2}$}}
\def\ls[#1 #2]{{$^{#1}${#2}}}
\def\tm[#1 #2 #3]{{$^{#1}${#2}$_{#3}$}}

\newcounter{ion}

\begin{document}

\title{Investigating the reliability of coronal emission measure distribution
  diagnostics using 3D radiative MHD simulations}

\author{Paola Testa$^{1}$}
\email{ptesta@cfa.harvard.edu}
\author{Bart De Pontieu$^{2}$}
\author{Juan Mart\'inez-Sykora$^{2,3}$}
\author{Viggo Hansteen$^{3}$} 
\author{Mats Carlsson$^{3}$}

\affil{$^1$ Smithsonian Astrophysical Observatory,60 Garden street, MS
  58, Cambridge, MA 02138, USA} 
\affil{$^2$ Lockheed Martin Solar and Astrophysics Laboratory, Org. A021S, 
  Bldg. 252, 3251 Hanover Street, Palo Alto, CA 94304, USA} 
\affil{$^3$ Institute of Theoretical Astrophysics, University of Oslo,
  P.O. Box 1029 Blindern, N-0315 Oslo, Norway} 

\begin{abstract}

Determining the temperature distribution of coronal plasmas can provide 
stringent constraints on coronal heating. Current observations with the 
Extreme ultraviolet Imaging Spectrograph onboard Hinode and the 
Atmospheric Imaging Assembly onboard the Solar Dynamics Observatory 
provide diagnostics of the emission measure distribution (EMD) of the 
coronal plasma. 
 
Here we test the reliability of temperature diagnostics using 3D radiative 
MHD simulations.   We produce synthetic observables from the models, 
and apply the Monte Carlo Markov chain EMD diagnostic. By comparing 
the derived EMDs with the ``true'' distributions from the model we assess 
the limitations of the diagnostics, as a function of the plasma parameters 
and of the signal-to-noise of the data.
 
We find that EMDs derived from EIS synthetic data reproduce some 
general characteristics of the true distributions, but usually show 
differences from the true EMDs that are much larger than the estimated 
uncertainties suggest, especially when structures with significantly 
different density overlap along the line-of-sight. When using AIA 
synthetic data the derived EMDs reproduce the true EMDs much less 
accurately, especially for broad EMDs. The differences between the 
two instruments are due to the: (1) smaller number of constraints 
provided by AIA data, (2) broad temperature response function of the 
AIA channels which provide looser constraints to the temperature 
distribution.
 
Our results suggest that EMDs derived from current observatories may 
often show significant discrepancies from the true EMDs, rendering 
their interpretation fraught with uncertainty. These inherent limitations 
to the method should be carefully considered when using these 
distributions to constrain coronal heating.

\end{abstract}


\section{Introduction}
\label{s:intro}

The heating mechanism that is responsible for the million degree solar
corona remains unknown, though several candidates exist. 
It is one of the most important open issues in
astrophysics. Constraining the properties of this heating mechanism is a
difficult task, but usually performed through spectral and imaging
observations of the solar corona, which provide
diagnostics of the plasma temperature distribution. The latter has
important implications for the energy balance of the corona
(see e.g., \citealt{Klimchuk06,Reale10} and references therein). 

The thermal distribution of the plasma - or emission measure
distribution, EMD (in section \ref{s:method} we define the 
relationship between EMD and the differential emission measure, 
DEM, which is also often used to describe the plasma thermal 
distribution) - is crucial to test heating models. For instance,
the EMD in coronal loops, strongly depends on the spatial and temporal 
properties of the energy release \citep[e.g.,][]{Klimchuk01,Cargill04,Testa05}. 
The presence of a high temperature ($T \gtrsim 5$MK) component, e.g.,
in the EMD of an active region, is a good tracer of the properties of
the heating \citep[e.g.,][]{Patsourakos06}, and it has recently been
addressed by several studies
\citep[e.g.,][]{Patsourakos06,Reale09,Reale09b,Ko09,Schmelz09a,Testa11,Testa12b}.
Also, the thermal distribution in the cross-field direction
can help discern between a
monolithic, single strand, loop (isothermal at a given location along
the loop; e.g., \citealt{Aschwanden00,Aschwanden05,Aschwanden11,
Delzanna03,Landi02,Landi06,Landi08}),
and a loop structure composed of several strands (multi-thermal
plasma; e.g., \citealt{Schmelz01,Schmelz05,Warren09,Brooks09}).

\begin{figure*}[!t]\vspace{0.5cm}
\centerline{\includegraphics[scale=0.45]{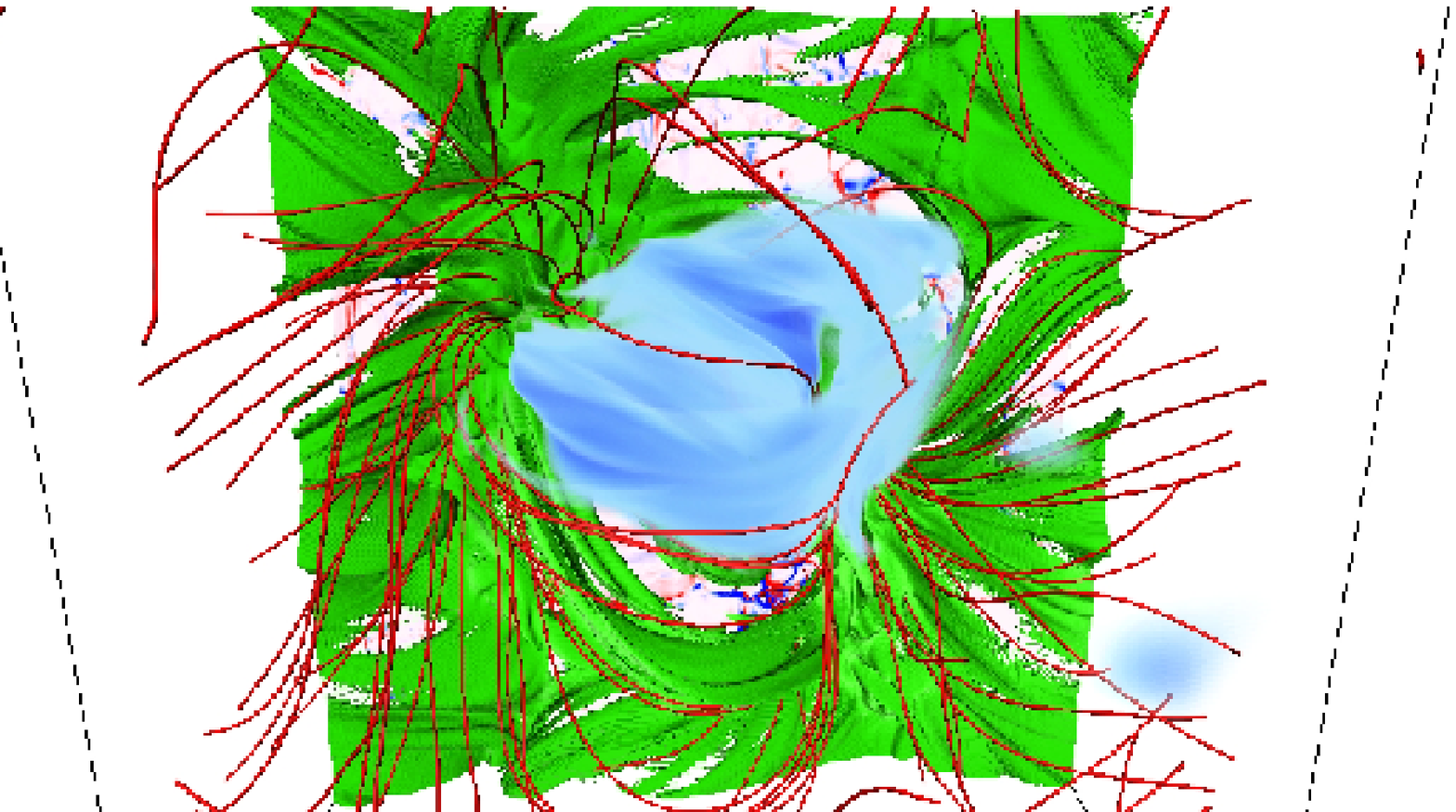}
  \includegraphics[scale=0.45]{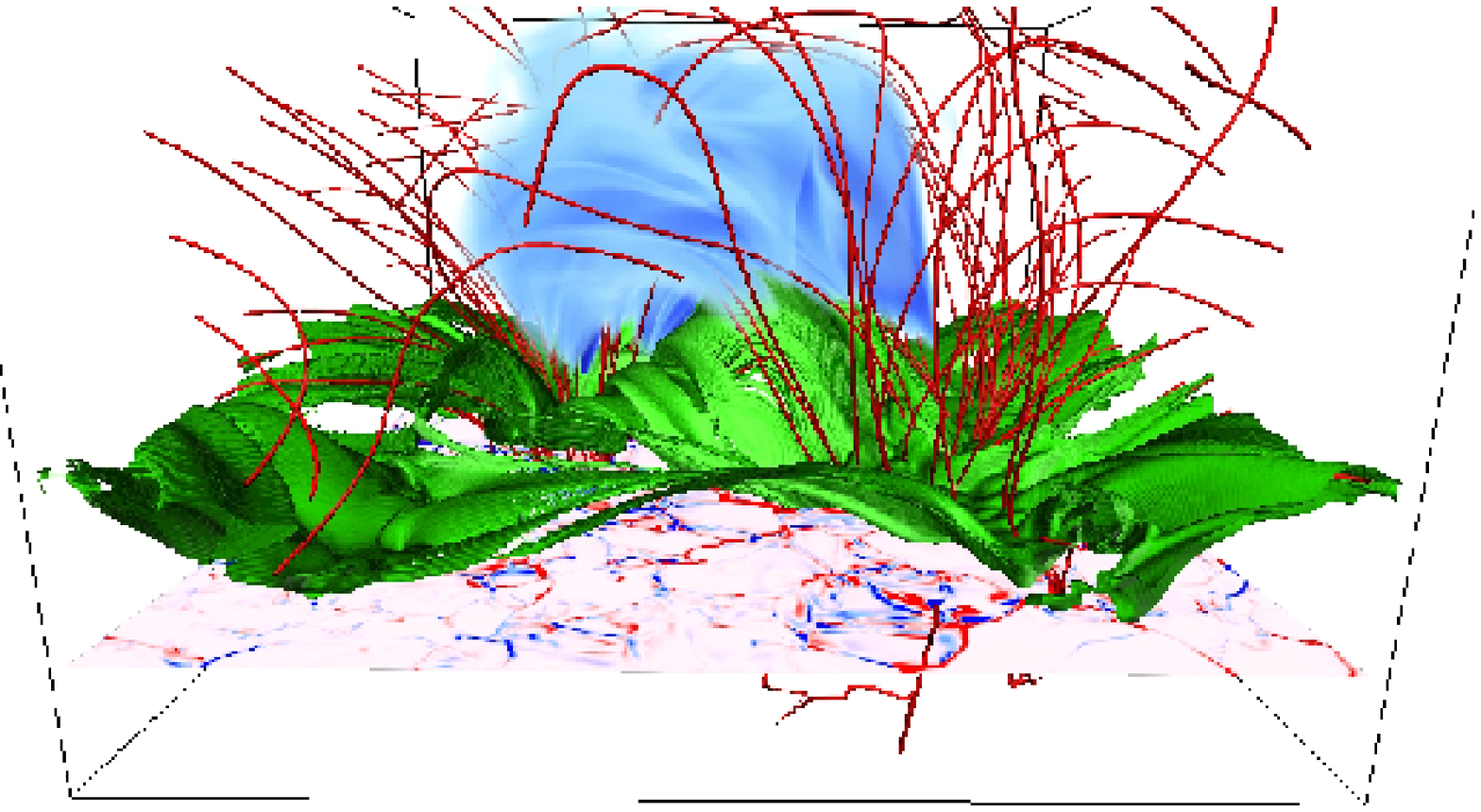}}
\caption{ Three-dimensional snapshot of one of the 3D rMHD models 
  (model H): top view (left panel) and side view (right panel).   Some 
  selected magnetic field lines are shown in red to give an indication of 
  the magnetic field topology in the corona. The isosurfaces of \fexii\ 
  and \caxvii\ emission are shown in green and blue respectively. 
  We also show at the photospheric layer (z=0~Mm) the strength of the 
  vertical magnetic field in blue-red color scale.
  \label{fig:sim_cb}} 
\end{figure*}

Ample efforts have been devoted to the accurate determination of the
thermal structuring of coronal plasma to derive robust observational
constraints on the coronal heating mechanism(s). 
The plasma temperature distribution of the quiet corona and of active
regions has been investigated through imaging data and spectroscopic
observations \citep[e.g.,][]{Brosius96,Landi98,Aschwanden00,Testa02,
DZM03,Reale07,Landi09,Shestov10,Sylwester10}. 
Several recent studies have focused on EUV spectra obtained with the 
\hinode\ Extreme Ultraviolet Imaging Spectrometer (\eis; \citealt{Culhane07})
which provides good temperature diagnostic capability, together with
higher spatial resolution and temporal cadence than previously 
available \citep[e.g.,][]{Watanabe07,Warren08loops,Patsourakos09,
Brooks09,Warren09,Testa11,Tripathi11}.
Imaging observations obtained by the {\em SDO} Atmospheric Imaging 
Assembly (AIA; \citealt{Lemen12}) in narrow EUV bands have also been
used for studying the plasma temperature distribution, especially in 
active region loops \citep[e.g.,][]{Schmelz10,Aschwanden11,
Aschwanden11b,Schmelz11},
also in conjunction with EIS data \citep[e.g.,][]{Brooks11,Warren11,Testa12b}.

The optically thin nature of the coronal EUV and X-ray emission
implies that the emission observed in the resolution element
(instrument pixel) is generally produced by several independent
structures along the LOS. Most plasma diagnostics rely on some
homogeneity assumptions (e.g., constant plasma density along the
LOS) whereas these overlapping structures can in principle have 
very different plasma parameters. This poses
significant challenges for interpreting the meaning of the derived
plasma parameters (which are by necessity weighted averages of the
distributions along the LOS) in terms of the actual physical
conditions of the plasma. Further assumptions typically made for
specific diagnostics (e.g., on the functional form and smoothness of
the EMD) also have a potentially significant impact on the accuracy of
the diagnostics.     
Several efforts have been carried out to test the accuracy of the
plasma temperature diagnostics. For instance, the notorious challenges
in determining the emission measure distribution and its confidence
limits have been addressed in several studies \citep[e.g.,][]{Craig76,
Judge97,McIntosh00,Judge10,LandiKlimchuk10,Landi12,Hannah12}. 
The main  limitation of these previous efforts lies in the necessarily
simplified test cases adopted.   

In this paper we use advanced 3D radiative MHD simulations of the
solar atmosphere with the Bifrost code \citep{Gudiksen11} to carry 
out detailed tests of coronal plasma temperature diagnostics.  
We focus on the plasma temperature diagnostics using current 
spectral ({\em Hinode/EIS}) and imaging ({\em SDO/AIA}) data. 
 We synthesize EIS and AIA data from 3D radiative MHD (rMHD)
simulations and analyze them like real data, and use the comparison of
the thermal distributions inferred from the synthetic data with the
``true'' distributions, which in the case of the simulations are
known, to carefully assess the accuracy of the diagnostics.
These 3D simulations provide us with the opportunity to improve upon
previous work by exploring more realistic configurations, with
significant superposition of different structures along the LOS,
allowing a statistical approach to determine the accuracy and
limitations of the plasma diagnostics, for a variety of spatial and
thermal structuring of the plasma.   
The three dimensional nature of the simulations also allows us to
explore a variety of realistic viewing angles, reproducing typical
distributions of structuring ranging from on disk to limb observations. 

In Section~\ref{s:method} we describe the analysis methods, we 
include a short description of the Bifrost code, and discuss the 
characteristics of the 3D rMHD models and of the corresponding 
synthetic observables used in this work. 
The analysis of the synthetic data and the results of the
determination of the plasma temperature distribution are presented and
discussed in Section~\ref{s:results}. We summarize our findings and
draw our conclusions in Section~\ref{s:conclusions}. 

\section{Analysis Methods}
\label{s:method}

In this work we investigate how accurately the thermal distribution
of the plasma can be inferred from coronal observations that are 
currently available. 
The observed intensities of a set of spectral emission lines
constrain the plasma temperature distribution, as they depend on the
abundance $A_Z$ of the emitting element, the plasma emissivity of the
spectral feature $G_{\lambda}(T,n_e)$ as a function of temperature $T$
and electron density $n_e$, and the differential emission
measure distribution $DEM (T)$:
\begin{equation}
I_{\lambda} = A_Z \int_{T} G_{\lambda}(T,n_e) DEM(T) \,dT
\label{eq:Iline}
\end{equation}
where $ DEM(T) = n_e^2 \,dV / dT$ $[{\rm cm}^{-3} {\rm K}^{-1}]$. 
Analogously, in the case of imaging observations in broad/narrow
passbands the observed intensity in a channel will depend on the
temperature response function, $R_{\rm chan}(T)$:
\begin{equation}
I_{\rm chan} =\int_{T} R_{\rm chan}(T) DEM(T) \,dT .
\label{eq:Ichan}
\end{equation}
Throughout the paper, instead of the DEM(T) we will discuss the
emission measure distribution \emt\, which is obtained by integrating
the differential emission measure distribution $DEM(T)$ in each
temperature bin; here we use a temperature grid with constant binning
in logarithmic scale ($\Delta \log T = 0.05$). 

\begin{figure*}[!t]
\centerline{\includegraphics[scale=0.42]{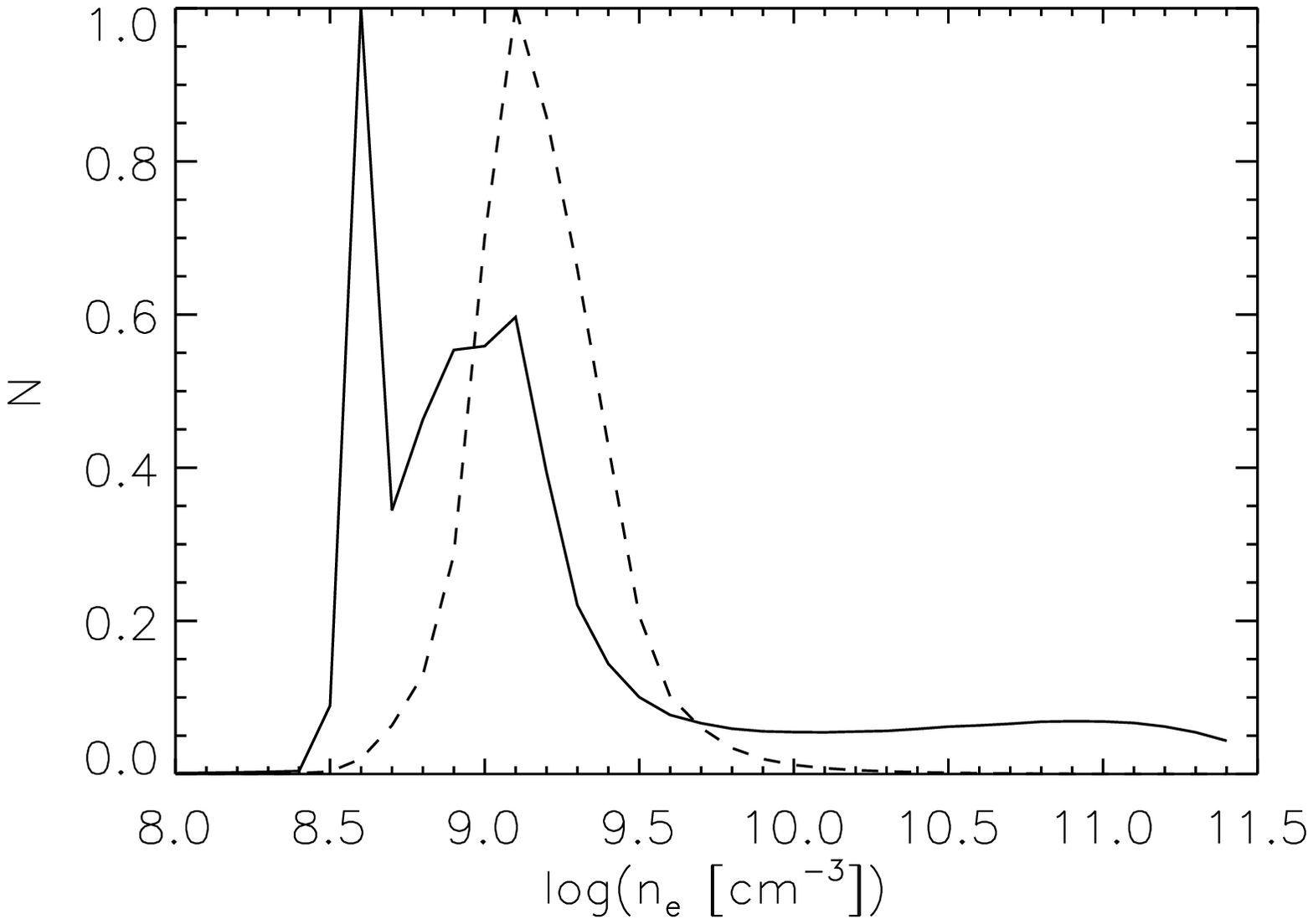}
  \includegraphics[scale=0.42]{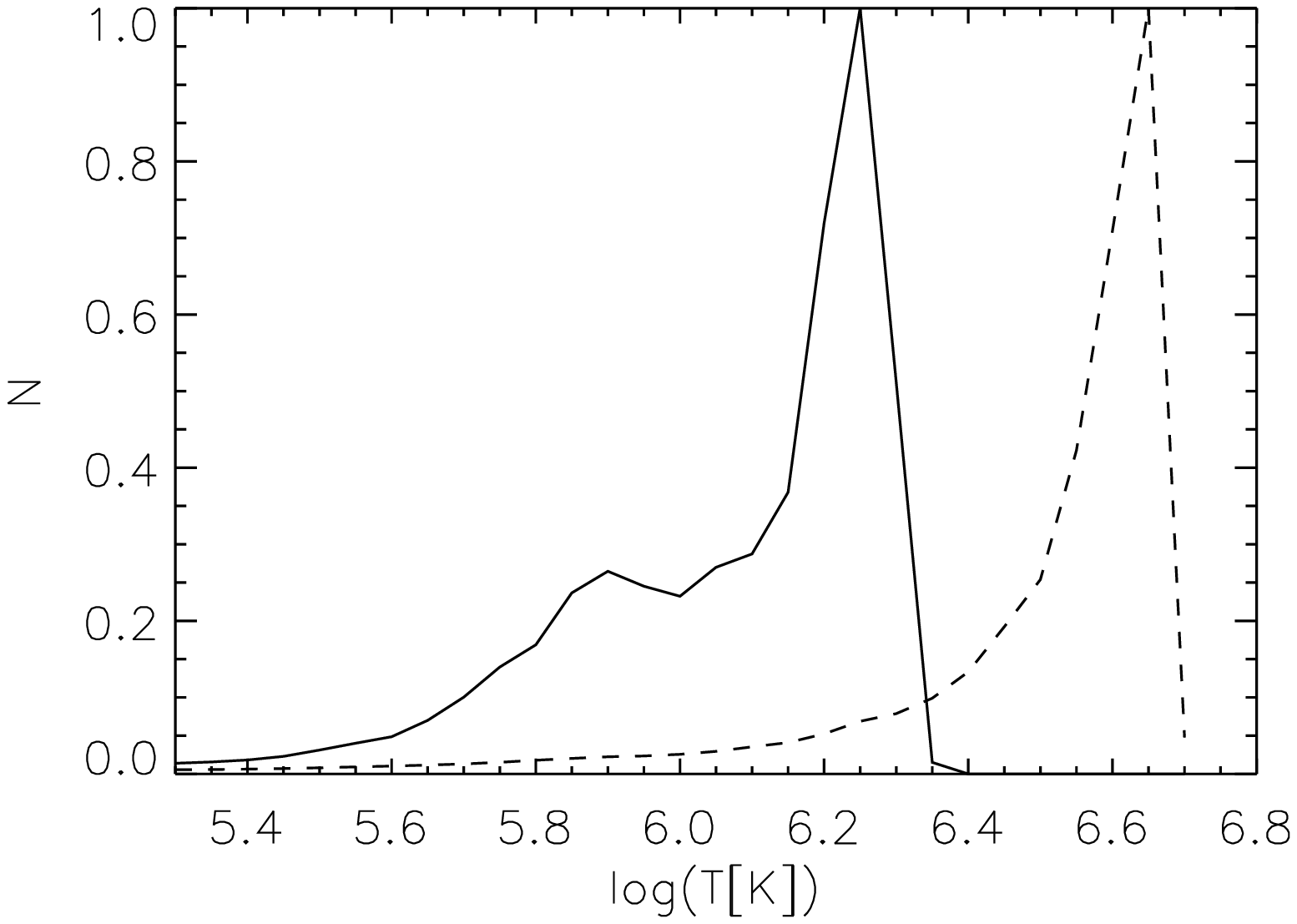}}
\caption{Histogram of the electron density (left panel) and temperature (right
  panel) for the two used snapshots (for the voxels with $\log T \geq
  5.3$). 
  The simulation ``C'' modeling the smaller region with emerging flux 
  (solid line) is characterized by lower coronal temperature than the larger 
  simulation ``H'' (dashed lines), and it has a broader distribution of 
  densities (where the tail at $\log n_e \gtrsim 10$ corresponds to dense 
  plasma in the emerging flux region). 
\label{fig:neT_hist_sims}}
\end{figure*}

Several methods have been developed to reconstruct emission
measure distributions from a set of observed intensities in lines, or
passbands in the case of imaging observations (see e.g., review
by \citealt{Phillips08}, and the discussion and references in 
\citealt{Hannah12}). 
Here we test the Monte Carlo Markov chain (MCMC, hereafter) forward 
modeling method \citep{Kashyap98}, which is widely used and 
considered to provide robust results (see e.g.,
\citealt{Landi12,Hannah12}; see also 
http://www.lmsal.com/$\sim$boerner/demtest/ for a recent comparative 
analysis of results from different methods, applied to AIA).
With respect to several other methods, the MCMC method has the 
advantages of not imposing a pre-determined functional form for the 
solution, and, most importantly, of estimating the uncertainties associated 
with the resulting emission measure distribution \citep[see e.g.,][for
additional details]{Kashyap98,Testa11}.  We use the Package for
Interactive Analysis of Line Emission (PINTofALE, \citealt{PoA}) 
which is available as part of SolarSoft. 
Though no pre-determined functional form is imposed, the 
MCMC method does apply some smoothness criteria which are 
locally variable and based on the properties of the temperature 
responses/emissivities for the used data, instead of being arbitrarily 
determined a priori.

\input{tab_lines}

\begin{figure*}[!t]\vspace{-0.5cm}
\centerline{\includegraphics[scale=0.45]{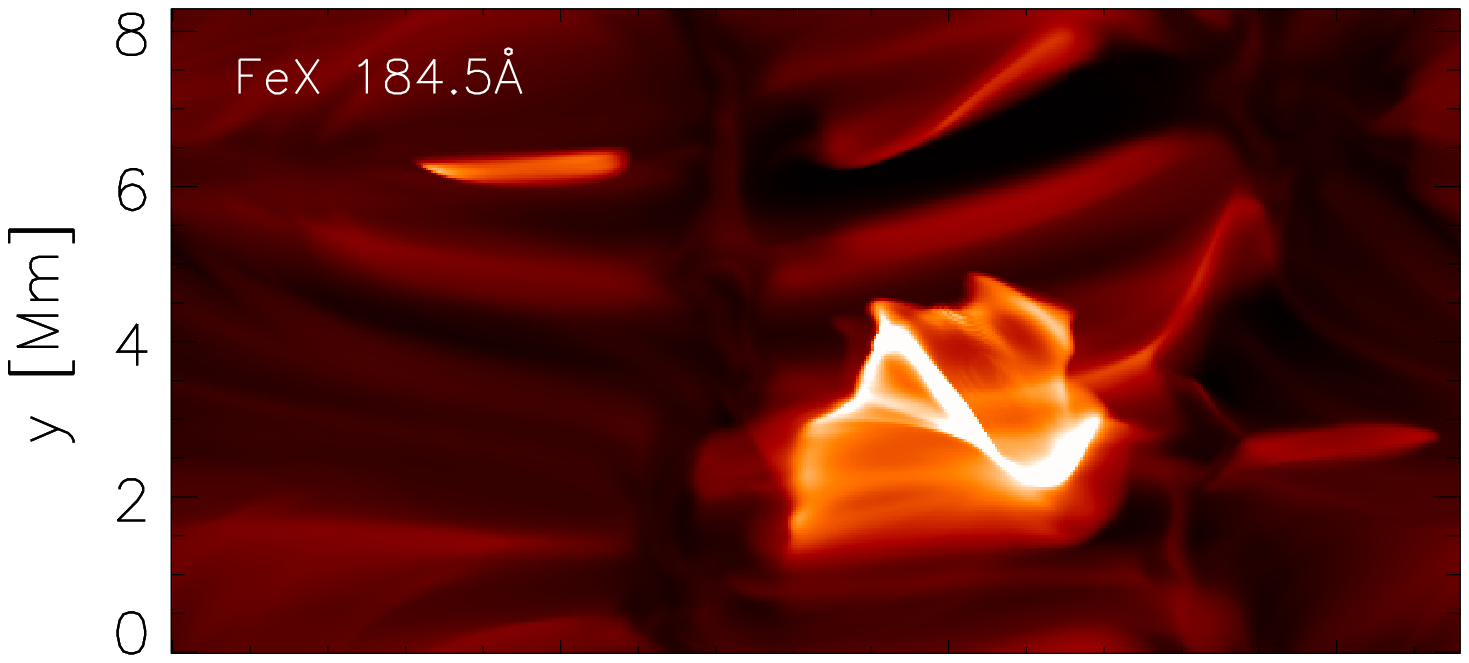}\hspace{-0.5cm}
  \includegraphics[scale=0.45]{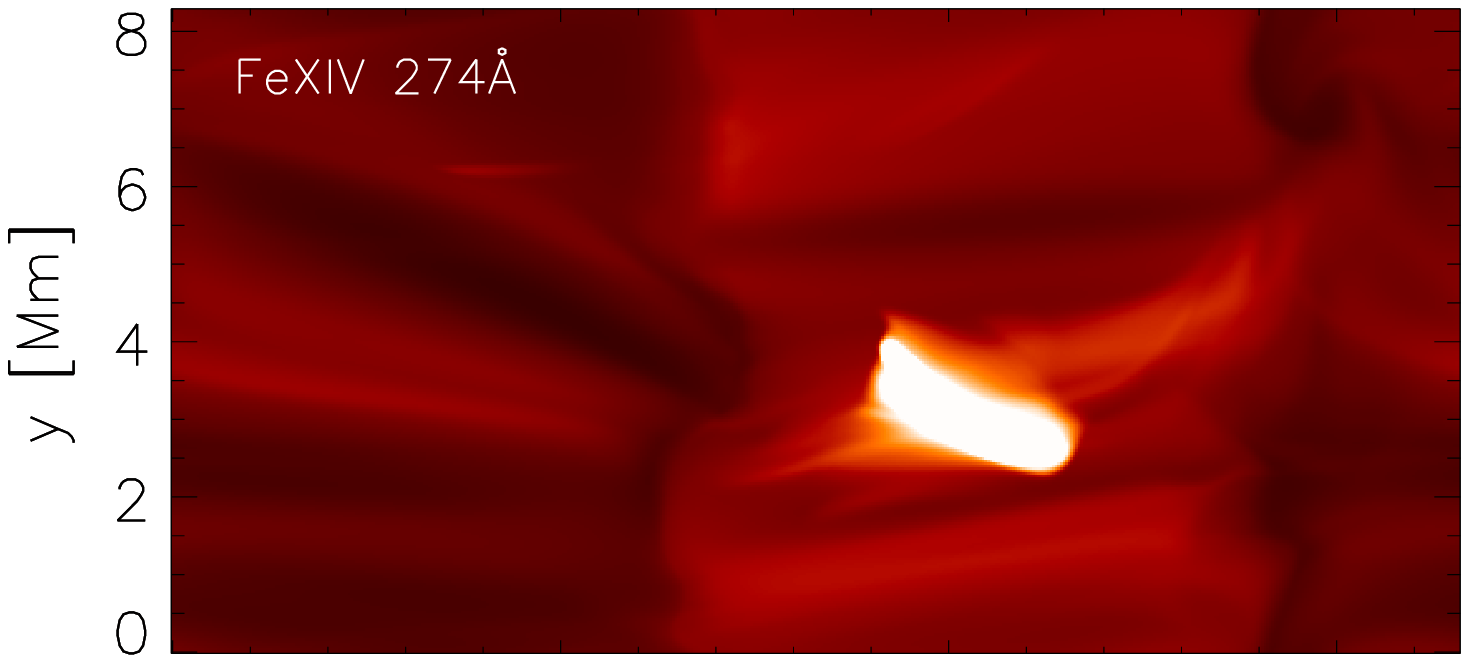}\vspace{-2.1cm}}
\centerline{\includegraphics[scale=0.45]{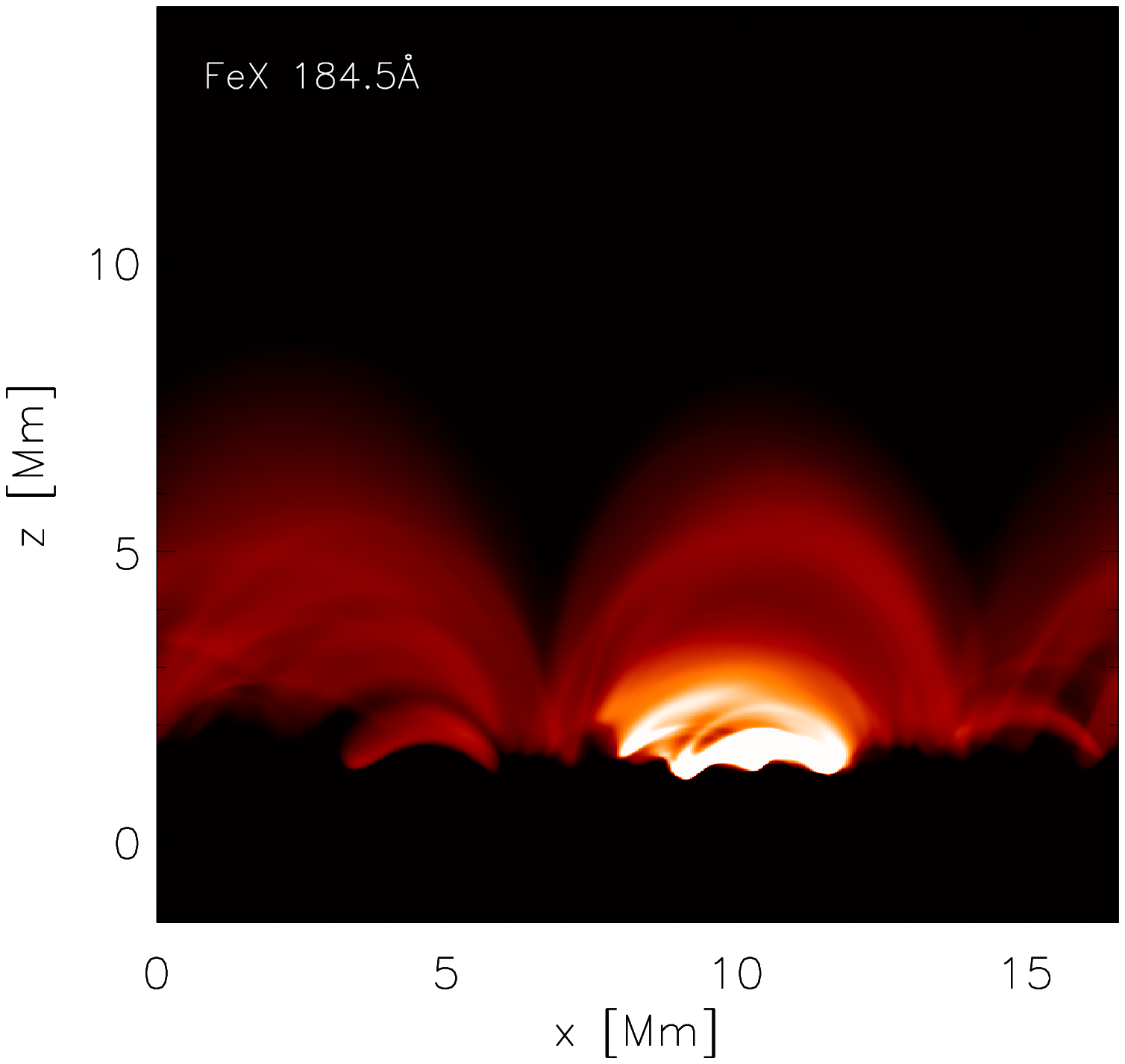}\hspace{-0.5cm}
  \includegraphics[scale=0.45]{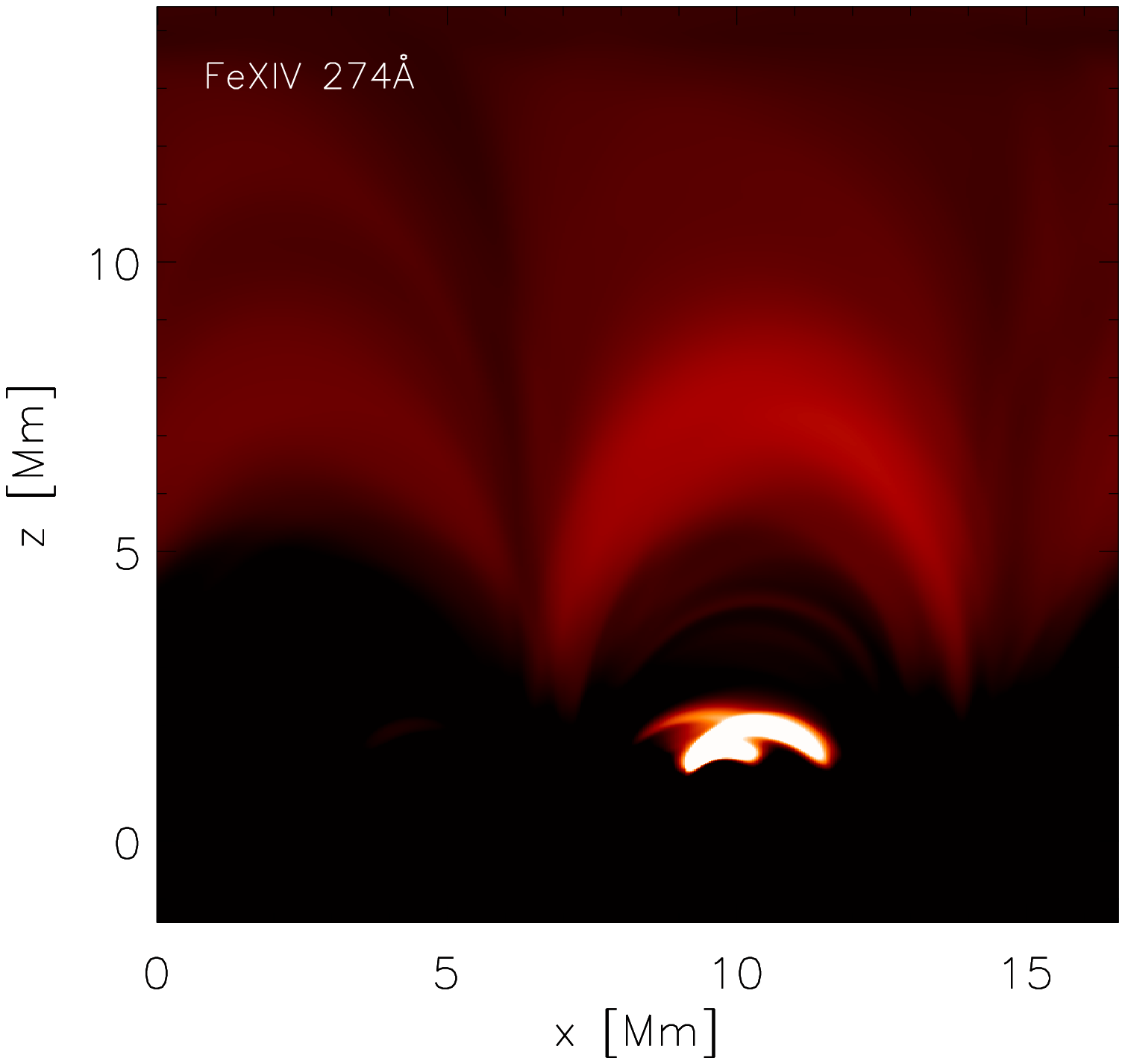}\vspace{-0.1cm}}
\caption{Synthetic top-view (top row) and side-view (bottom row) \fex\ 
  (184.5\AA) and \fexiv\ (274.2\AA)  images from the small cooler snapshot,
  at the intrinsic resolution of the simulation.  \label{fig:syndat_j}}
\end{figure*}

For a given simulation snapshot, which provides the electron density 
\dens\ and temperature values for each grid point (voxel) in a three 
dimensional box (see section \ref{ss:model} for a description of the
characteristics of the simulations we analyzed), and a selected
LOS we proceed as follows: 
\begin{itemize}
\item Using the \dens\ and T values, and the emissivities from CHIANTI
  \citep{chianti,chianti6} we synthesize intensities of a set of EIS
  lines and in the 6 AIA coronal channels, using the optically thin 
  approximation and statistical ionization equilibrium;
\item by integrating through the box along the LOS, and degrading the
  spatial resolution to the instrument resolution, we obtain synthetic
  coronal images in the different lines/passbands;
\item we consider two cases: with or without photon (Poisson)
  noise; i.e., in the latter case we randomize the intensities according to
  the photon counting statistics.
  The noise level and the uncertainties associated with the intensities are
  calculated by assuming signal-to-noise ratios typical of actual observations;
\item we use the intensities and uncertainties, in each pixel, as
  input for the MCMC routine to calculate the \emt\ solution, pixel by pixel;
\item we analyze the results, using several parameters to assess the
  ability of the method to reproduce the input intensities and the
  ``true'' \emt,  which in these test cases are known.
\end{itemize}
 In the following subsection, section \ref{ss:model}, we describe in detail
 the choices and assumptions made, and the characteristics of the
 selected simulations and of the resulting synthetic data.

\subsection{3D Models and Synthetic Observables}
\label{ss:model}

The model considered spans from the upper layer of the convection 
zone up to the low corona, and self-consistently produces a 
chromosphere and hot corona, through Joule dissipation of electrical 
currents.
In order to investigate the robustness of the \emt\ reconstruction
method for a wide range of plasma conditions, we selected snapshots
from two simulations that are characterized by significantly different 
parameters (strength and spatial distribution of the seed magnetic 
field, and dimensions of the box), which leads to significantly different
distributions of temperature and density throughout the box. 

We analyze one of the snapshots from the simulation that was 
previously used to investigate the relative contribution of different 
lines to the emission in the AIA passbands \citep{MartinezSykora11}.
This simulation covers a volume ($x,y,z$, where $z$ is the vertical
direction) of $16 \times 8 \times 16$~Mm$^3$ ($512 \times 256 \times
365$ grid points; with $\Delta x = \Delta y \sim 32$km, and $\Delta
z$ non-uniform and smaller where gradients are large with
values $\Delta z \approx 28$km up to a height of 4Mm and increasing to 
$\approx 150$km at the top of the computational box), and is highly
dynamic yielding coronal temperatures that increase with time. We used an
intermediate snapshot ($t=1200$s), where the distribution of
temperature peaks around $\log (T[K]) \sim 6.25$, and $\log T_{\rm
  max} \sim 6.4$. As we discussed in \cite{MartinezSykora11}, the 
plasma conditions of this simulation represent a good comparison for 
quiet Sun/coronal hole conditions, with some small emerging flux 
regions, e.g., small bright points. In the following we use the label 
C to refer to this cooler and smaller snapshot.

The other snapshot we consider (H, hereafter), is from a simulation 
modeling a larger region ($24 \times 24 \times 16$~Mm$^3$; 
$768 \times 768 \times 768$ grid points; $\Delta x = \Delta y \sim 31$km,
$\Delta z \approx 14$km up to a height of 4Mm and increasing to 
$\approx 80$km at the top of the computational box; see 
Figure~\ref{fig:sim_cb}), 
with a magnetic field configuration consisting of two small regions of 
opposite polarities similar to a small active region (Carlsson et al., in 
preparation). 
The distribution of coronal temperatures is also similar to typical 
active region values, with a peak around $\log (T[K]) \sim 6.6$, 
and $\log T_{\rm max} \sim 6.7$.

Figure~\ref{fig:neT_hist_sims} shows the distributions of electron density
and temperature throughout the simulation box, for the two selected
snapshots, including only plasma with $\log T$ higher than 5.3.
These distributions show that simulation ``C'', modeling the smaller 
region with emerging flux, is characterized by lower coronal temperature 
than the larger simulation ``H'', and it has a broader distribution of 
densities. 

\begin{figure*}[!t]
\centerline{\includegraphics[scale=0.45]{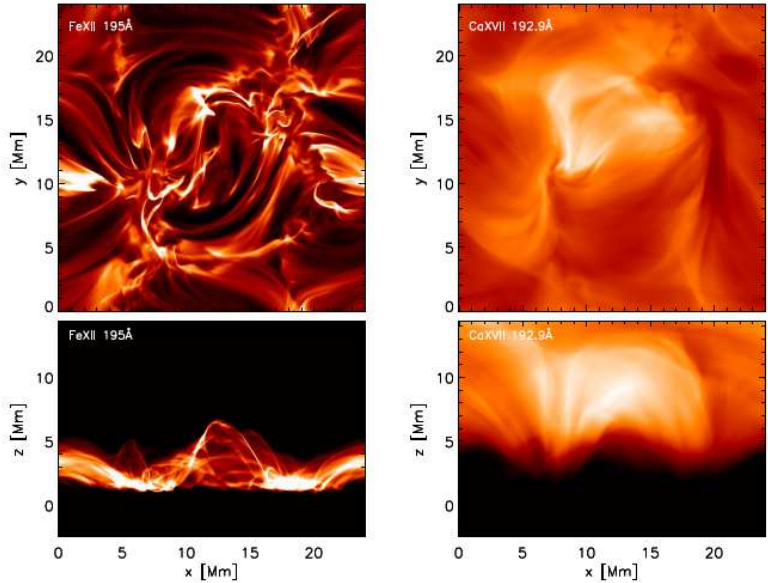}\vspace{-0.1cm}}
\caption{Synthetic top-view (top row) and side-view (bottom row) \fexii\ 
  (195.1\AA) and \caxvii\ (192.8\AA) images, at the intrinsic resolution of 
  the simulation, from the larger and hotter snapshot (H) of the numerical 
  simulation containing a plage-like region, representative of the core of 
  a small active region. 
  \label{fig:syndat_m}}
\end{figure*}

Intensities of a selection of EIS lines and AIA passbands were
computed for each voxel of the 3D simulation box $V_{xyz}$, using the
plasma electron density and temperature values and using the density
and temperature dependent contribution functions $G(T,n_e)$ from
CHIANTI \citep{chianti,chianti6}, assuming coronal abundances
\citep{Feldman92}, and the ionization fractions of chianti.ioneq. 
For AIA we compute synthetic intensities in the coronal passbands
(94\AA, 131\AA, 171\AA, 193\AA, 211\AA, 335\AA), by calculating the
spectra (line by line, taking into account the temperature and density
dependence of the contribution function for each line, and adding the 
continuum emission) and folding
them through the effective area of each channel \citep{Boerner12}.
In the case of EIS we selected a set of lines that provide a good
coverage of the temperature range $\log T \sim 5.6-6.7$, but using a
relatively limited number of lines, representative of a typical EIS
observation: we selected 14 lines, which are listed in
Table~\ref{tab:lines}. 

Images are then derived for each snapshot, and for two different
LOS, by integrating the emission through the simulated box along
the LOS: along the horizontal direction $y$ for the ``side view'',
and along the vertical direction $z$ in the case of the ``top
view'', analogously to the case presented in
\cite{MartinezSykora11}. Figures \ref{fig:syndat_j} and
\ref{fig:syndat_m} show examples of such images, at the intrinsic 
resolution of the simulations.
In addition, we degrade the spatial resolution to the instrumental spatial
resolution: 0.6~arcsec/pixel for the AIA synthetic data, and
1~arcsec/pixel for EIS. 

\begin{figure*}[!t]\vspace{0.3cm}
\centerline{\includegraphics[scale=0.45]{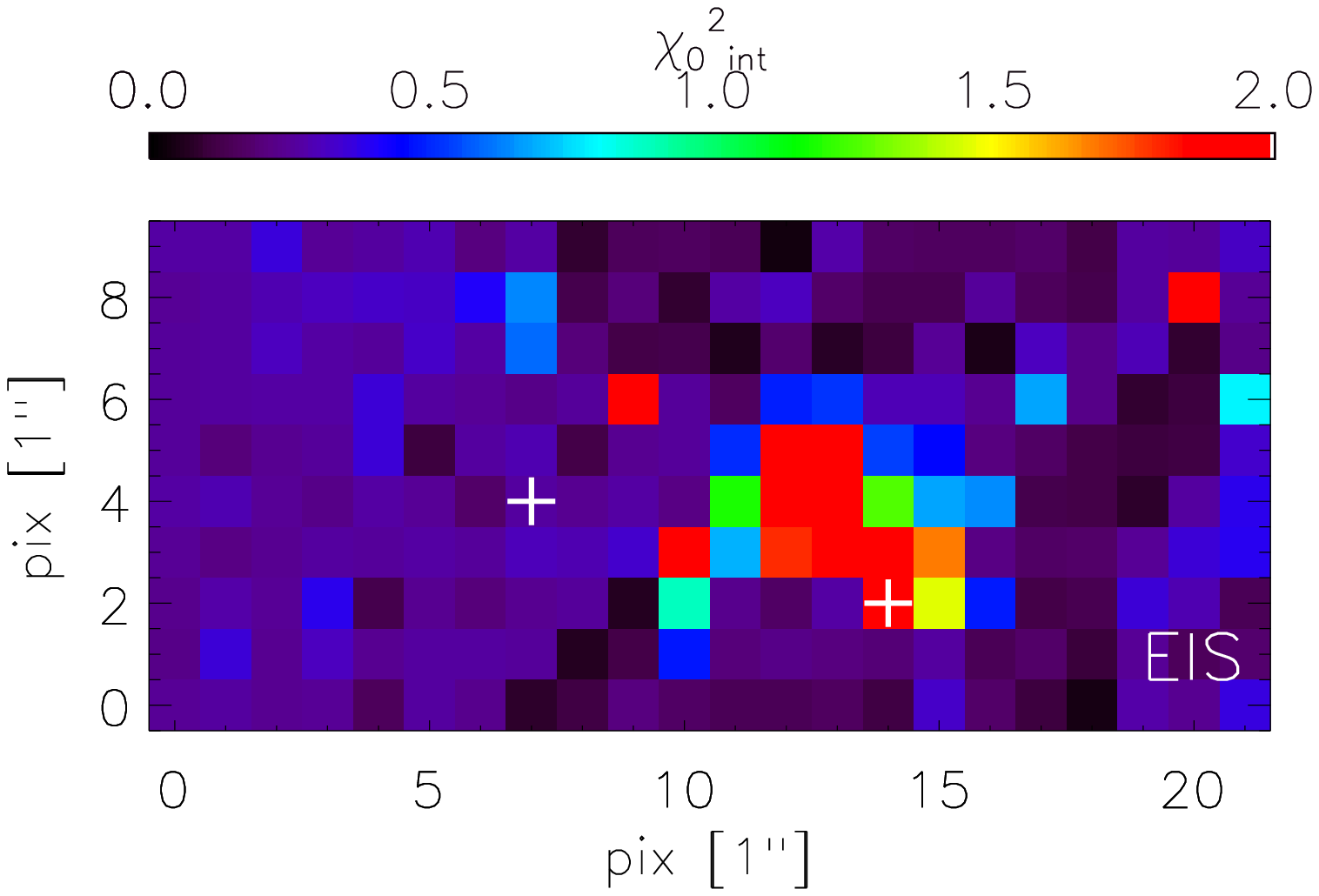}\hspace{-0.5cm}
  \includegraphics[scale=0.45]{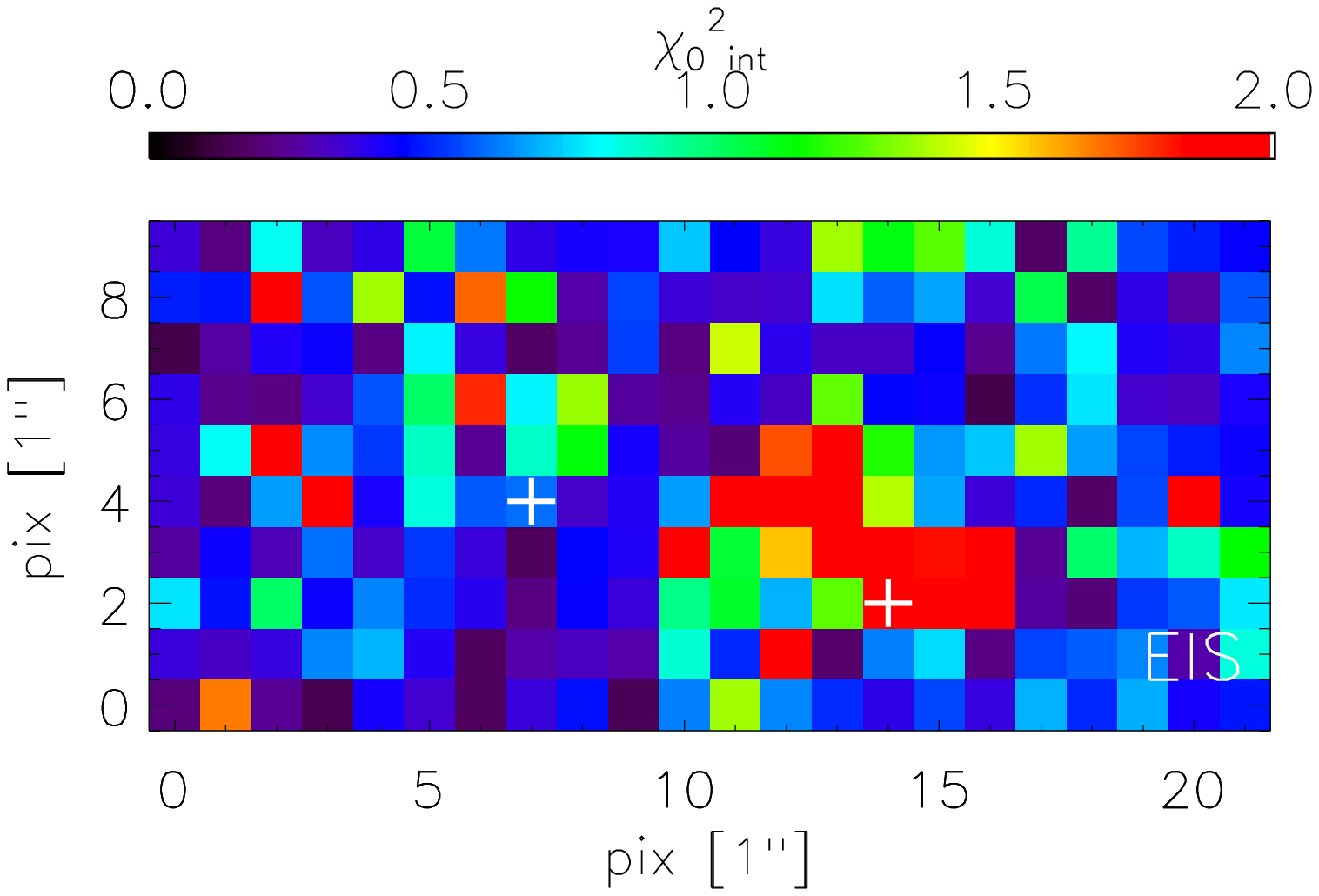}}
\centerline{\includegraphics[scale=0.45]{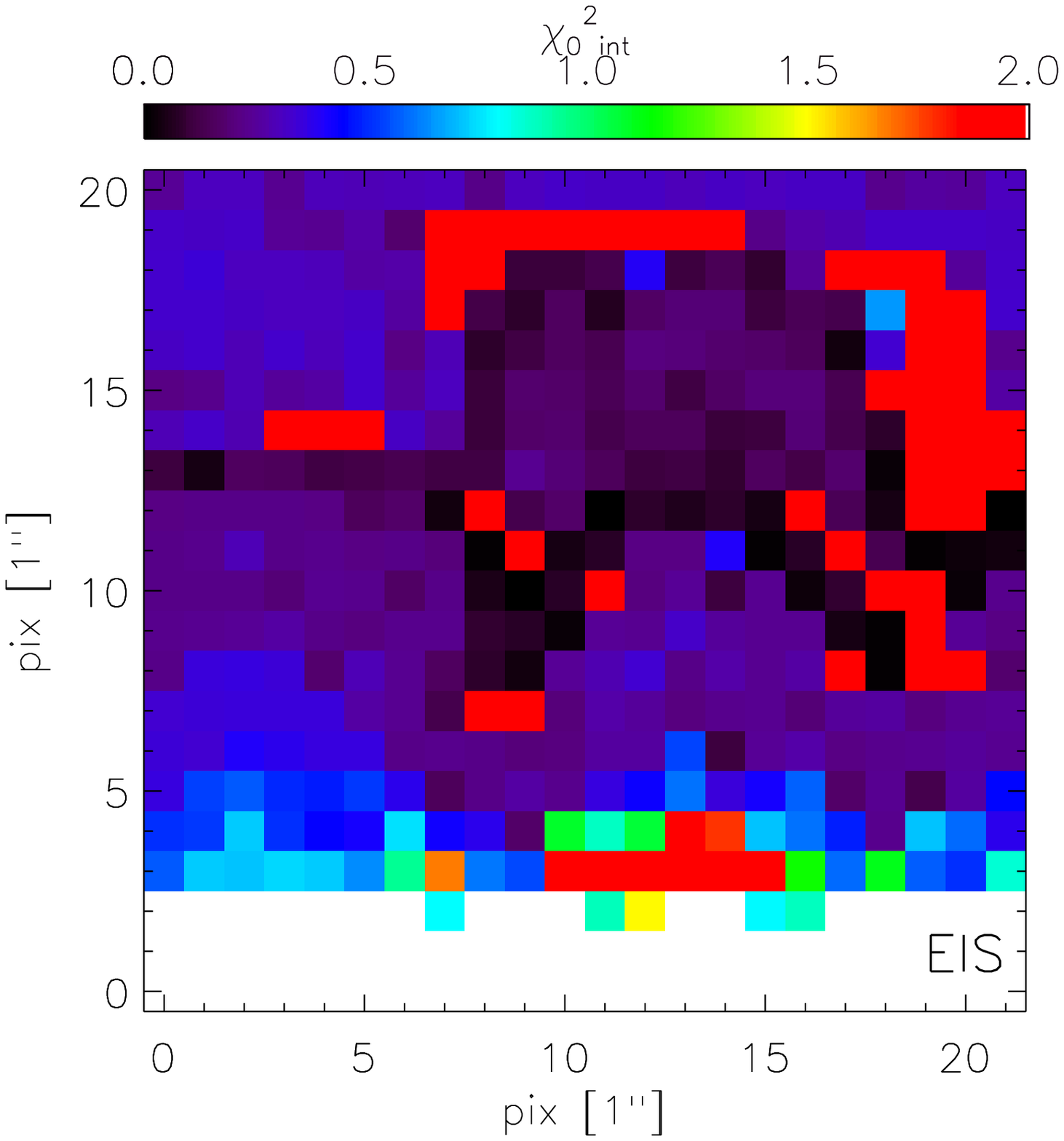}\hspace{-0.5cm}
  \includegraphics[scale=0.45]{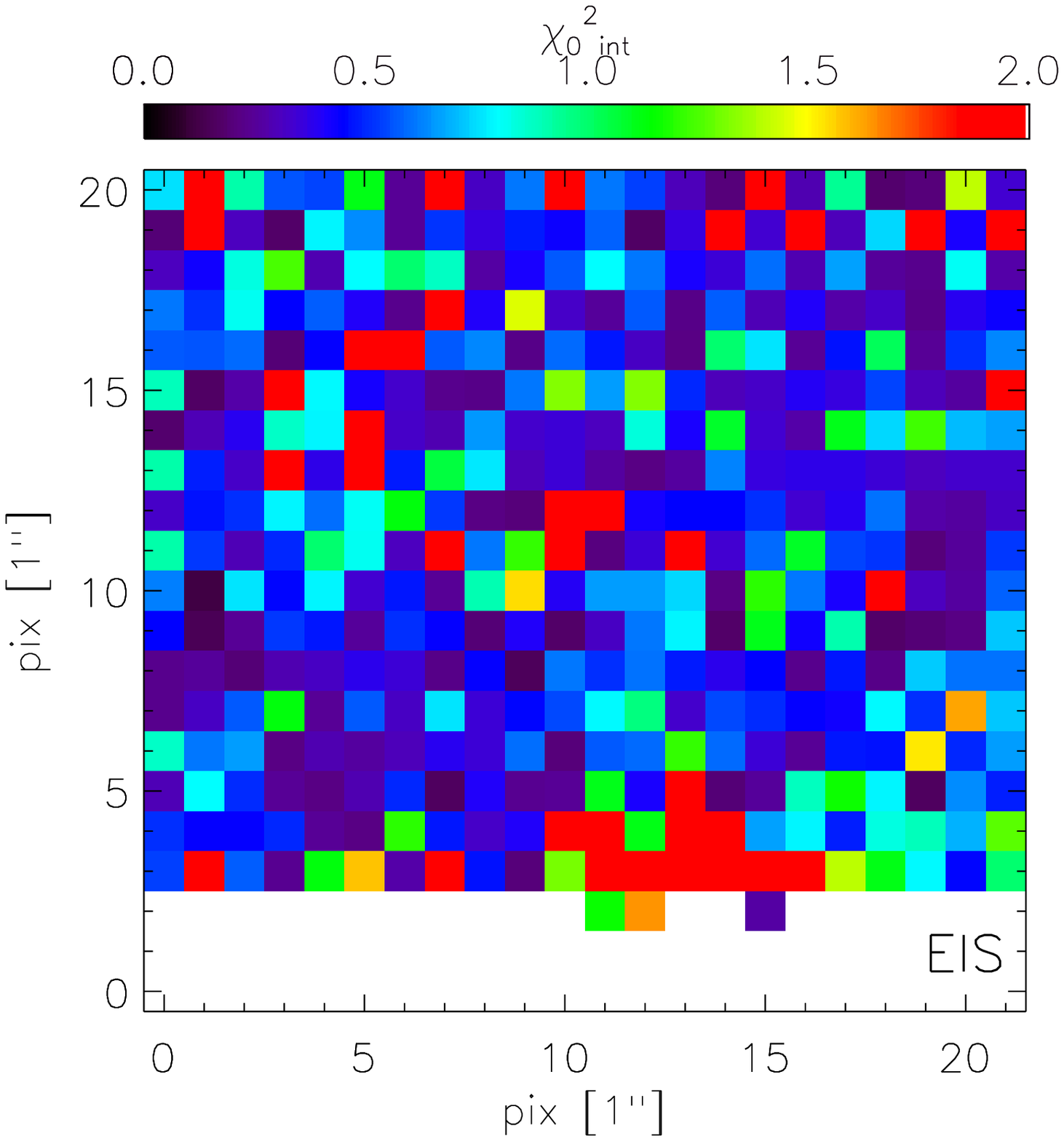}}\vspace{-0.3cm}
\caption{Maps showing the $\chi^2_0$ for the EMD derived
  using the EIS lines listed in Table~\ref{tab:lines}, for snapshot C, for the
  two LOS, ``xy'' (top), and ``xz'' (bottom). The right column shows the case 
  where Poisson noise has been included. The crosses mark two pixels with 
  significantly different $\chi^2_0$ that will be analyzed in more detail later 
  in the paper (see Figure~\ref{fig:neT_hist}). 
  Pixels where the EMD is not calculated - because less than 5 lines/channels
  have non-negligible intensity - are plotted in white color.
  \label{fig:chi2_e_dyn}}
\end{figure*}

In order to compute the uncertainties of the simulated intensities we
assume a signal-to-noise ratio (S/N) typical of well exposed active region
observations \citep[e.g.,][]{Reale11}. 
For the bright channels of AIA (171\AA, 193\AA, 211\AA) 
we use the DN pix$^{-1}$ assuming typical exposure times in a single 
image ($t_{\rm exp} \sim 2$s).
For the other channels (94\AA, 131\AA, 335\AA) which typically have 
significantly fainter emission we use the S/N calculated assuming 
the exposure of 10 summed images (corresponding to 
$t_{\rm exp} \sim 30$s) .
In the case of EIS, we calculate the S/N by scaling the images so that
the \fexii\ 195\AA\ line, which is usually one of the brightest emission 
lines in \eis\ observations, has 500 counts, as in typical observations, 
and computed the uncertainties from photon counting statistics.  
We first considered the case without Poisson noise, and then included 
the Poisson noise by randomizing the intensity values assuming the 
photon counting statistics of typical AIA and EIS observations as 
described above. 

In appendix~A we present a comparison of the synthetic 
intensities derived from the simulations, as described above, with 
measured intensities from recent SDO/AIA observations. This comparison
shows that the simulations here considered yield emission values of 
similar order of magnitude of real observations, and therefore provide
a sensible test case for the coronal diagnostics.

\section{Results}
\label{s:results}

Following the procedure described in detail in section \ref{s:method}, we analyzed the 
synthetic data to diagnose the plasma temperature distributions, pixel by
pixel, using the MCMC reconstruction method.  We will first discuss the results 
obtained using the EIS synthetic data, and then describe the results obtained 
with the AIA intensities.

\begin{figure*}[!t]\vspace{0.3cm}
\centerline{\includegraphics[scale=0.45]{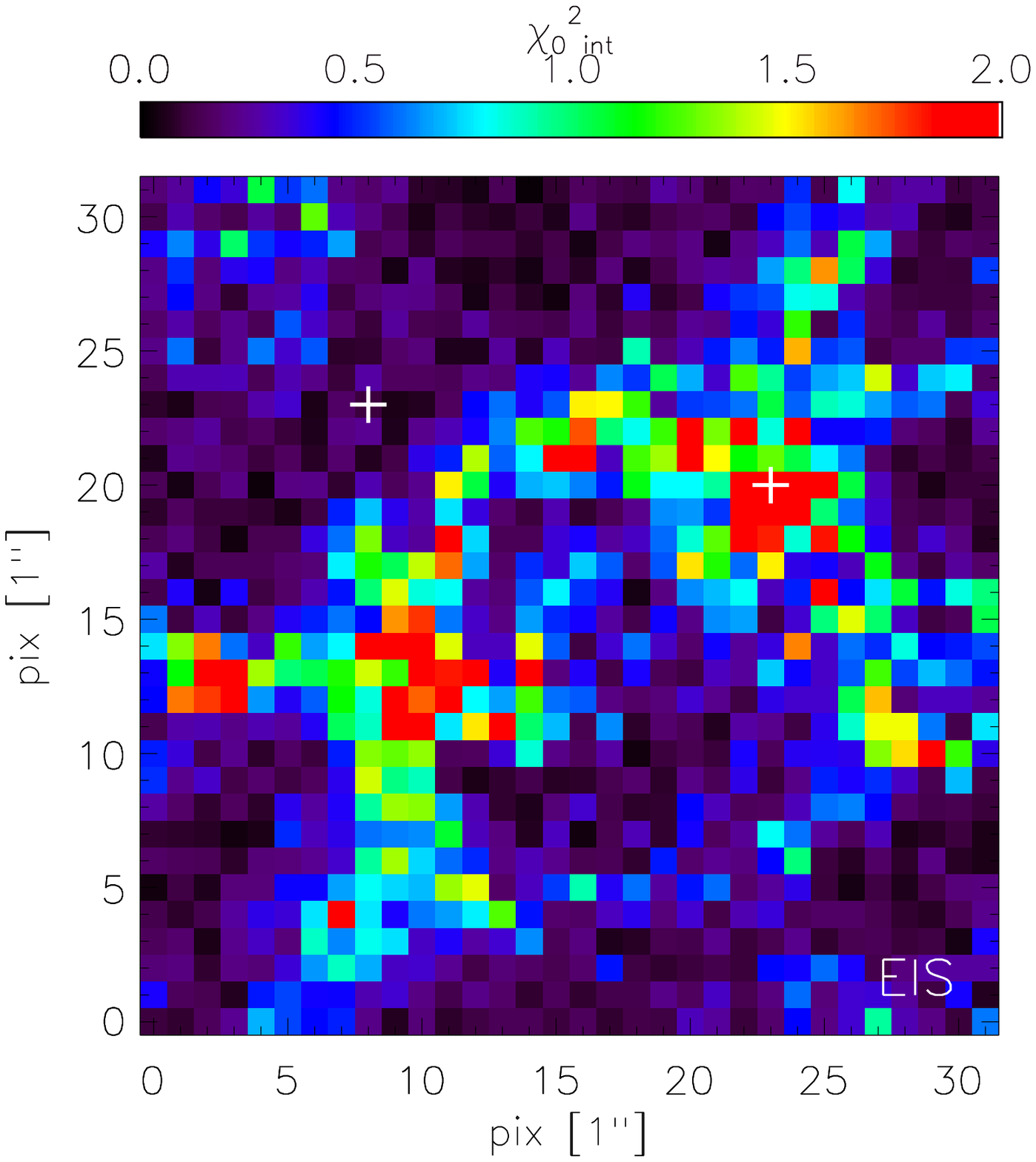}\hspace{-0.5cm}
  \includegraphics[scale=0.45]{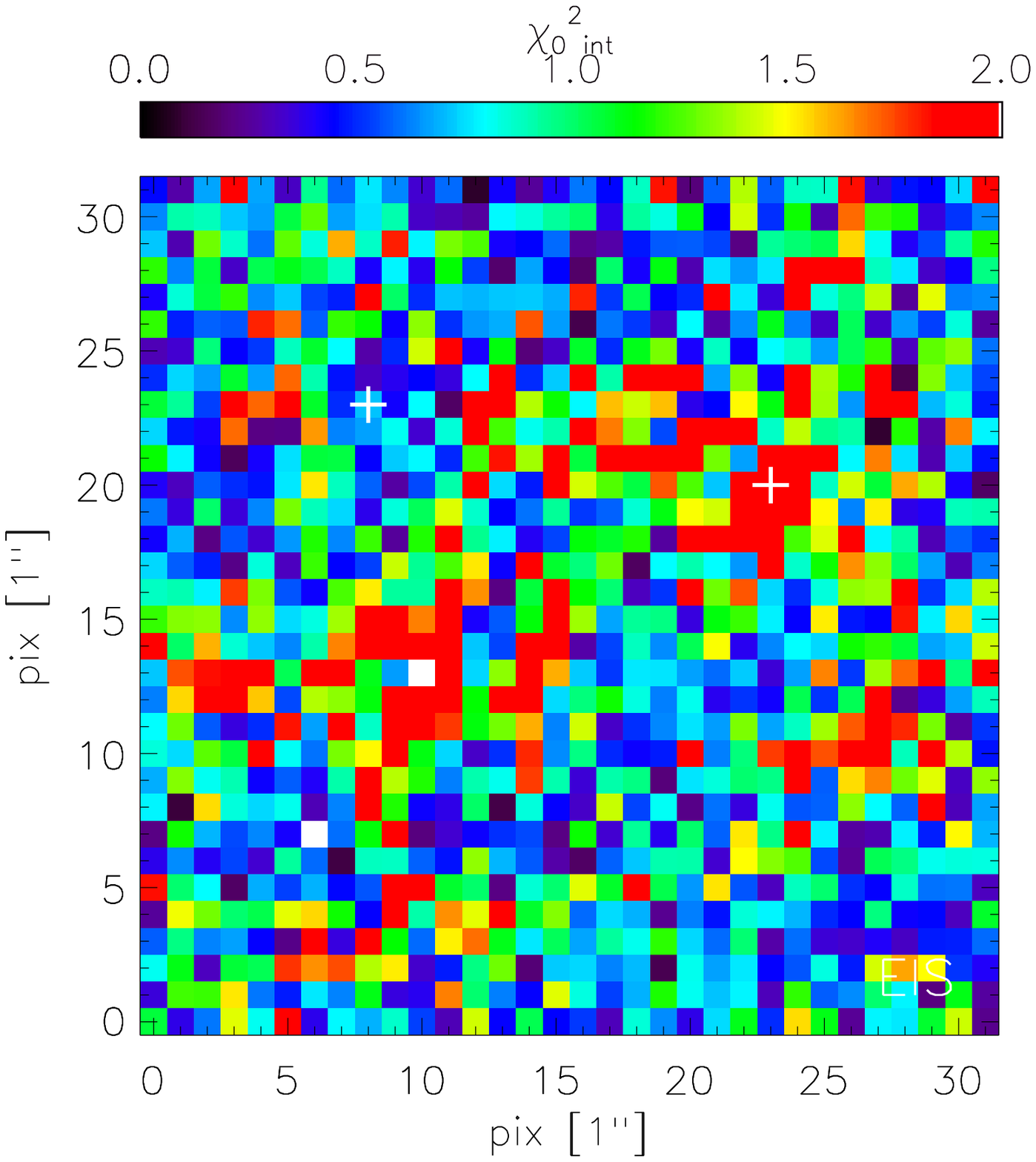}}
\centerline{\includegraphics[scale=0.45]{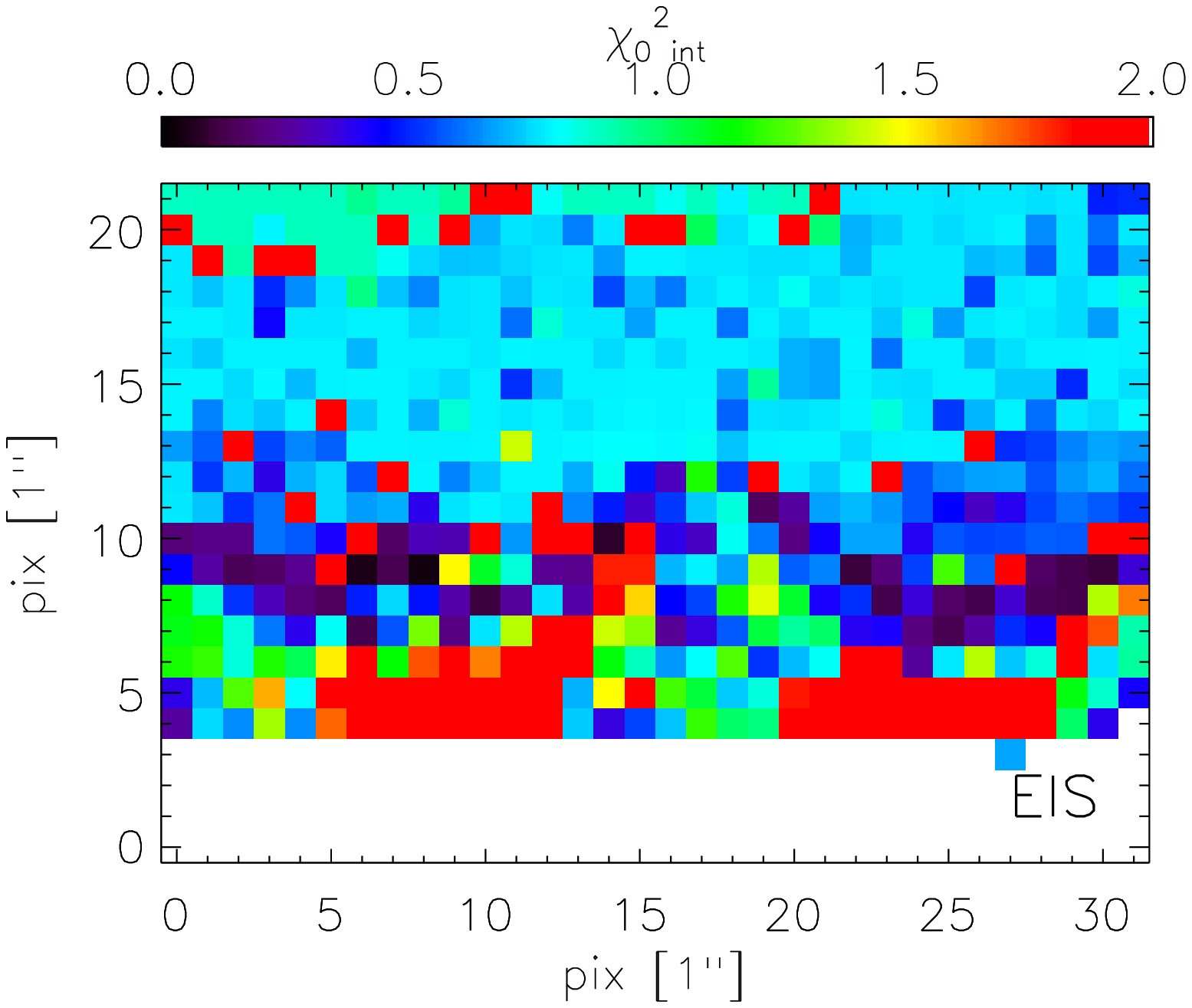}\hspace{-0.5cm}
  \includegraphics[scale=0.45]{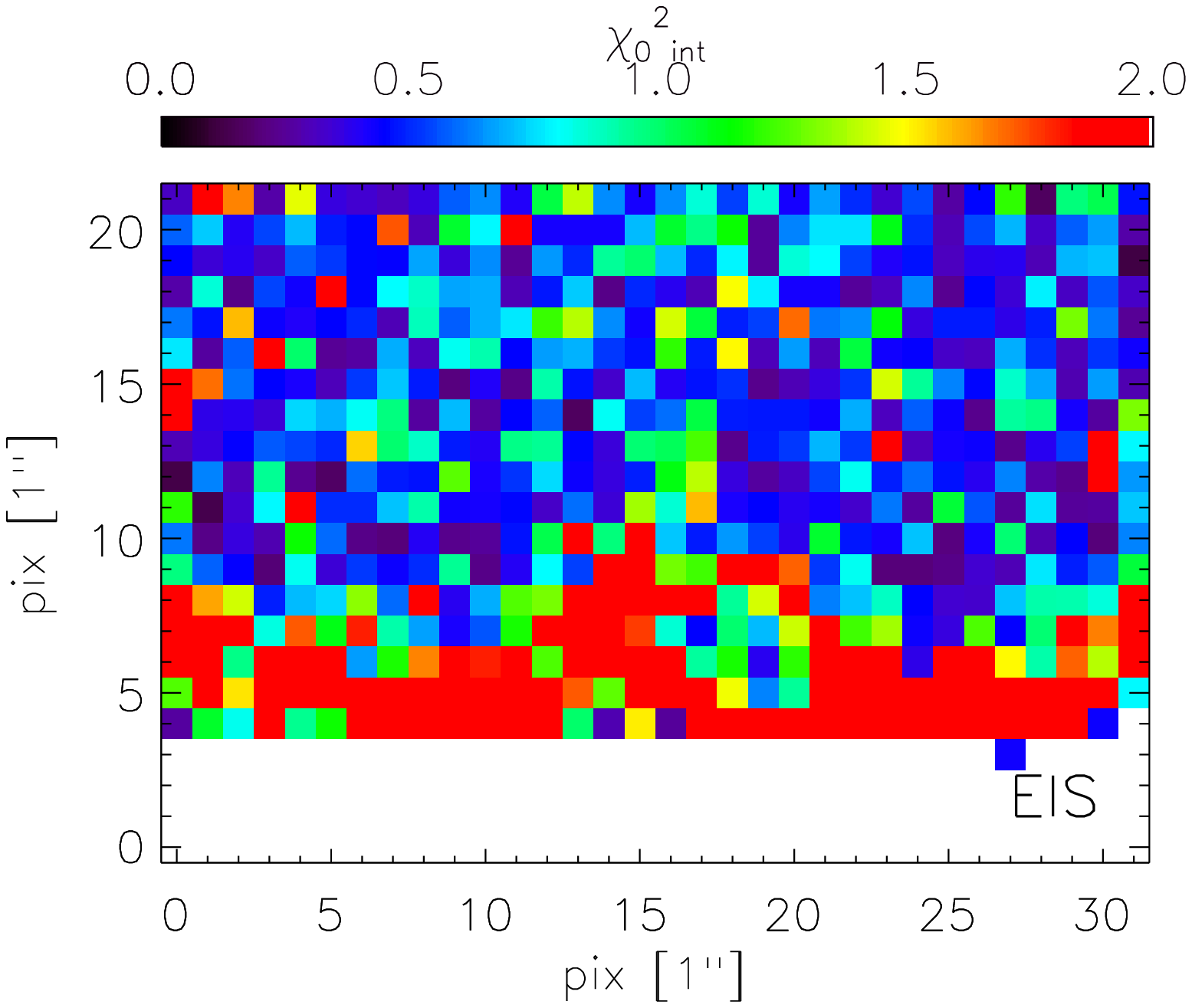}}\vspace{-0.3cm}
\caption{$\chi^2_0$ maps analogous to Figure~\ref{fig:chi2_e_dyn} but for 
  snapshot H. \label{fig:chi2_e_cb}}
\end{figure*}

When analyzing real data, the only measure of the goodness of the inferred
emission measure distribution consists in the agreement between the 
measured intensities and the emissions predicted using the best fit EMD.
Therefore, as a first step, we show the maps of $\chi^2_0$
( $= \Sigma_j [(I_{j, pred}-I_{j, obs})/\sigma_j]^2/df$, where $I_{j,
  obs}$ and $\sigma_j$ are the synthetic intensity and associated
error for the line/channel j, $I_{j, pred}$ is the intensity predicted
by the best fit EMD and $df$ are the degrees of freedom\footnote{The
  degrees of freedom are calculated as described in the PINTofALE documentation
http://hea-www.harvard.edu/PINTofALE/doc/MCMC\_DEM.html\#out.}), for both
snapshots and both LOS, using the EIS synthetic intensities 
(snapshots C and H in Figure~\ref{fig:chi2_e_dyn}, and \ref{fig:chi2_e_cb}, respectively). 
We show both the case without Poisson noise ({\em left}), and including 
Poisson noise ({\em right}). 
In both cases the errors on the intensities, $\sigma_j$, are calculated
as the Poisson error on the counts values. 
For both snapshots the $\chi^2_0$ maps 
indicate that the inferred EMD on average reproduce the ``measured'' fluxes 
adequately. The case without noise is characterized by $\chi^2_0$ lower 
than 1 over large areas, while areas with large $\chi^2_0$ present clear
structuring, which is still present in the case including the effect of noise.

For the top view case, the poorer results in those areas appear to be related 
to the plasma electron density. To investigate this effect, we use the
diagnostic based on the \fexii\ line ratio 186.88\AA/195.12\AA, 
which is one of the most useful density diagnostics accessible with 
EIS spectra \citep[e.g.,][]{Young09}, and included in many EIS 
observing programs.

\begin{figure}[!t]
\centerline{\includegraphics[scale=0.42]{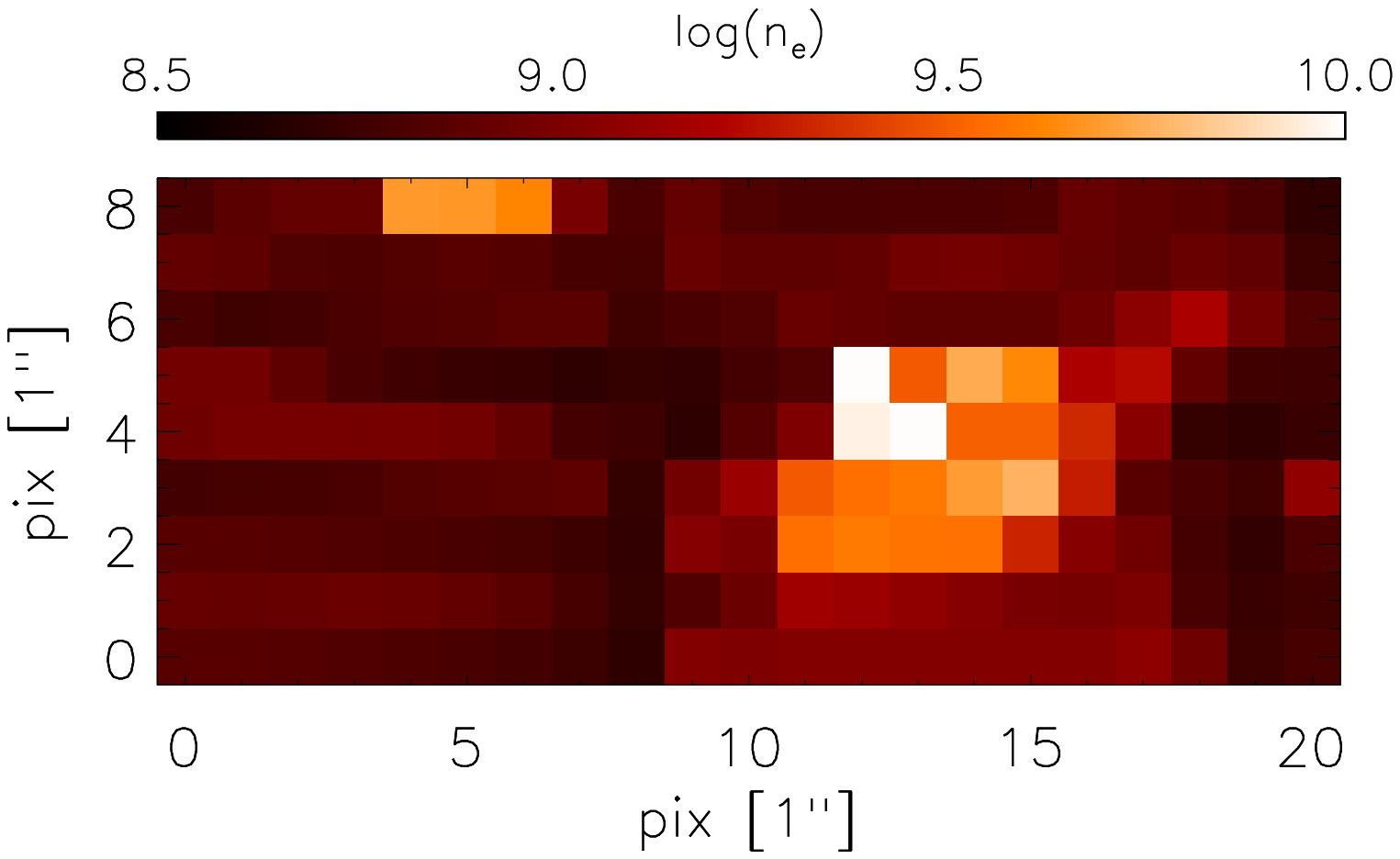}}\vspace{-0.3cm}
\centerline{\includegraphics[scale=0.42]{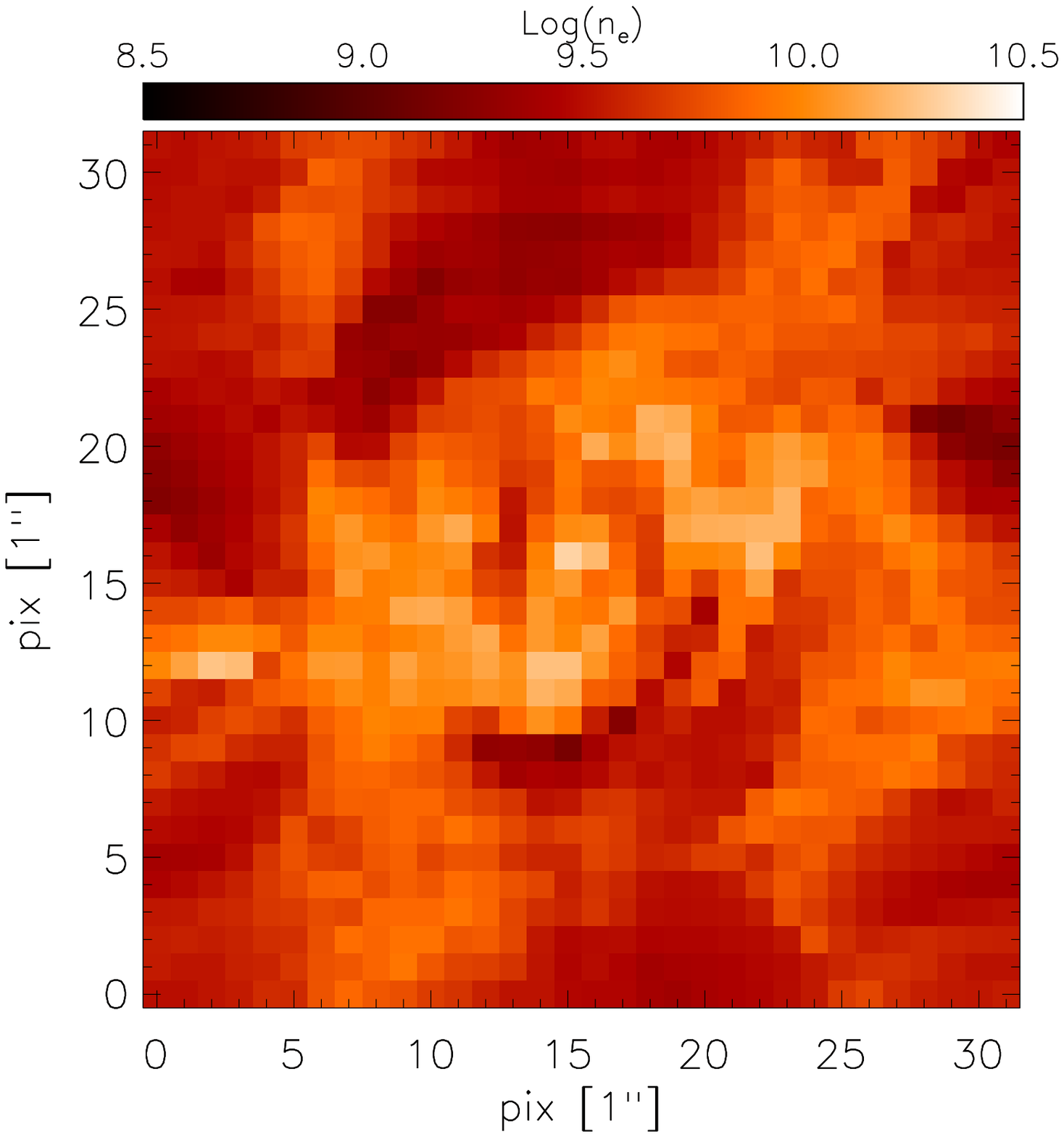}}
\caption{Maps of electron density derived from synthetic data
  for the two snapshots (C, top, H, bottom), for the top view, 
  using the \fexii\ diagnostics (186\AA/195\AA) available with 
  EIS observations (see e.g., \citealt{Young09}).  \label{fig:ne_e}}
\end{figure}

In Figure~\ref{fig:ne_e} we show the top view map of electron density 
obtained from the ratio of the synthetic \fexii\ intensities, computed 
from the  two snapshots.   By comparing these density maps to the 
$\chi^2_0$ plots (top row, Figures~\ref{fig:chi2_e_dyn} and 
\ref{fig:chi2_e_cb}), it is clear that the high density regions, such as for 
instance the emerging flux region of snapshot C or footpoint (``moss'') 
regions of snapshot H, generally correspond to high $\chi^2_0$ areas. 

\begin{figure*}[!t]\vspace{0.5cm}
\centerline{\includegraphics[scale=0.33]{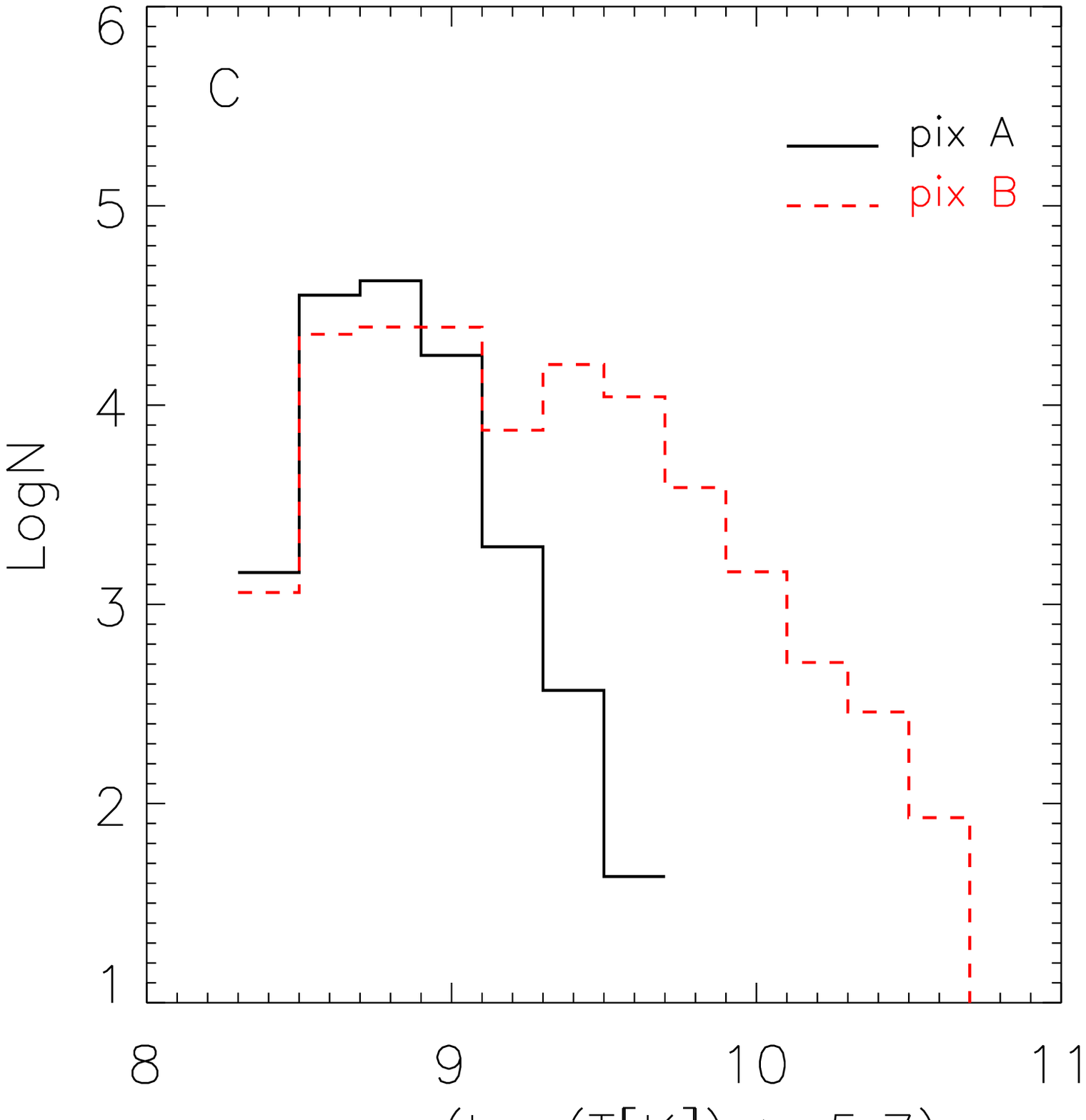}
  \includegraphics[scale=0.33]{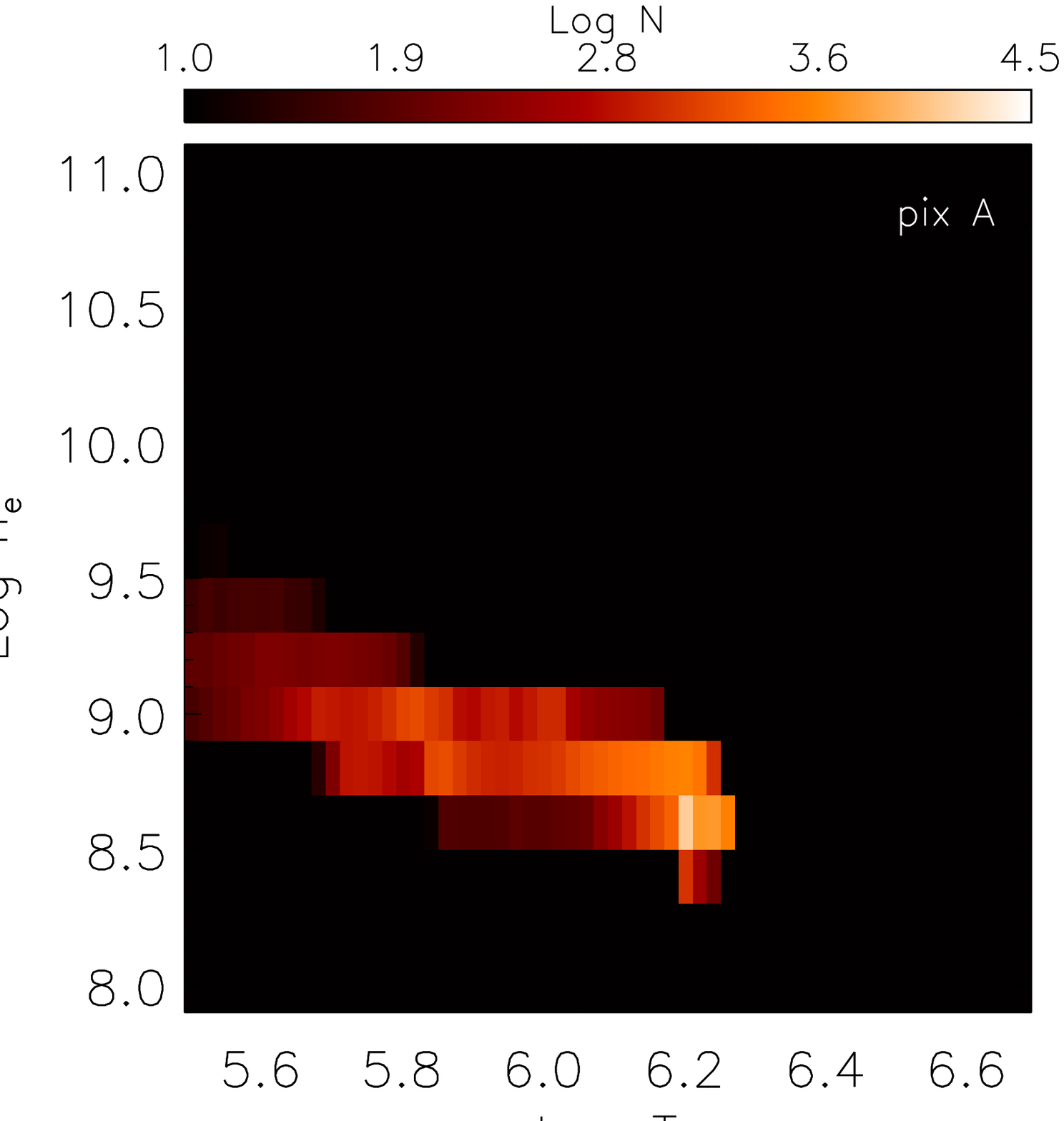}
  \includegraphics[scale=0.33]{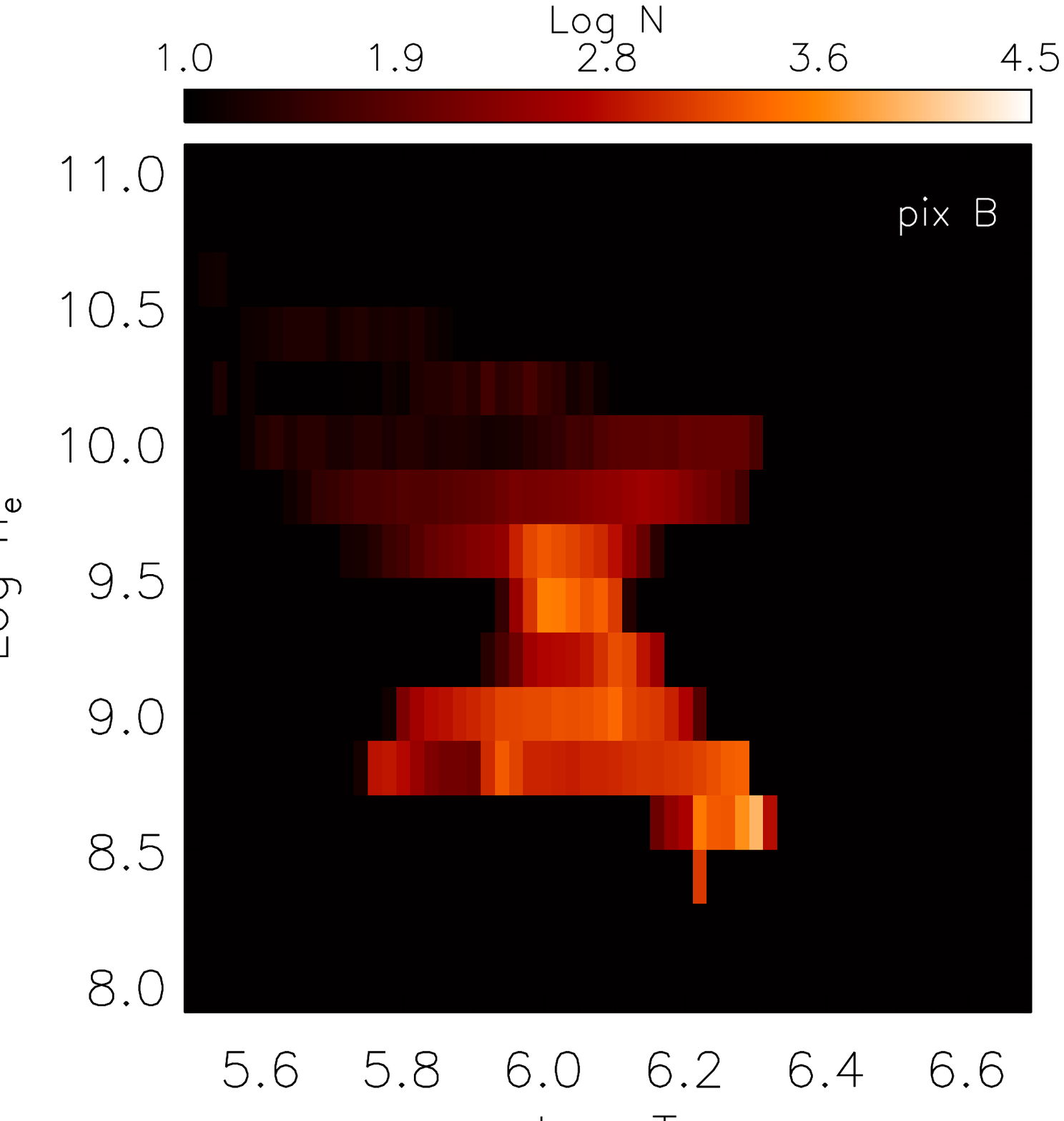}\vspace{0.5cm}}
\centerline{\includegraphics[scale=0.33]{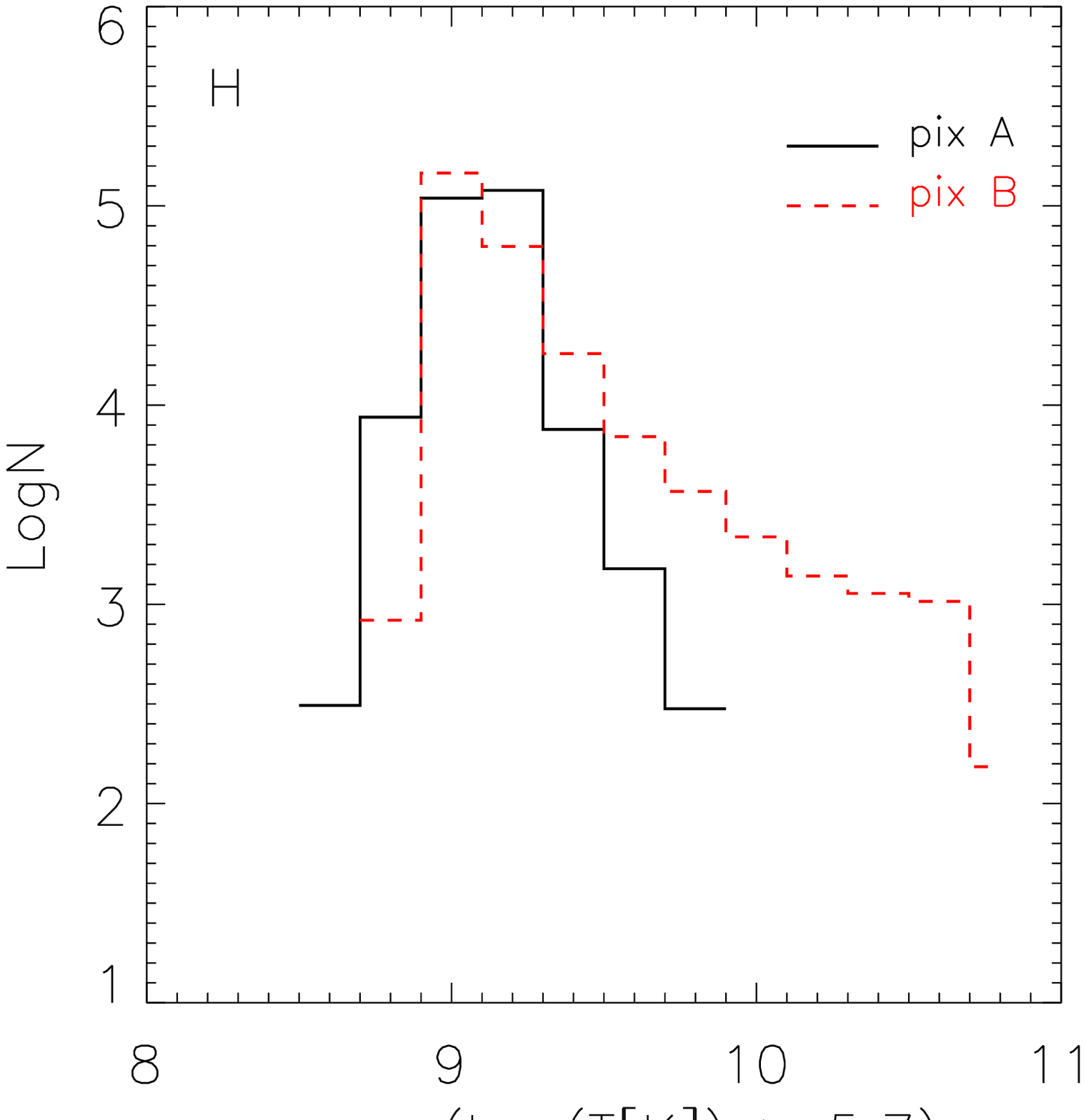}
  \includegraphics[scale=0.33]{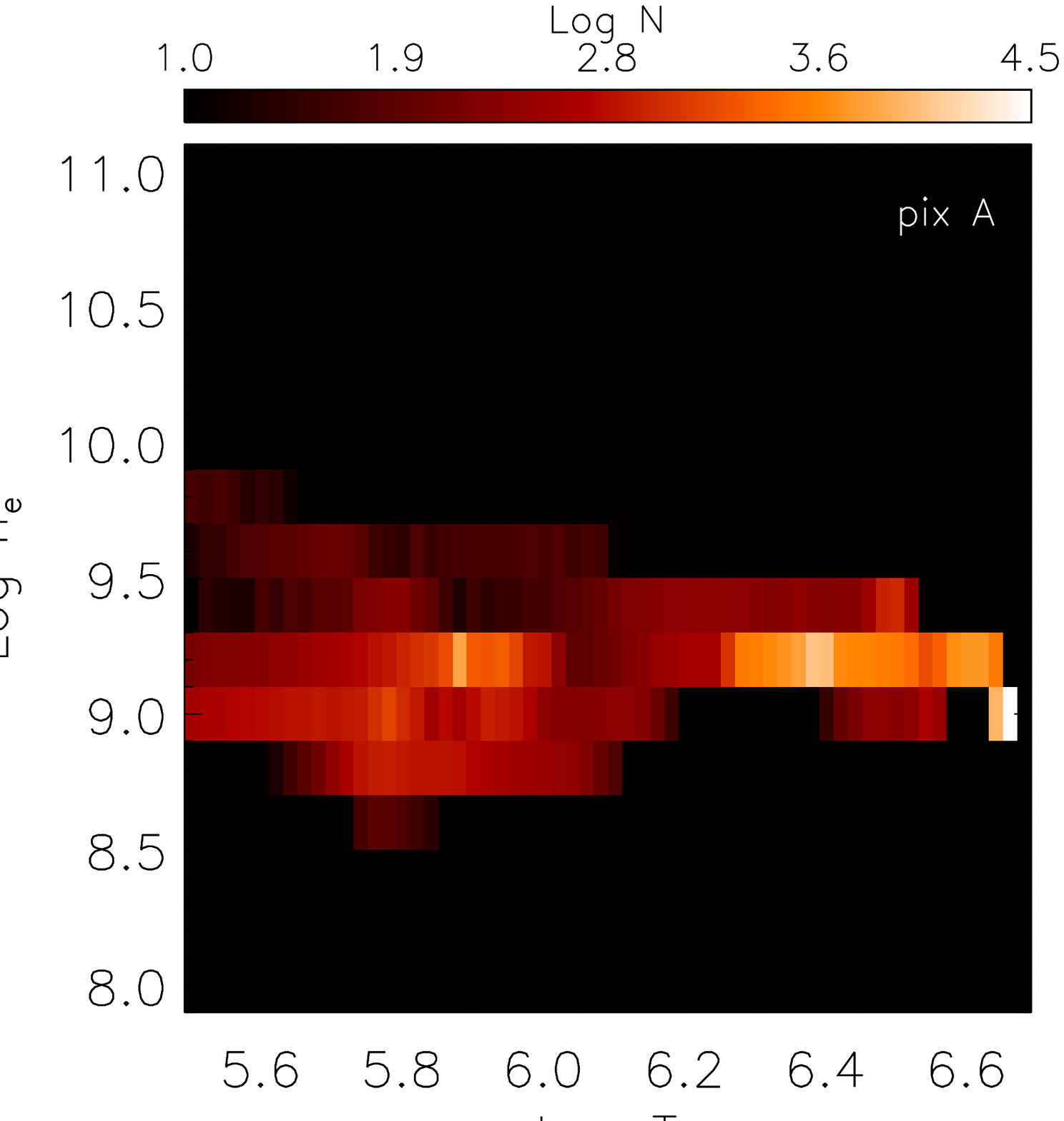}
  \includegraphics[scale=0.33]{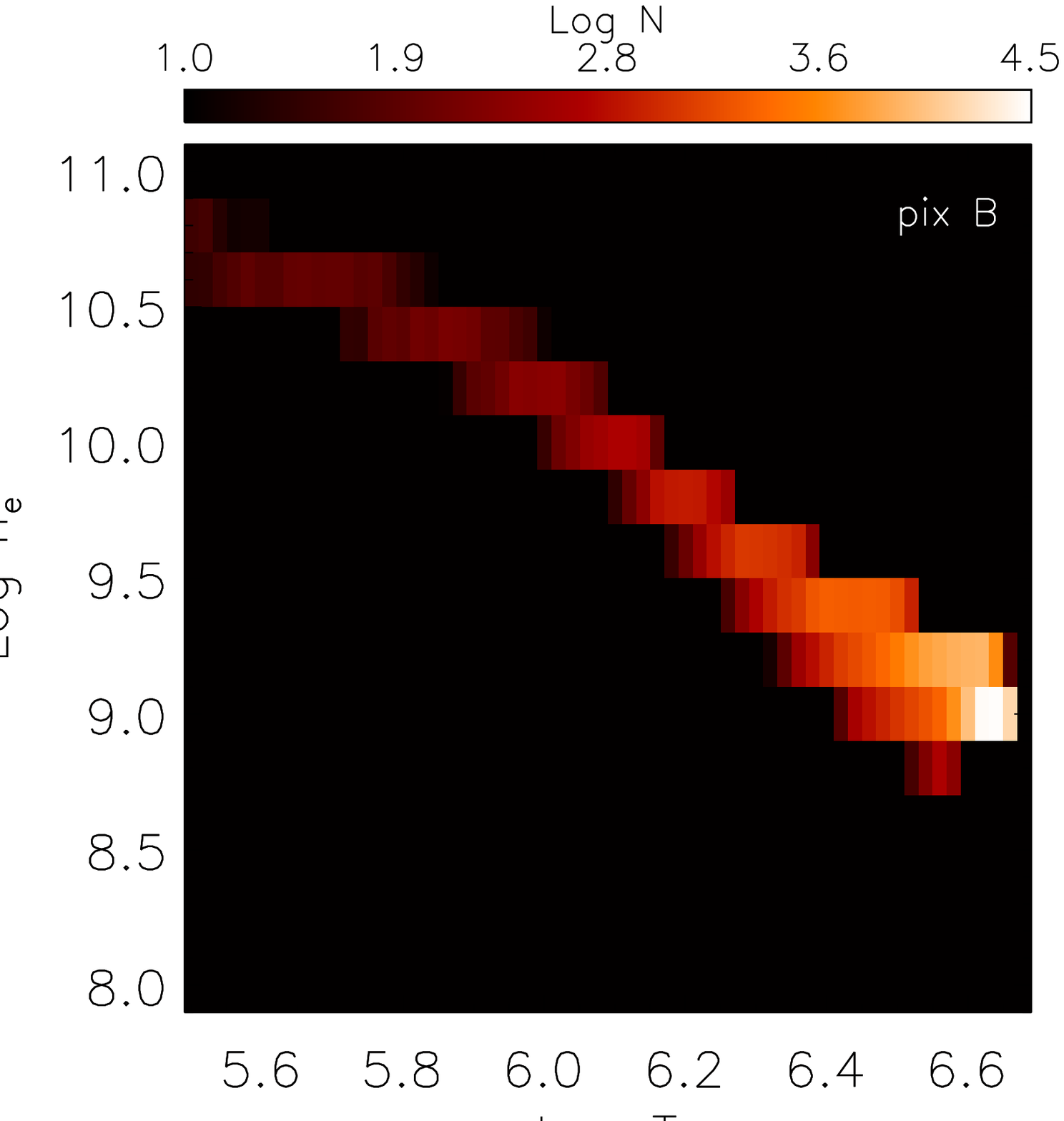}}\vspace{0.5cm}
\caption{Distribution of electron density, from the simulations,
  in the voxels contributing to two pixels with different $\chi^2_0$ (pixel 
  A with $\chi^2_0 < 1$; pixel B with $\chi^2_0 \gtrsim 2$) in the top view 
  for each snapshot (simulation C in the top row and simulation H in the 
  bottom row) to illustrate the effect of the mixing of plasma volumes 
  characterized by significantly different plasma parameters.
  The selected pixels are marked in the plots showing the  $\chi^2_0$ maps
  (Figure~\ref{fig:chi2_e_dyn} and ~\ref{fig:chi2_e_cb} for C and H respectively).
  Histograms of densities in the two pixels are shown in the {\em left panel}; 
  2D histograms, as a function of both temperature and densities are also shown 
  ({\em middle}: pixel A; {\em right}: pixel B). \label{fig:neT_hist}}
\end{figure*}

As an example, in Figure~\ref{fig:neT_hist} we show the electron density 
distribution from the simulations, within two pixels with significantly 
different $\chi^2_0$
(with pixel A showing lower $\chi^2_0$ than B), in the top view case 
for each of the two snapshots. The selected pixels are marked in 
Figure~\ref{fig:chi2_e_dyn} and ~\ref{fig:chi2_e_cb} (top row panels). 
We find that the plasma in pixel B ($\chi^2_0 \gtrsim 2$) has a distribution 
of density along the LOS much broader than the plasma in pixel A. 
This leads to a worse determination of the emission measure and worse
match of the observed fluxes: even if the selected lines, similar to typical 
analyses of real data, are mostly insensitive to electron density, their 
small density dependency significantly affects the diagnostics of emission measure
distributions when there is significant mixing of plasmas with different 
electron densities within the LOS. 
The same effect of density mixing is causing the large $\chi^2_0$ regions 
at the loop footpoints in the side views of both snapshots (bottom
panels of Figure~\ref{fig:chi2_e_dyn} and~\ref{fig:chi2_e_cb}).

\begin{figure*}[!t]\vspace{0.5cm}
\centerline{\includegraphics[scale=0.45]{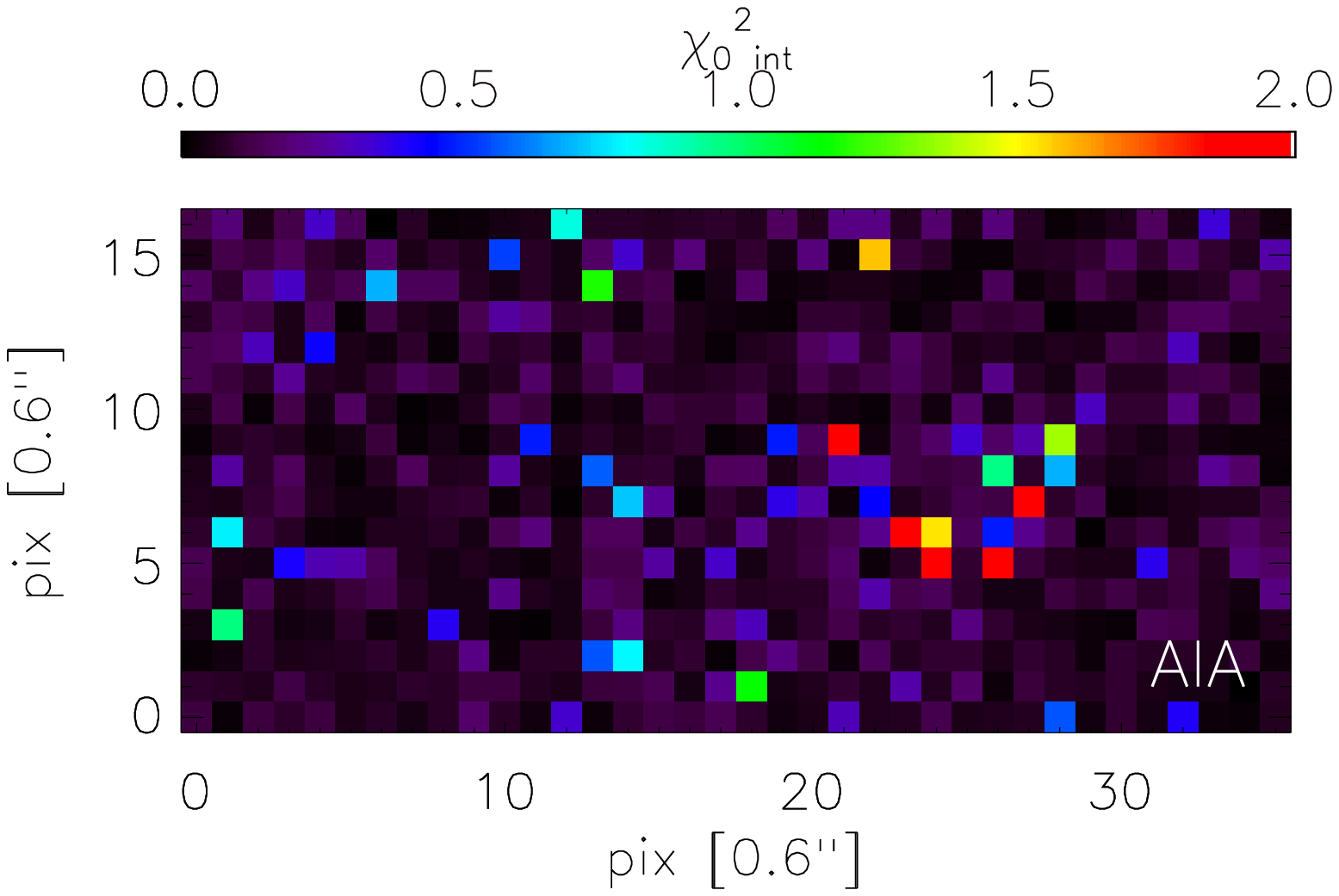}\hspace{-0.5cm}
  \includegraphics[scale=0.45]{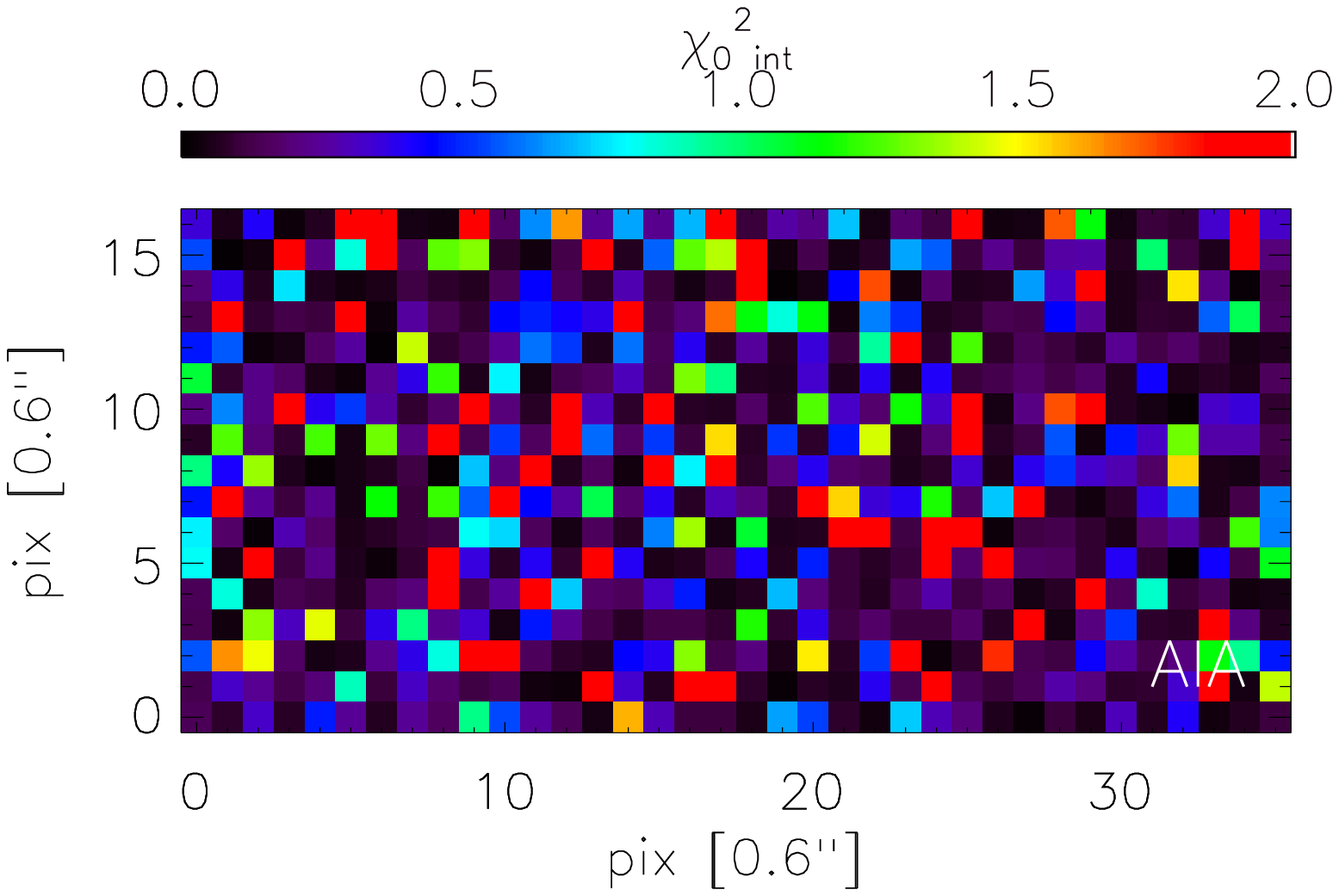}}\vspace{0.3cm}
\centerline{\includegraphics[scale=0.45]{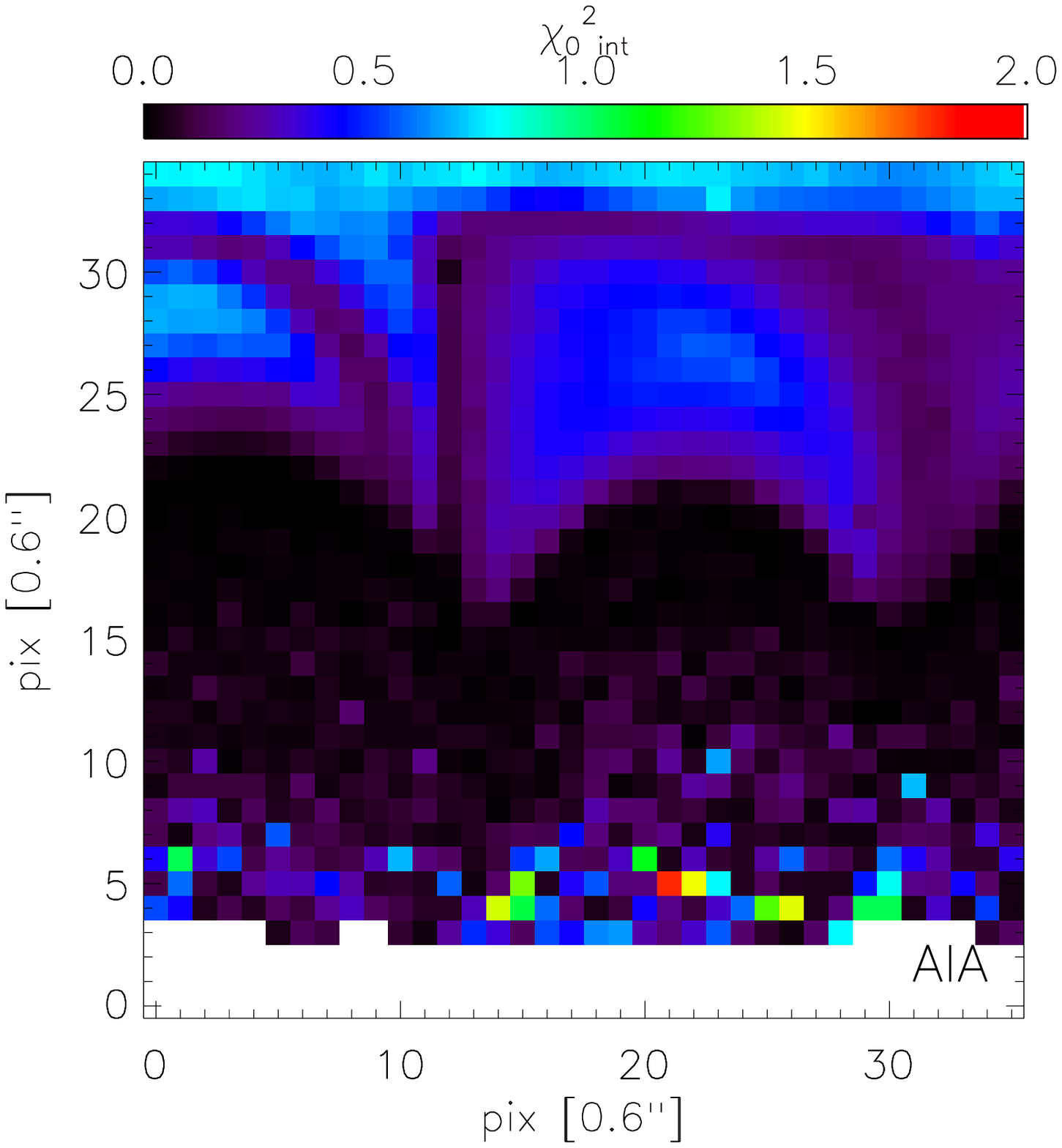}\hspace{-0.5cm}
  \includegraphics[scale=0.45]{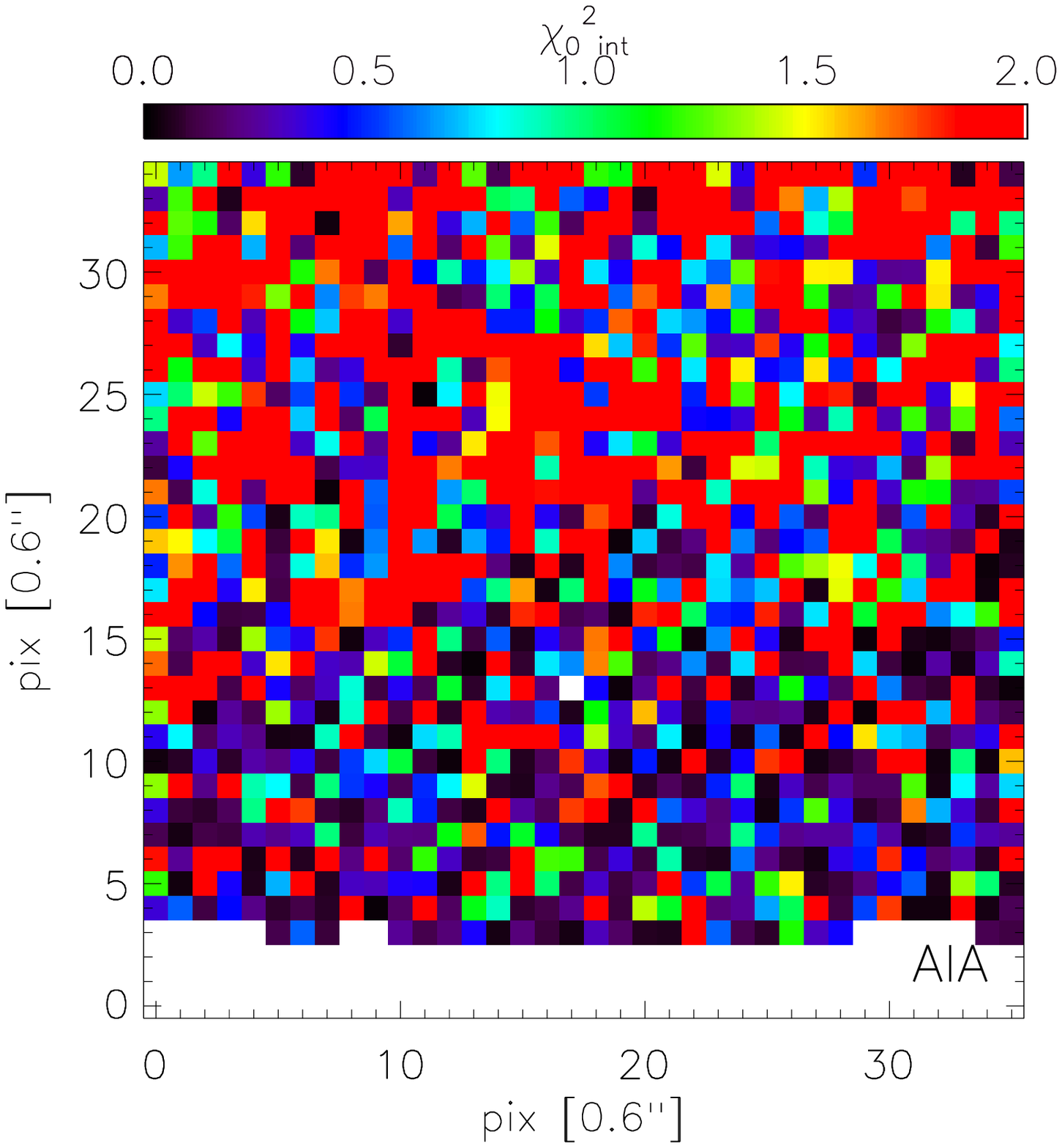}}\vspace{-0.5cm}
\caption{Maps showing the $\chi^2_0$ for the reconstruction of the EMD 
  using the AIA intensities, for snapshot C. As in Figure~\ref{fig:chi2_e_dyn}, we
  show the results for the two LOS, ``xy'' (top), and ``xz'' (bottom), and
  also the case including the Poisson noise (right column).  \label{fig:chi2_dyn}}
\end{figure*}

For the side view we note that for snapshot H, in the higher atmosphere,
where the temperature distribution in each pixel is essentially isothermal
at high temperatures (see also appendix~B), the case including 
the noise tends to have slightly {\em lower} $\chi^2_0$ values 
(Figure~\ref{fig:chi2_e_cb}).  We interpret this 
effect as due to the extremely narrow (isothermal) distribution: the MCMC
method applies some physically based, locally variable, smoothness criteria
based on the properties of the temperature responses/emissivities for 
the used data (see also discussion in \citealt{Kashyap98,Testa11}), and 
therefore it does not perfectly recover purely isothermal distributions;
the effect of the noise mimics slight departures from isothermal distributions
and therefore the MCMC methods finds better matches to the intensities.
A similar effect is also observed for small regions of snapshot C (e.g., the 
high $\chi^2_0$ strip at z=19~pix and x from pixel 7 to 14).

\begin{figure*}[!t]\vspace{0.5cm}
\centerline{\includegraphics[scale=0.45]{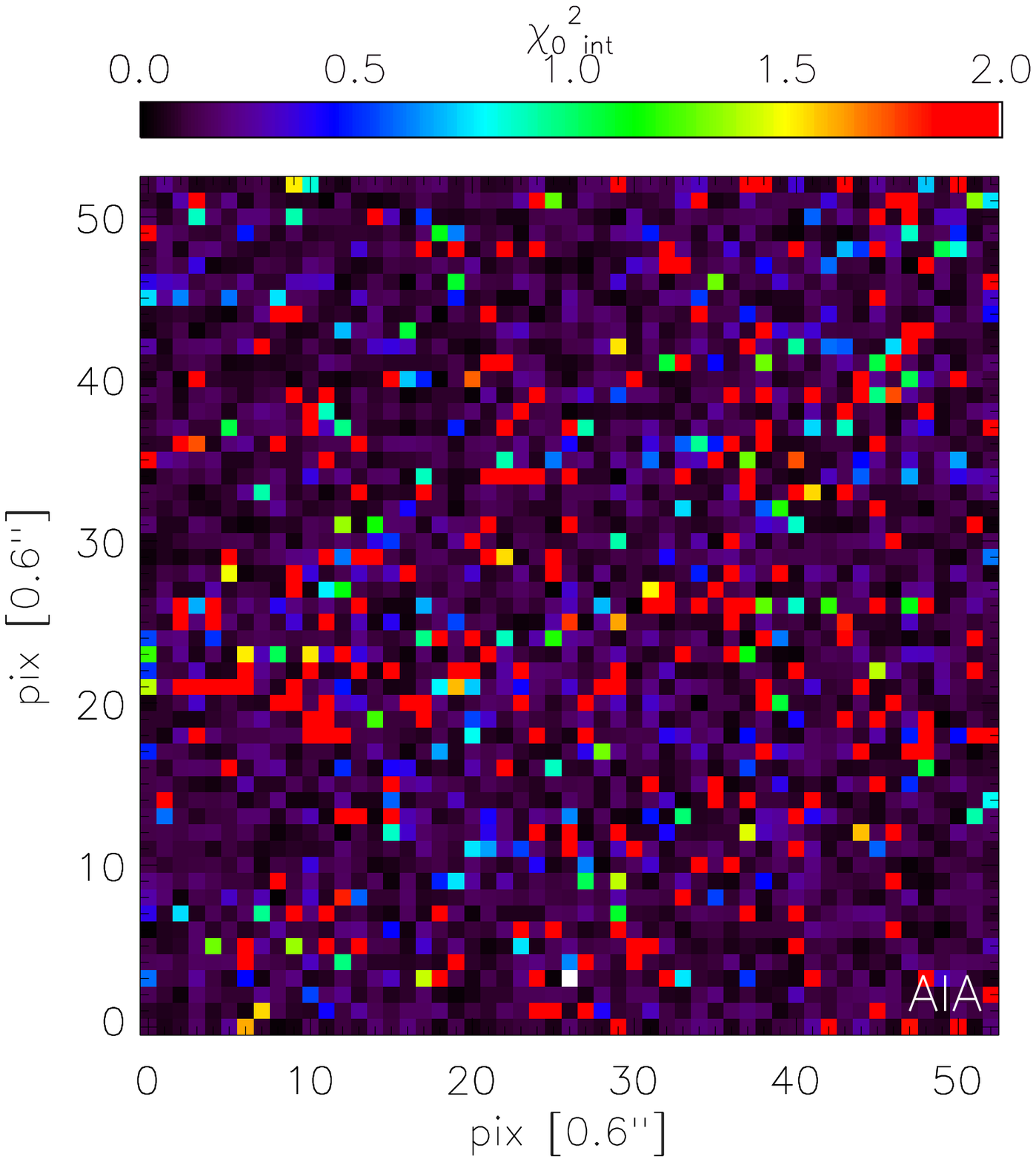}\hspace{-0.5cm}
  \includegraphics[scale=0.45]{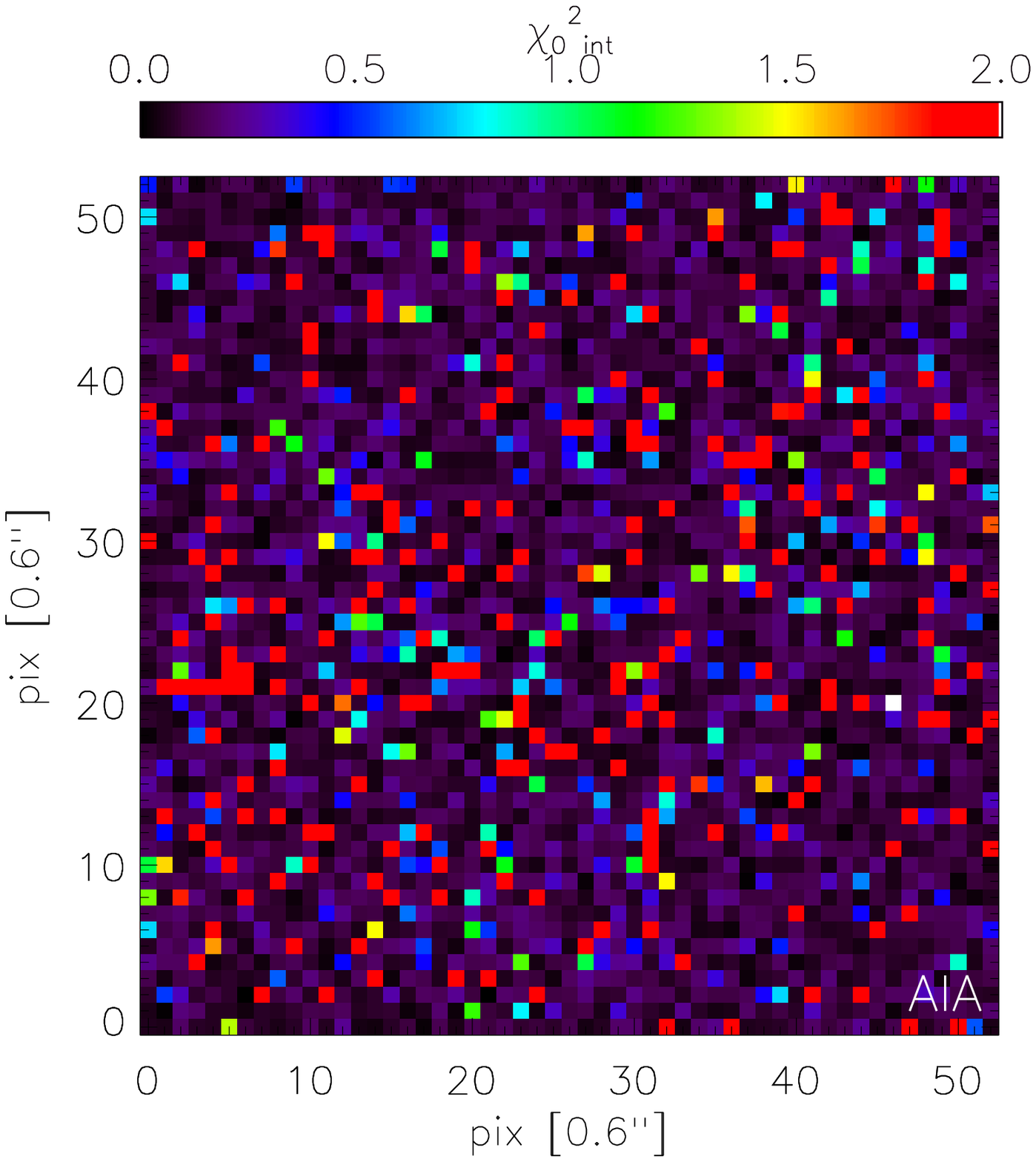}}\vspace{0.3cm}
\centerline{\includegraphics[scale=0.45]{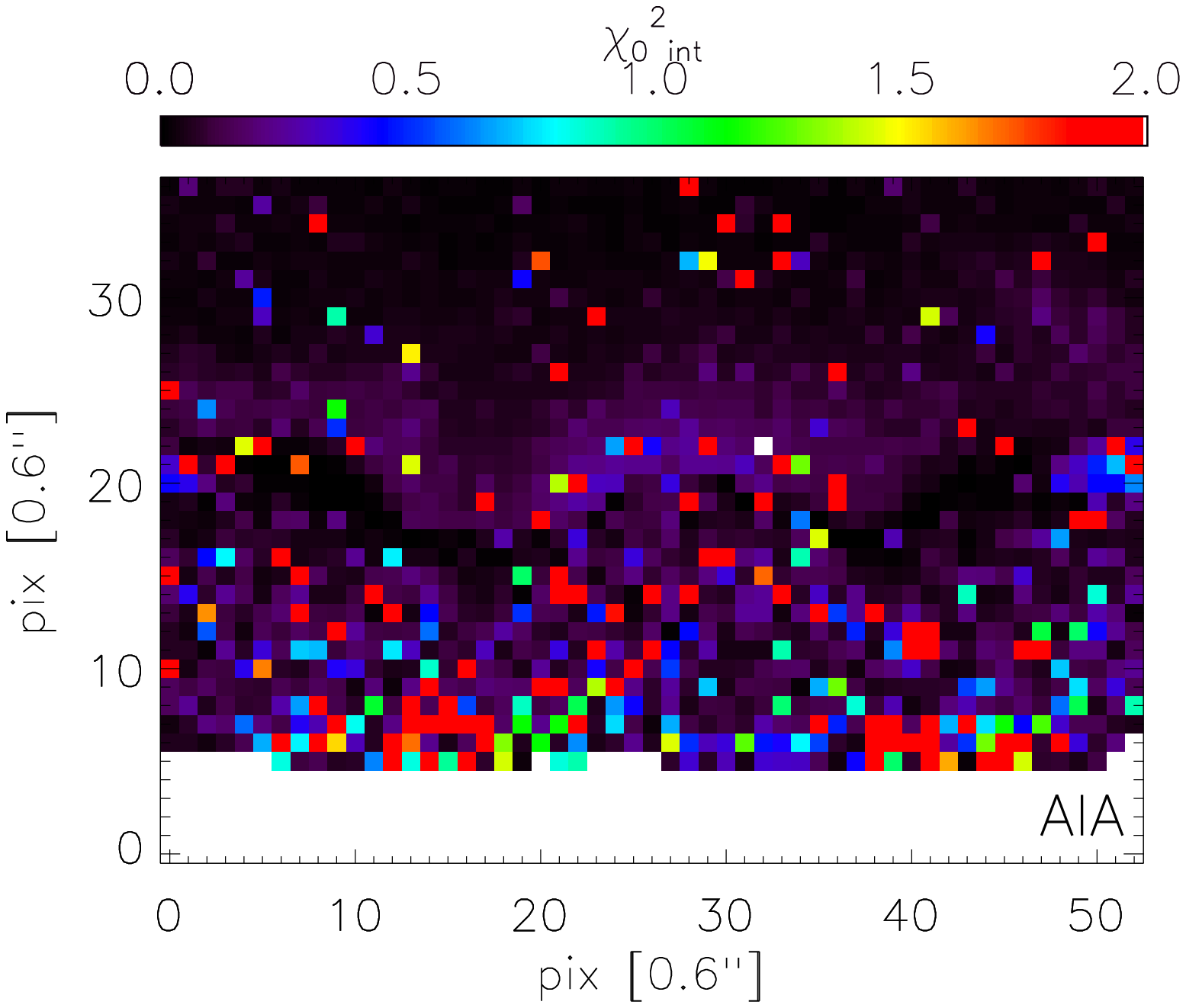}\hspace{-0.5cm}
  \includegraphics[scale=0.45]{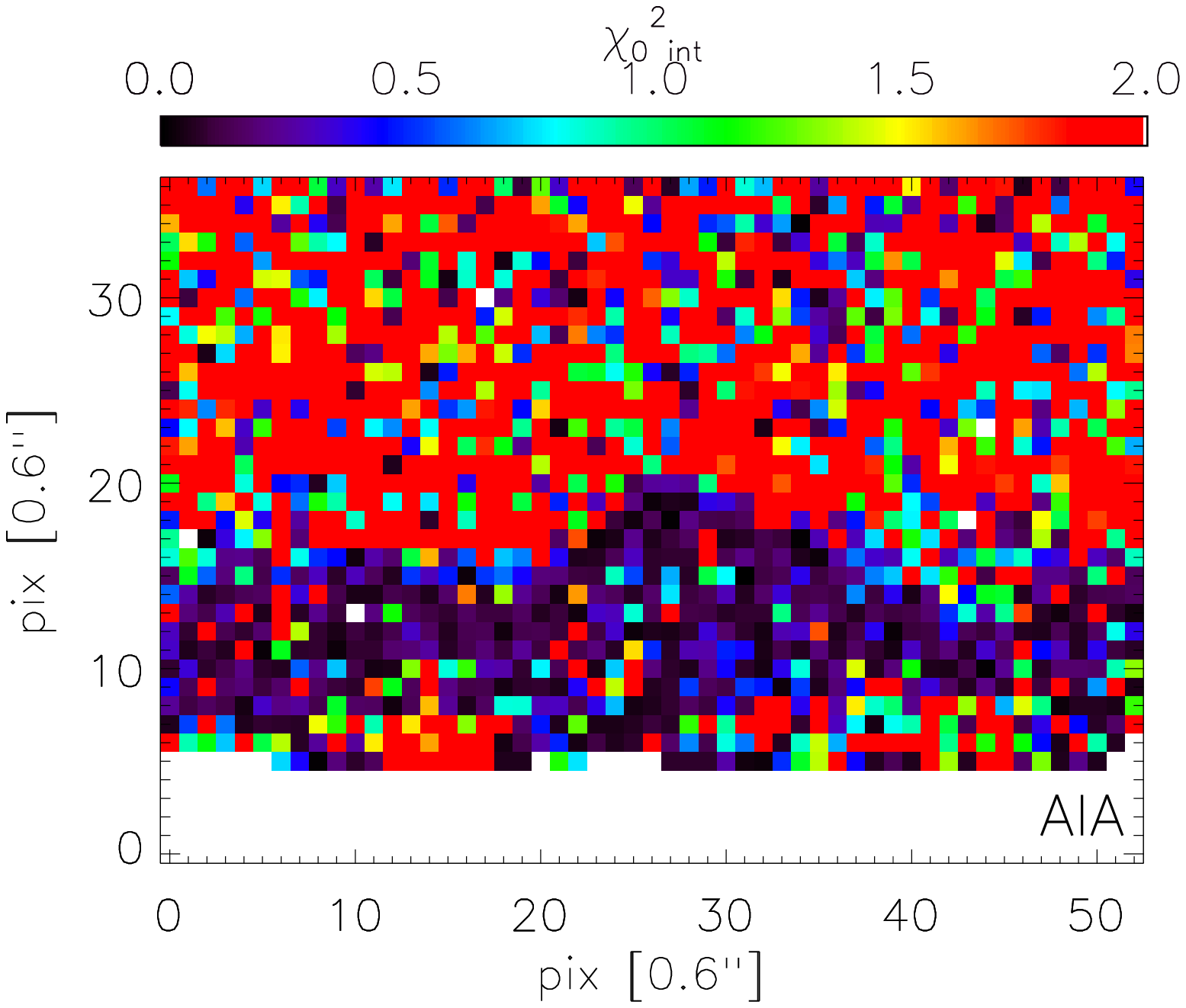}}\vspace{-0.5cm}
\caption{$\chi^2_0$ maps analogous to Figure~\ref{fig:chi2_dyn}, for 
  snapshot H. \label{fig:chi2_cb}}
\end{figure*}

\begin{figure*}[!t]
\centerline{\includegraphics[scale=0.85]{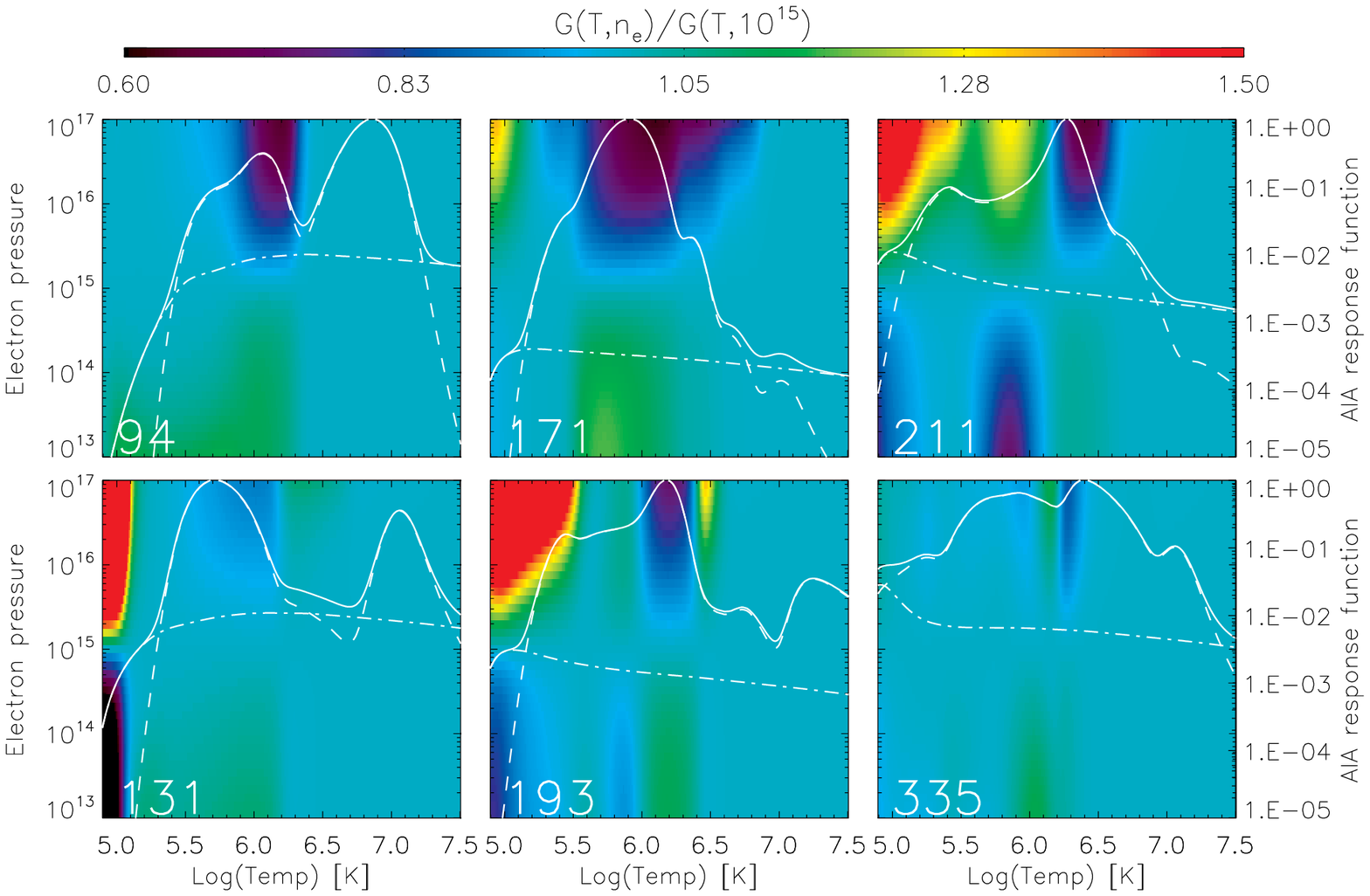}}\vspace{-0.2cm}
\caption{Dependence of AIA temperature responses on plasma pressure 
  ($n_e \times T$ [cm$^{-3}$~K]).
  In the pressure-temperature space, we plot maps of the ratio of the 
  AIA response to the response calculated for a pressure of 
  $10^{15}$~cm$^{-3}$~K, which is the pressure assumed for the standard 
  AIA temperature responses available in SolarSoft. From left to right: 
  {\em top -} 94\AA, 131\AA, 171\AA; {\em bottom -} 193\AA, 211\AA, 
  335\AA\ passbands. We overplot the default AIA temperature responses 
  (white lines;  in units of $[$DN$~s^{-1}$~pix$^{-1}$~cm$^{5}]$, and 
  normalized to their peak), also showing the contribution of the lines 
  (dashed lines) and the continuum emission (dot-dashed lines).
\label{fig:aia_tresp_press}}
\end{figure*}

We now consider the $\chi^2_0$ maps obtained for the analysis of 
the AIA synthetic intensities; Figures~\ref{fig:chi2_dyn} and \ref{fig:chi2_cb}
show the maps for snapshot C and H respectively (analogous to 
the EIS results of Figures~\ref{fig:chi2_e_dyn}, and \ref{fig:chi2_e_cb}).
For snapshot C the top view shows a few pixels with higher $\chi^2_0$ 
in the region corresponding to the dense emerging flux region, similarly 
to the EIS case. For snapshot H there is no clear correspondence between
the $\chi^2_0$ value and the regions with high density mixing. 
This can be explained by looking at the density sensitivity of the AIA
bands, which we have already discussed in \cite{MartinezSykora11}
and summarize in Figure~\ref{fig:aia_tresp_press}. The plots in 
Figure~\ref{fig:aia_tresp_press} show that most channels have limited
density sensitivity, which becomes negligible at high temperature. 
Therefore we expect less of a density effect for snapshot H where
high temperature plasma ($\log T > 6.5$; see also appendix~B) 
largely contributes to the observed intensities.
In the side view case, the intensities are reproduced quite well in the
cases that do not include Poisson noise, whereas the cases including noise
show large areas of high $\chi^2_0$ values, especially at large heights.
The larger effect of the noise for AIA is due to the fact that for the side
view, the AIA intensities are rather low (see also Figure~\ref{fig:his_sim})
in some channels, especially in 131\AA\ for both snapshots and 94\AA\ for 
snapshot C. These channels are more sensitive to the cool plasma
(though, for snapshot H the 94\AA\ intensity is largely due to the 
Fe\,{\sc xviii} emission from the high temperature plasma) which has 
very small or zero contribution higher up in the atmosphere, where the 
plasma is close to isothermal (as we will show in detail in section 
\ref{ss:EM_maps}).
The low statistics in few channels makes the fit sensitive to the noise. 
As a result, the fits become worse when including the effect of 
Poisson noise.
We have investigated the results for significantly larger  
S/N ratio for side view case of snapshot H. We re-run the case 
including noise, but assuming exposure times that correspond 
to 50 images for all AIA channels.  
We find that for very large S/N values the $\chi^2_0$ values for 
the intensities tend to get large, because of the very small errors 
associated with the intensities (down to the level of $\sim 0.1\%$ for 
the brightest channels). We find that in order to get reasonable 
$\chi^2_0$ values the MCMC procedure needs to be run with a 
significantly larger number of simulations to reproduce the intensities 
to that level of accuracy.  
However, as we will discuss later in section~\ref{ss:EMD}, results 
show that the effect of the noise is not the main factor that determines 
the ability of the MCMC method to reconstruct EMDs from AIA data.

For snapshot H, for both LOS, for the case without noise the $\chi^2_0$ 
values are generally worse than the corresponding C snapshot 
(Figure~\ref{fig:chi2_cb}). 
We interpret this as an effect of the broader temperature distributions
(especially in the top view) extending to rather high temperatures 
($\log T[K] \gtrsim 6.6$; see also following 
Figures~\ref{fig:emt_map_cb_e_xy} and~\ref{fig:emt_map_cb_e_xz}, 
and appendix~B and section~\ref{ss:EM_maps}). In this temperature 
regime AIA provides less reliable temperature diagnostics than at lower 
temperatures ($\log T[K] \sim 6-6.3$) where several AIA channels have 
high sensitivity and their relative shapes in temperature response provide 
better constraints to the plasma temperature distributions.

We note that the uncertainties we assume here are generally significantly 
smaller than the errors typically associated to intensity measurements
in real data for the EMD reconstruction. Those errors are typically larger
than the uncertainties associated with photon counting statistics because
they also take into account uncertainties in atomic data which are difficult 
to quantify.
In the test presented in this paper, there are no uncertainties to be associated
with the atomic data, since we use the same atomic data to synthesize the
intensities and to infer from them the EMD. Nevertheless, the small errors 
we assume can in part explain the sometimes large values of $\chi^2_0$ 
we obtained here.

The results discussed so far provide us with a measure of how well 
the temperature distributions inferred using the MCMC method
are able to reproduce the ``measured'' intensities. 
We now discuss in detail the comparison of the derived EMD with 
the true emission measure distributions. 
We looked at different parameters to assess the robustness and 
limits of the reconstruction method in recovering the plasma 
temperature distributions: (1) the temperature at which the EMD peaks 
($T(EMD_{max})$) within the temperature range $\log T[K] =[5.5-6.7]$,
(2) the EM values integrated in broad temperature 
bins, and (3) the full EMD (i.e., at the fine temperature resolution).
The temperature of the EMD peak is not necessarily a good 
indicator to evaluate the goodness of the MCMC solution, especially
for broader and more structured EMDs. We therefore present the 
analysis of $T(EMD_{max})$ in appendix~A, 
summarizing here only the main results, while in the main text 
we will focus on the other two comparisons
(sections \ref{ss:EM_maps} and \ref{ss:EMD}).

The results presented in appendix~A show that the
MCMC method performs well in recovering the $T(EMD_{max})$
when the EMD is close to isothermal. For broad EMDs the method
tends to generally underestimate $T(EMD_{max})$, due to several
factors, including poor constraints beyond the $\log T[K] =[5.5-6.7]$
range, and limited capability of the method to recover sharp features.

\subsection{Emission measure maps \label{ss:EM_maps}}

One of the approaches we use to determine the accuracy of the EMD 
diagnostics is by looking at the maps of emission measure in different 
temperature ranges. 

\begin{figure}[!ht]\vspace{0.5cm}
\centerline{\hspace{-0.8cm}\includegraphics[scale=0.43]{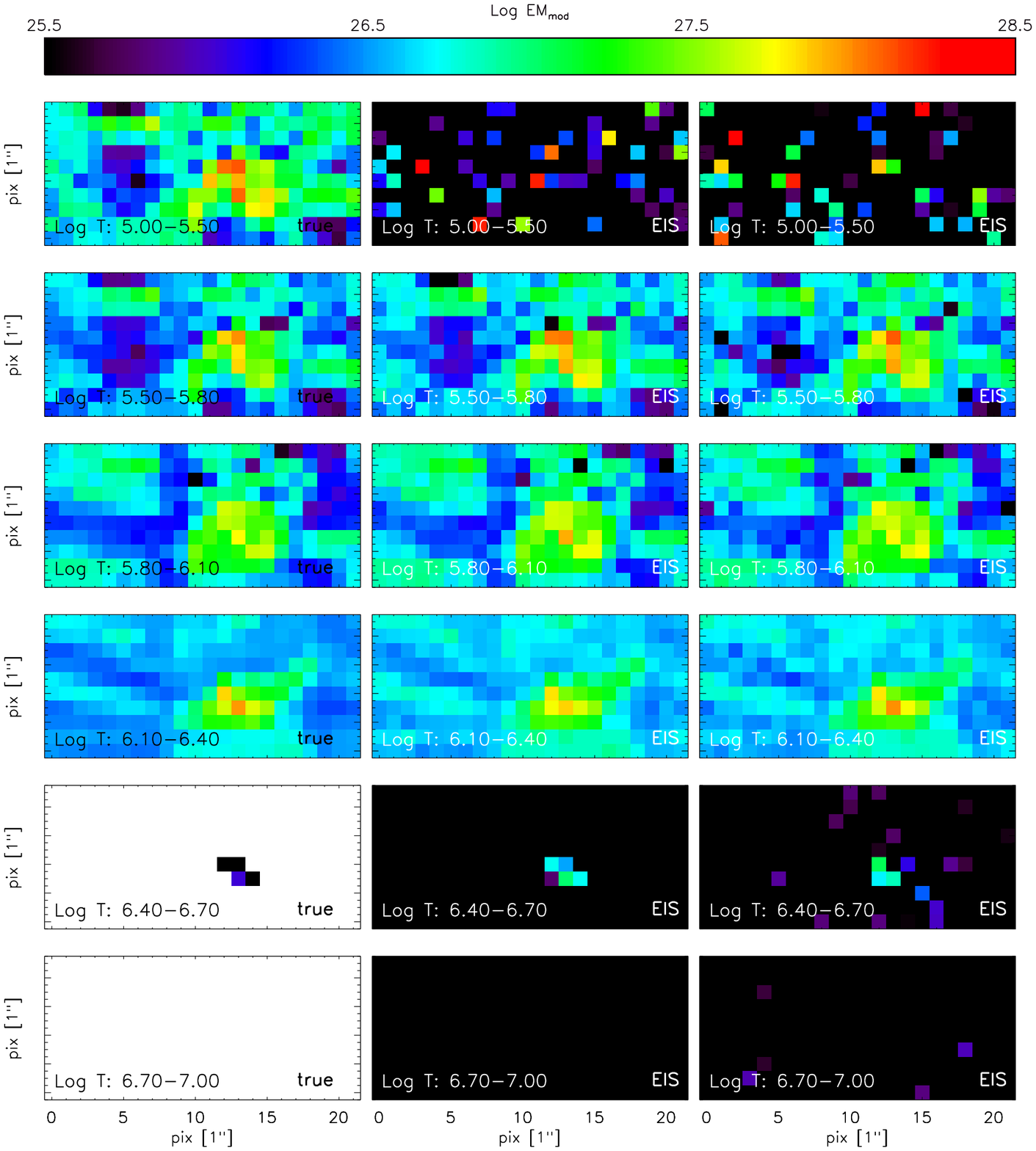}}\vspace{-0.6cm}
\caption{Emission measure maps for the top view of snapshot C, from the model 
  (left panels) and from EIS synthetic data (without noise or including noise, in the 
  middle and right panels respectively), integrated in six temperature bins:
  $\log T[K]:$ 5.0-5.5, 5.5-5.8, 5.8-6.1, 6.1-6.4, 6.4-6.7, 6.7-7.  
  White corresponds to values of EM equal to zero.
  \label{fig:emt_map_dyn_e_xy}}
\end{figure}

\begin{figure}[!ht]
\centerline{\includegraphics[scale=0.4]{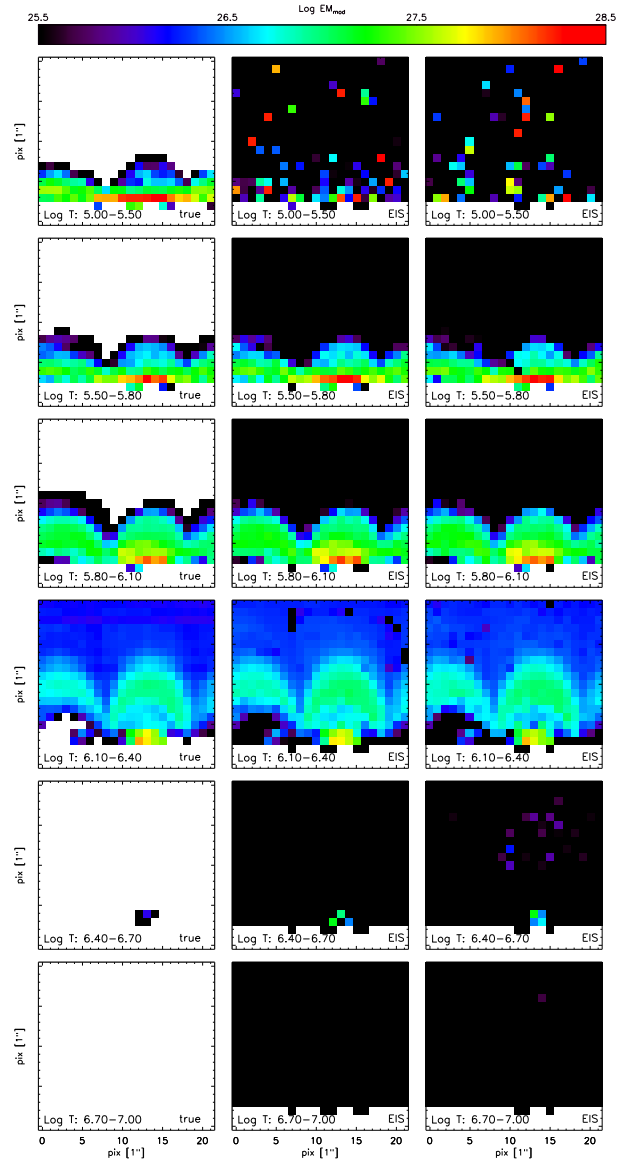}}\vspace{-1cm}
\caption{Comparison of emission measure maps obtained from the analysis
  of EIS synthetic data of snapshot C, analogous to Figure~\ref{fig:emt_map_dyn_e_xy}
  but for the side view.
  \label{fig:emt_map_dyn_e_xz}}
\end{figure}

\begin{figure}[!t]\vspace{0.5cm}
\centerline{\hspace{-0.8cm}\includegraphics[scale=0.43]{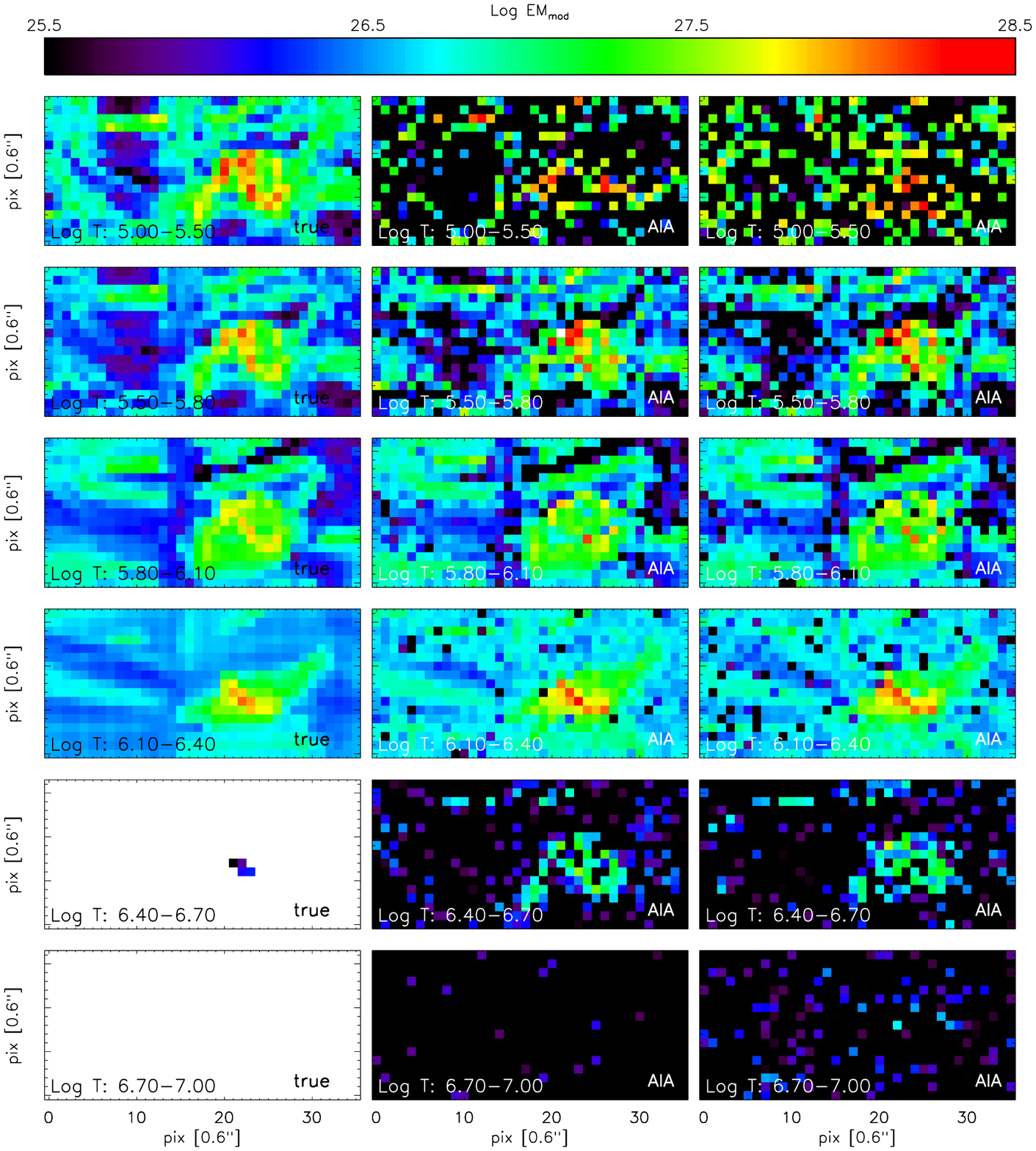}}\vspace{-0.6cm} 
\caption{Comparison of emission measure maps obtained from the analysis
  of the top view AIA synthetic data of snapshot C, analogous to 
  Figure~\ref{fig:emt_map_dyn_e_xy}.
  \label{fig:emt_map_dyn_a_xy}}
\end{figure}

\begin{figure}[!t]
\centerline{\includegraphics[scale=0.4]{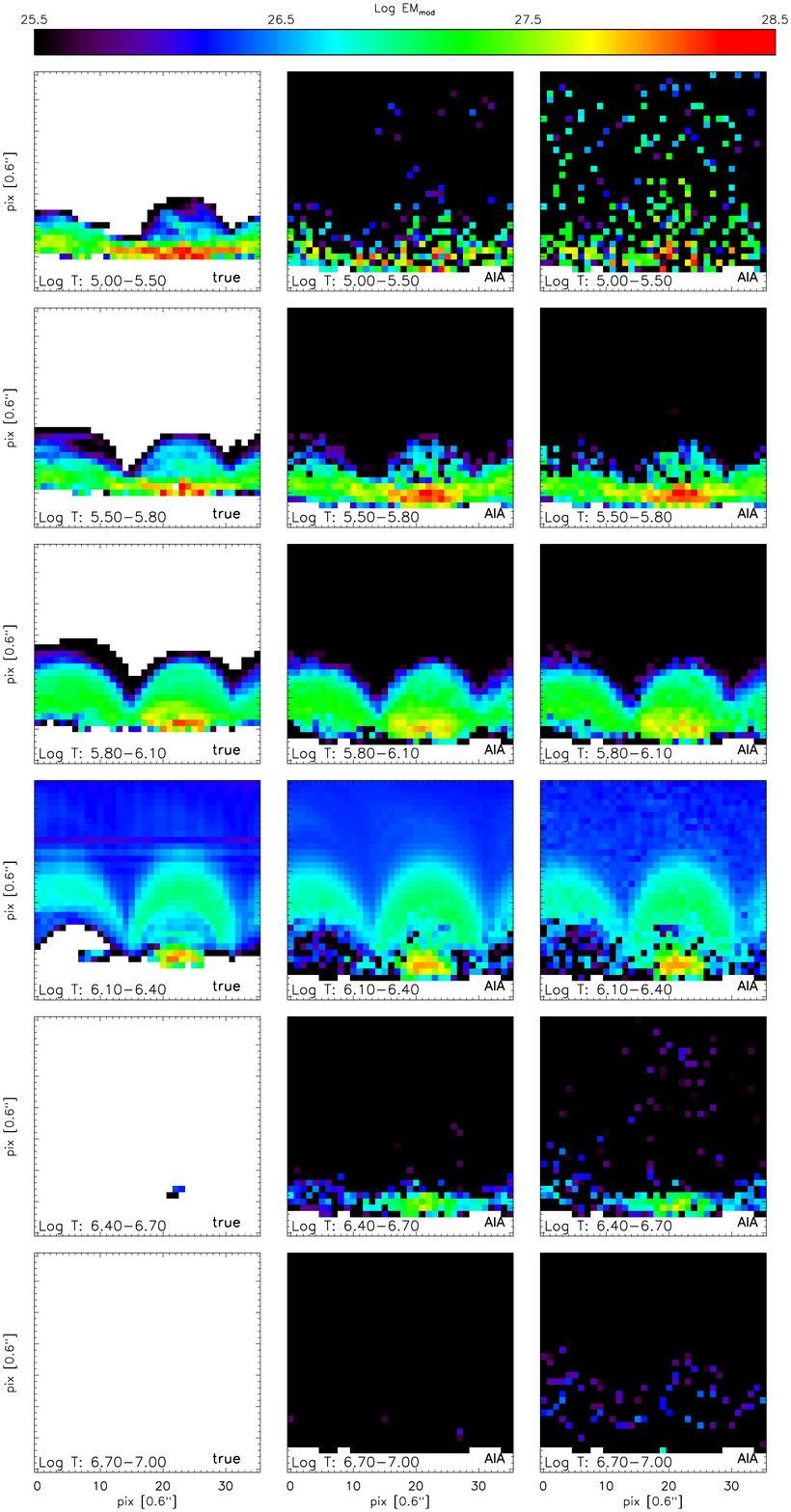}}\vspace{-1cm} 
\caption{Comparison of emission measure maps obtained from the analysis
  of the side view AIA synthetic data of snapshot C, analogous to 
  Figure~\ref{fig:emt_map_dyn_e_xy}.
  \label{fig:emt_map_dyn_a_xz}}
\end{figure}

\begin{figure}[!t]
\centerline{\includegraphics[scale=0.4]{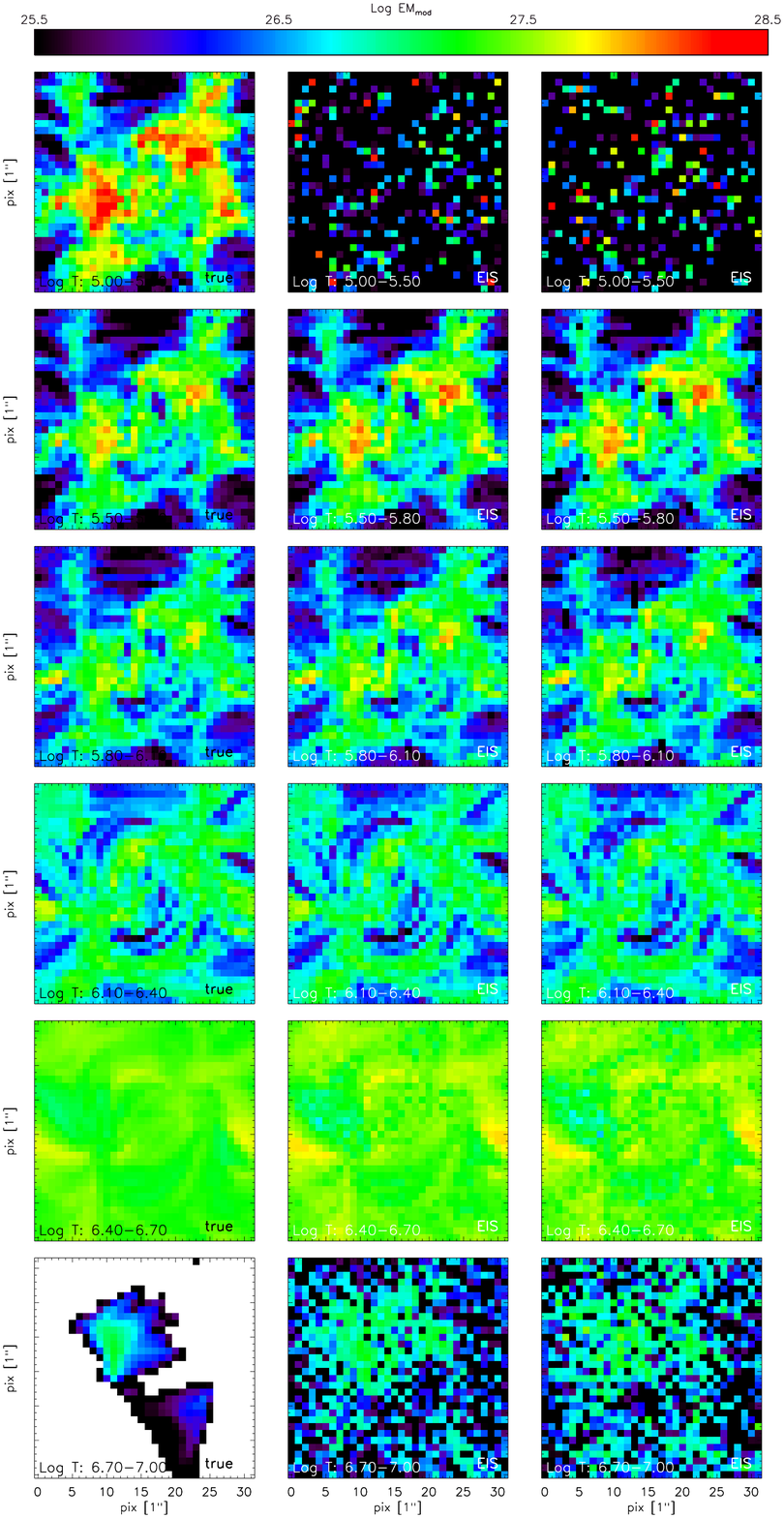}}\vspace{-1cm}
\caption{Emission measure maps for the top view of snapshot H, from the model 
  (left panels) and from EIS synthetic data (without noise or including noise, in the 
  middle and right panels respectively), integrated in six temperature bins:
  $\log T[K]:$ 5.0-5.5, 5.5-5.8, 5.8-6.1, 6.1-6.4, 6.4-6.7, 6.7-7. 
  Figure analogous to Figure~\ref{fig:emt_map_dyn_e_xy}, but for snapshot H.
  \label{fig:emt_map_cb_e_xy}}
\end{figure}

\begin{figure}[!t]\vspace{0.2cm}
\centerline{\includegraphics[scale=0.43]{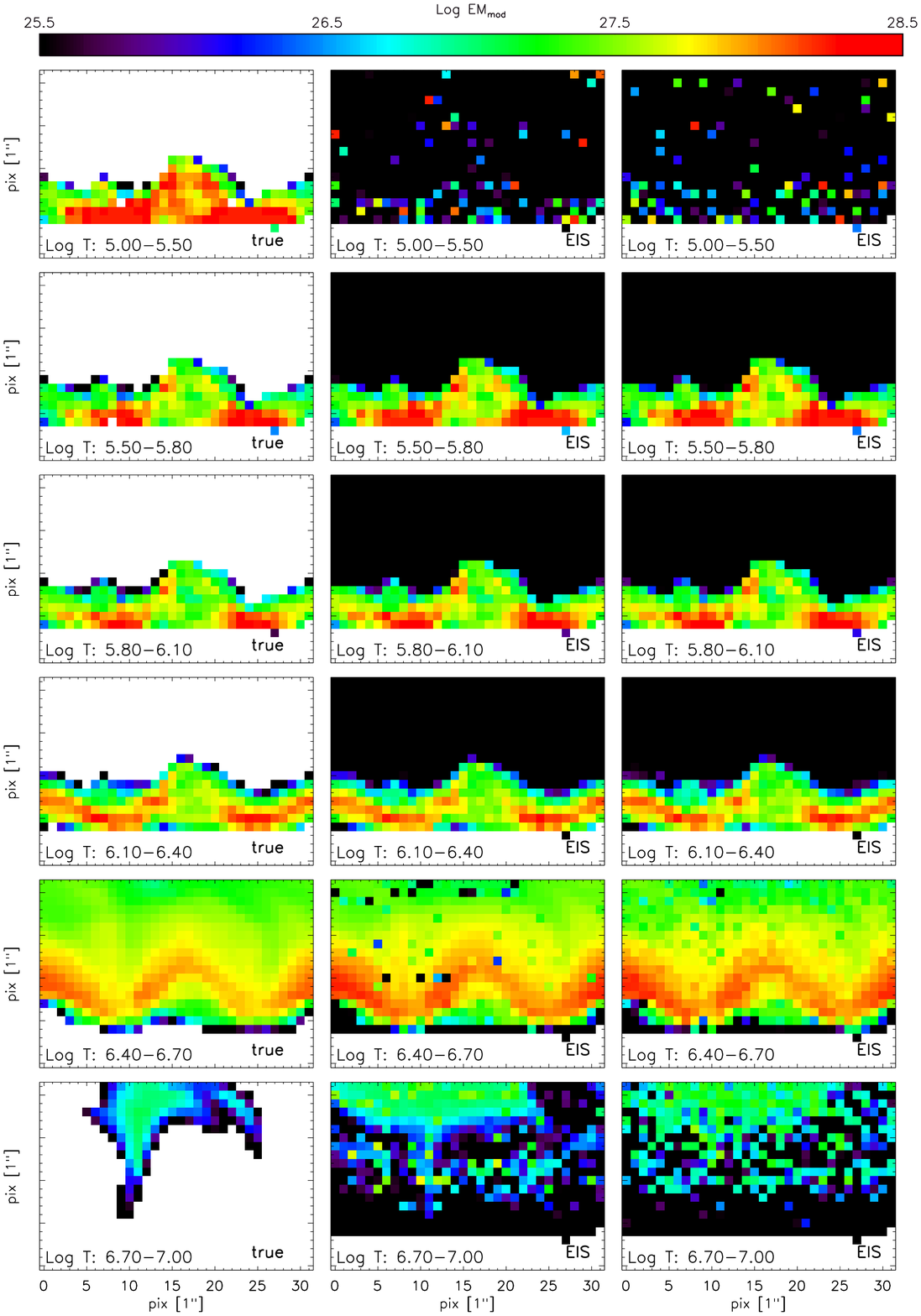}}\vspace{-0.6cm}
\caption{Comparison of emission measure maps obtained from the analysis
  of the side view EIS synthetic data of snapshot H, analogous to 
  Figure~\ref{fig:emt_map_cb_e_xy}.
  \label{fig:emt_map_cb_e_xz}}
\end{figure}

\begin{figure}[!t]
\centerline{\includegraphics[scale=0.4]{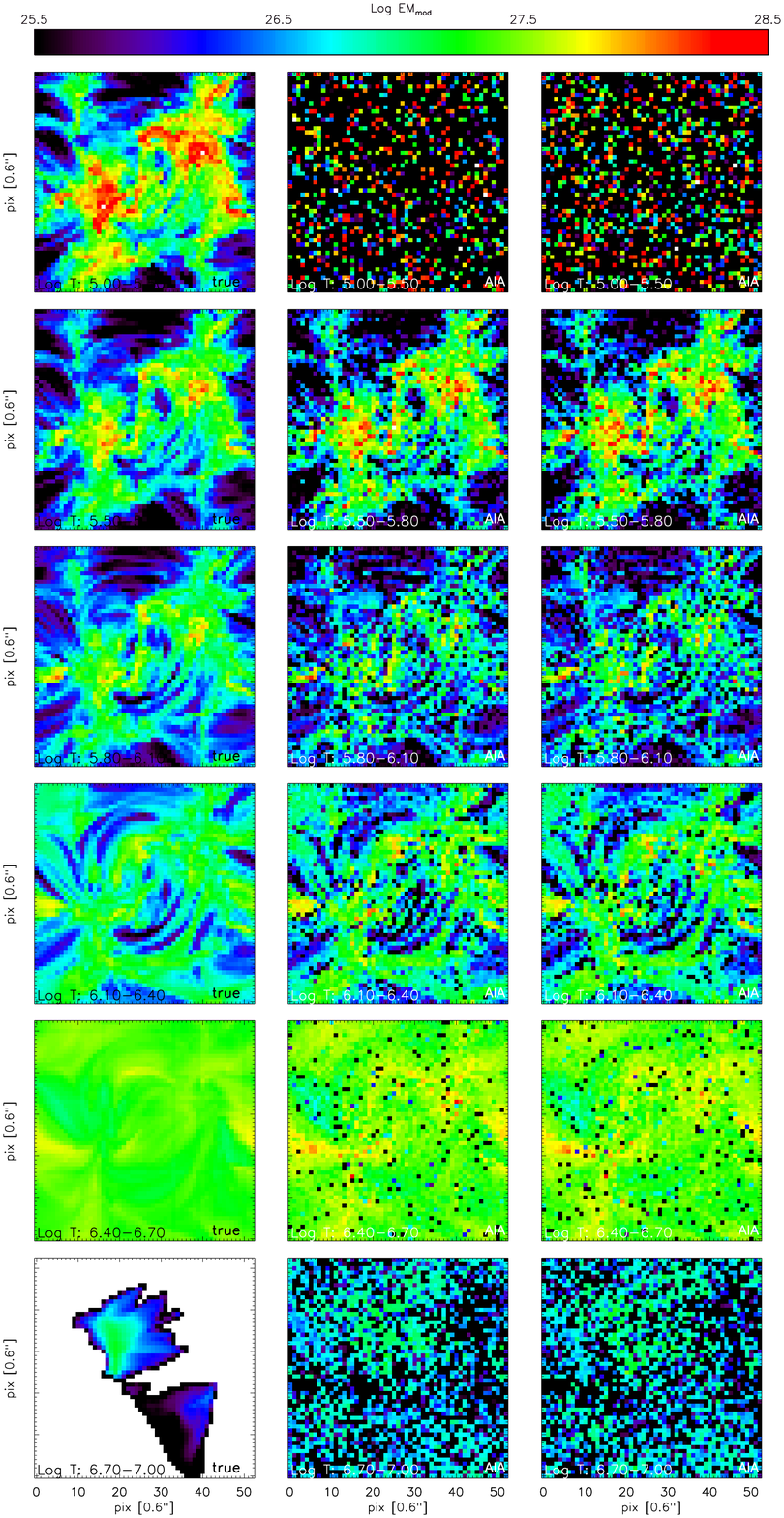}}\vspace{-1cm}
\caption{Comparison of emission measure maps obtained from the analysis
  of the top view AIA synthetic data of snapshot H, analogous to 
  Figure~\ref{fig:emt_map_cb_e_xy}.
  \label{fig:emt_map_cb_a_xy}}
\end{figure}

\begin{figure}[!t]\vspace{0.2cm}
\centerline{\includegraphics[scale=0.43]{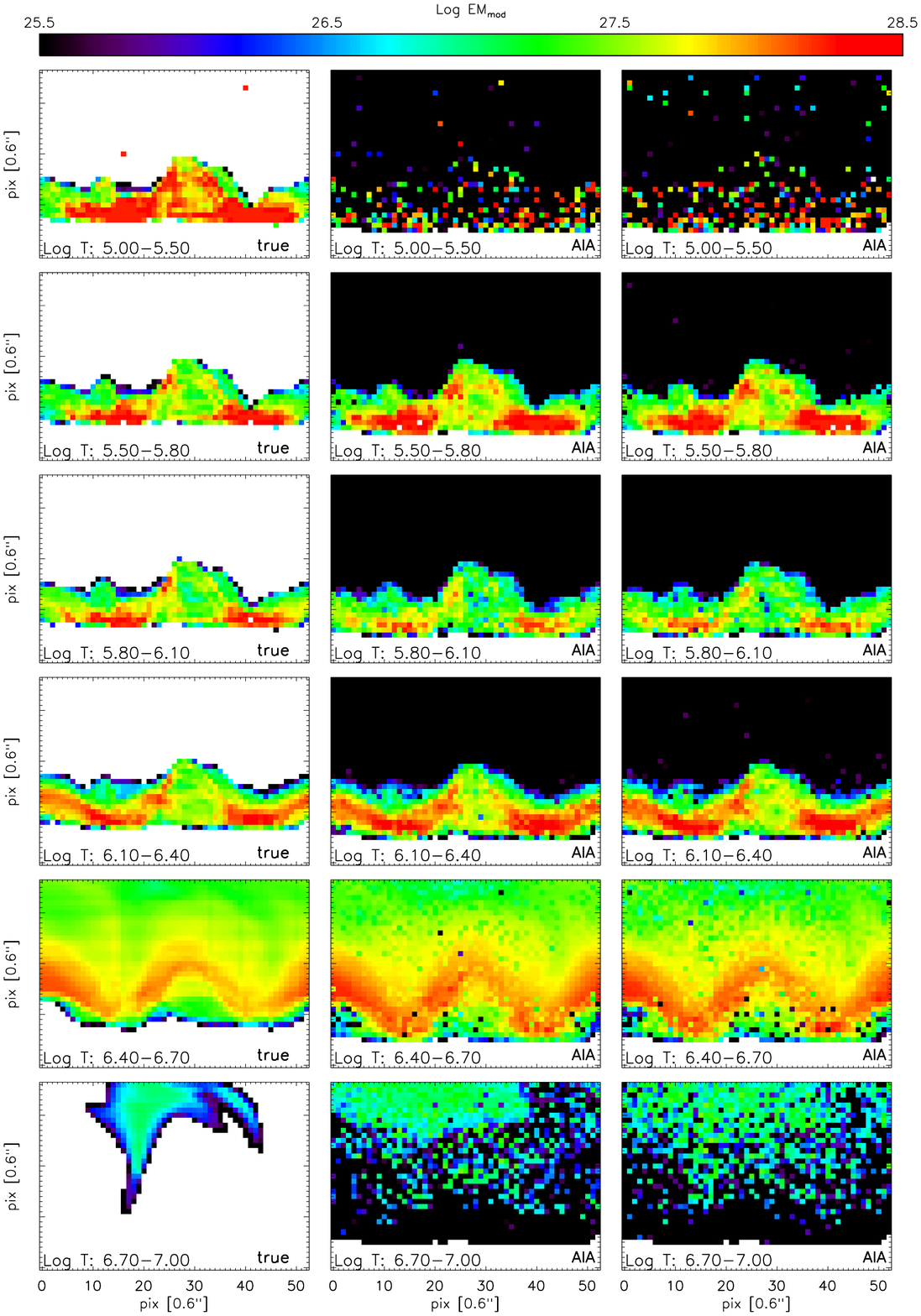}}\vspace{-0.6cm}
\caption{Comparison of emission measure maps obtained from the analysis
  of the side view AIA synthetic data of snapshot H, analogous to 
  Figure~\ref{fig:emt_map_cb_e_xy}.
  \label{fig:emt_map_cb_a_xz}}
\end{figure}

In Figure~\ref{fig:emt_map_dyn_e_xy} and \ref{fig:emt_map_dyn_e_xz} 
we show, for top and side view of snapshot C respectively, the maps of 
the true emission measure (left column), in six relatively broad 
temperature bins, spanning the temperature range $\log T[K] = 5-7$.
Figures~\ref{fig:emt_map_cb_e_xy} and \ref{fig:emt_map_cb_e_xz}  
show the analogous plots for snapshot H.
Figures~\ref{fig:emt_map_dyn_a_xy} and \ref{fig:emt_map_dyn_a_xz} 
(for snapshot C), and Figures~\ref{fig:emt_map_cb_a_xy} and 
\ref{fig:emt_map_cb_a_xz}  (for snapshot H) also show the EM maps 
derived at the AIA spatial resolution, for direct comparison with 
the corresponding maps derived by the analysis of the synthetic data.
These plots show, for the top view, how the plasma is characterized by
rather broad temperature distributions in most locations, especially 
in locations associated with the emerging flux region for snapshot C,
and with the loop footpoints (``moss'') for snapshot H 
(see Figures~\ref{fig:syndat_j} and \ref{fig:syndat_m}).
The simulated corona in snapshot C lacks significant plasma volumes
at temperatures above $\log T[K] =6.4$, while for snapshot H the coronal
plasma reaches temperature even higher than $\log T[K] = 6.7$ in a hot 
loop-like feature, which emits brightly in e.g., \caxvii\ (Figure~\ref{fig:syndat_m}).
The side view shows loop-like features in the lower corona of both snapshots, 
with rather broad temperature distribution. Higher up in the corona the plasma 
is characterized by narrow temperature distributions peaking around 
$\log T[K] \sim 6.2$ in snapshot C, and $\log T[K] \sim 6.6$ in snapshot H.

The corresponding maps for the EMD derived from the analysis of 
EIS and AIA synthetic data are shown next to the true
maps for all the cases, without or including noise (in center
and right column respectively of 
Figures~\ref{fig:emt_map_dyn_e_xy}-\ref{fig:emt_map_dyn_a_xz}, 
and \ref{fig:emt_map_cb_e_xy}-\ref{fig:emt_map_cb_a_xz}).

For EIS, the general features of the emission measure distribution in 
the central temperature bins (i.e., between $\log T[K] \sim 5.5$ and 
$\sim 6.4$ for snapshot C and in the range $\log T[K] \sim 5.5$-$6.7$ 
for snapshot H), are recovered quite well, also when including the 
effect of the noise.  In the top view case, for both 
snapshots, in the temperature bin at the low end of the constrained 
range ($\log T[K] \sim 5.5-5.8$) the EM are overestimated in several 
pixels because of the same effects discussed in appendix~B, 
because of the lack of constraints at lower temperatures.
The noisy results at the low and high temperature end are expected
considering the poor constraints provided by the selected EIS lines at 
those temperatures.

With AIA, though some structures such as the side view loops are still 
apparent in the reconstructed EM maps to a lesser extent, the EM maps
derived from the analysis of the synthetic data are not reproducing the 
true maps very accurately, especially for the top view case.
For instance for the top view case of snapshot C 
(\ref{fig:emt_map_dyn_a_xy}) the maps derived from AIA present 
significant EM, especially in the emerging flux region, in the bin 
temperature $\log T[K] \sim 6.4-6.7$, where the real EM is zero
almost everywhere.

\begin{figure*}[!t]
\centerline{\hspace{-0.8cm}\includegraphics[scale=0.7]{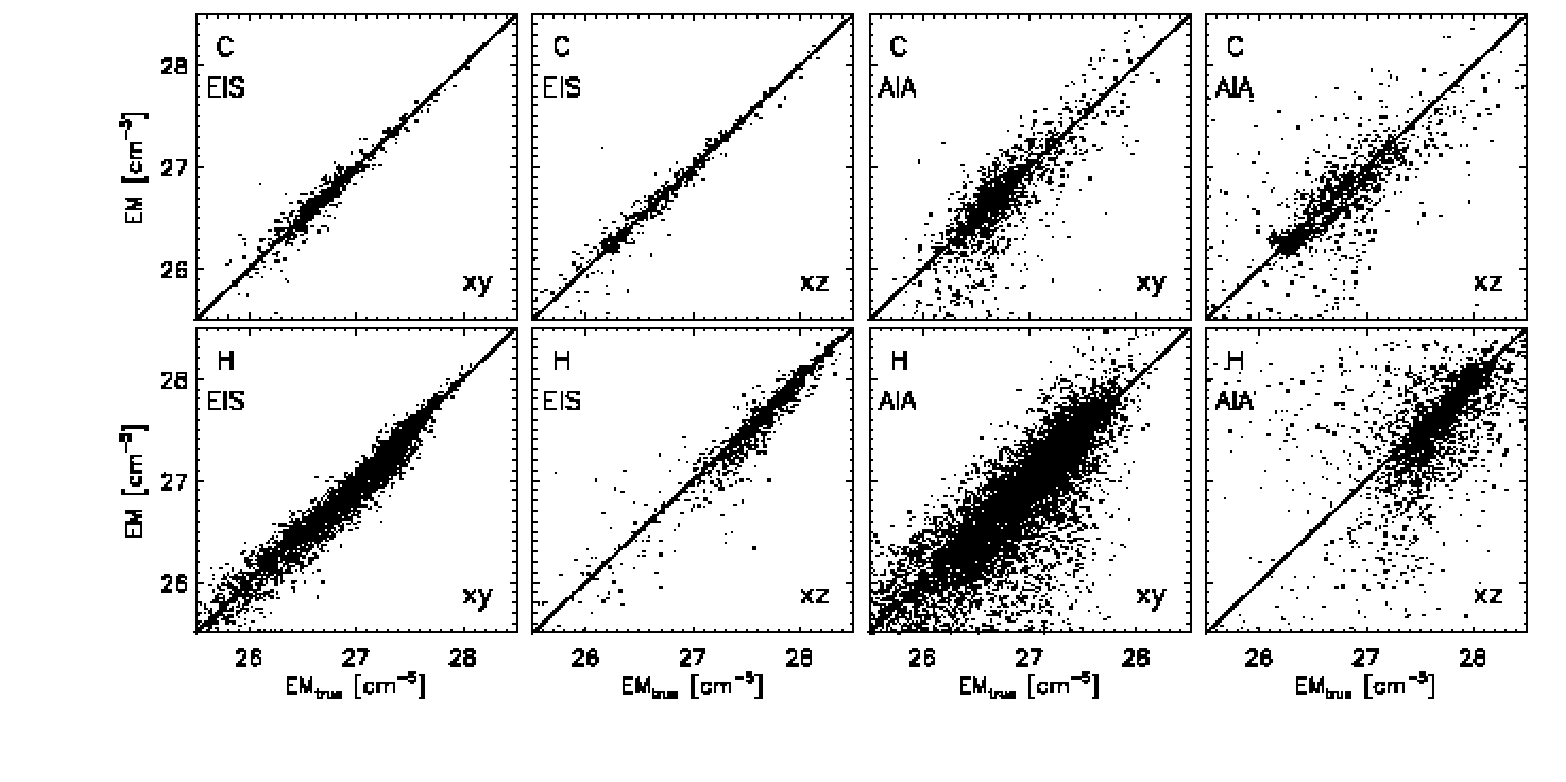}}
\caption{ Scatter plots of EM derived from EIS (first two columns) and 
  AIA (right columns) synthetic data from the two snapshots (cool 
  simulation ``C'' in the top row, and hot simulation ``H'' in the bottom
  row), vs.\ the corresponding true EM values. For each of the panels, 
  we plot the values for the central four temperature bins
  ($\log T[K]:$ [5.5-5.8],[5.8-6.1],[6.1-6.4],[6.4-6.7]), and excluded 
  the EM values for the two temperature bins at the extremes of the 
  range ($\log T[K]:$ [5.0-5.5],[6.7-7.0]).
  \label{fig:em_scatterplot}}
\end{figure*}

In Figure~\ref{fig:em_scatterplot} we show, for all the cases including the effect
of the noise, the scatter plots of the derived vs.\ true EM, for the central four
temperature bins, i.e.,  $\log T[K]:$ 5.5-5.8, 5.8-6.1, 6.1-6.4, 6.4-6.7.
These plots show that, when integrating in broad temperature ranges 
($\Delta \log T =0.3$), the EM derived from EIS reproduce rather well
the true EM, while the EM derived from AIA have significantly larger scatter
for the whole range of EM values. We note that these bins are significantly
wider than typically used by observers when deriving EMDs.

 \subsection{Full emission measure distributions \label{ss:EMD}}

\begin{figure*}[!t]
  \centerline{\hspace{-0.8cm}\includegraphics[scale=0.7]{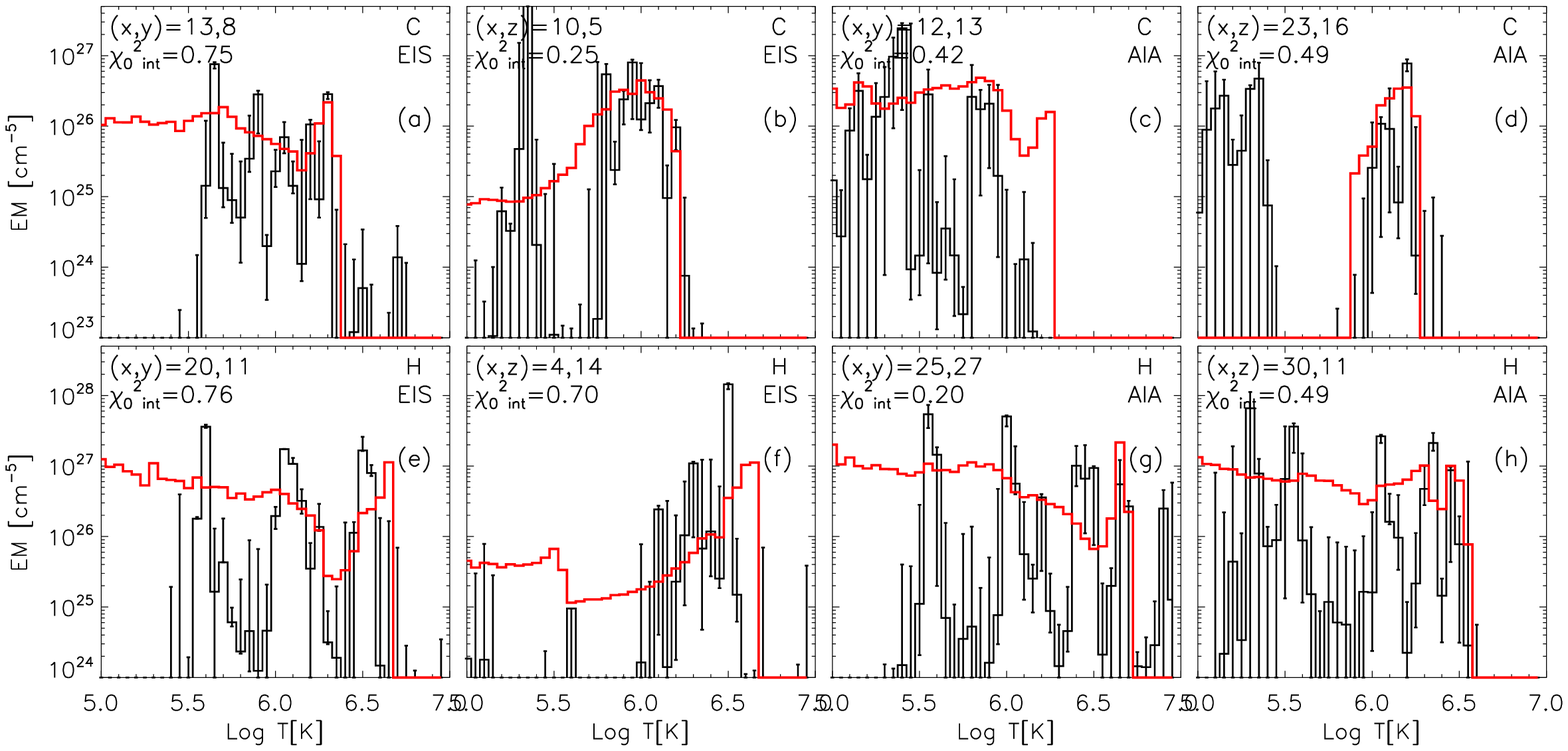}}\vspace{-0.7cm}
\caption{Sample of comparisons of true (red curves) and derived EMD
  (including noise),
  for one pixel for each case. The results from EIS synthetic data are
  shown in the left two columns, and the ones from AIA in the right 
  two columns.  The results for snapshot C are shown in the top row, while 
  the bottom four panels show the results for snapshot H.
  \label{fig:emt_sample}}
\end{figure*}

In the previous section we have addressed a comparison of the general 
properties of the EMDs derived from synthetic EIS and AIA data 
with the input emission measure distributions from 3D rMHD simulations.
In this section we will assess how well the derived EMDs reproduce the 
true EMDs in their details, on the finest temperature scale, and 
explore to what level of detail the MCMC EMD reconstruction method
is reliable.

In Figure~29 we have already shown the full 
EMDs for a selection of pixels chosen to explain the systematic 
discrepancies in $T(EMD_{max})$.
That set of plots already shows that, while as discussed above some 
general features of the EMD are recovered by the MCMC method, when 
comparing the EMDs on the fine temperature scale (i.e., here we used 
$\Delta \log T[K] = 0.05$ which is the typical intrinsic resolution of 
atomic data, and also the typical resolution used by observers), the 
derived EMD can significantly depart from the true distributions.

In Figure~\ref{fig:emt_sample} we show the comparison of the true
and derived EMD for a pixel selected for each case (for every snapshot 
and LOS, for the cases including noise), to show examples where,
even if the intensities are reproduced
within the errors ($\chi^2 \lesssim 1$), the derived and true EMD 
present significant discrepancies.

First we note that in most of the examples of 
Figures~29 and ~\ref{fig:emt_sample}, there 
are several temperature bins within the $\log T = [5.5-6.7]$ range 
where the true EMD is not compatible with the derived EMD within 
the uncertainties. 
In general we find that the EMDs resulting from the analysis of EIS
synthetic data reproduce the general properties of the EMDs - such as 
bulk of the emission and general shape - significantly better than
the corresponding AIA EMDs, as also shown in the previous 
section~\ref{ss:EM_maps}. 
However, even in the temperature range $\log T[K] \sim [5.5-6.7]$ 
where the emission measure distribution should be well constrained by 
the selected spectral lines, the best fit EMD can be very noisy and 
present peaks and valleys which are not present in the true EMD.

We note that, as discussed by \cite{Kashyap98} and \cite{Testa11}, the 
uncertainties associated with the EMD solution are correlated in the 
different temperature bins, and therefore, what are shown as error bars 
in the single temperature bins cannot strictly be interpreted as error 
bars for the EMD value in that bin, and instead they describe the range 
containing 68\% of the sets of solutions.
Nevertheless, since they are typically used as error bars we investigate
here also the limitations of this assumption.

\begin{figure*}[!ht]\vspace{0.5cm}
 \centerline{ \includegraphics[scale=0.37]{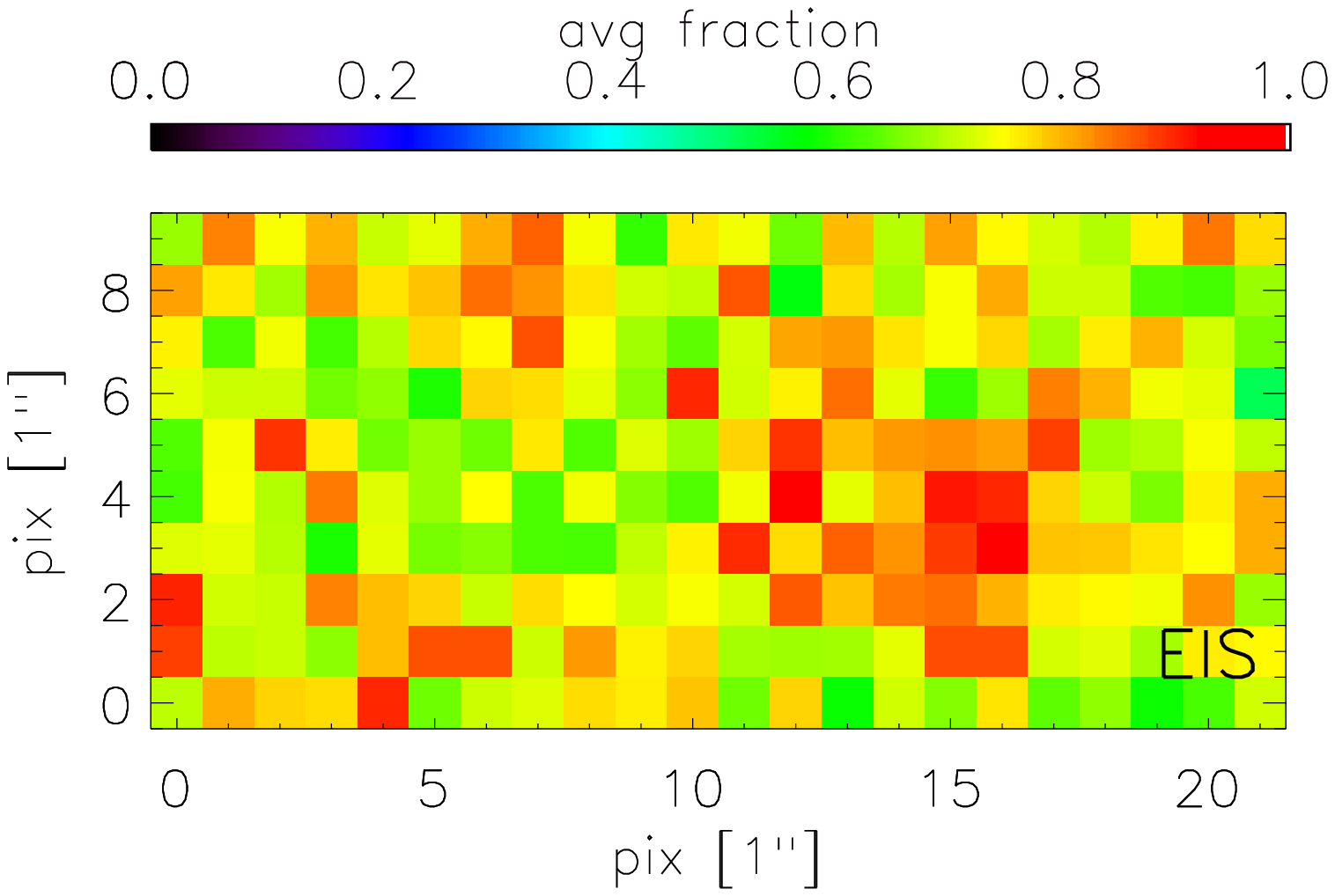}\hspace{-1cm}
  \includegraphics[scale=0.37]{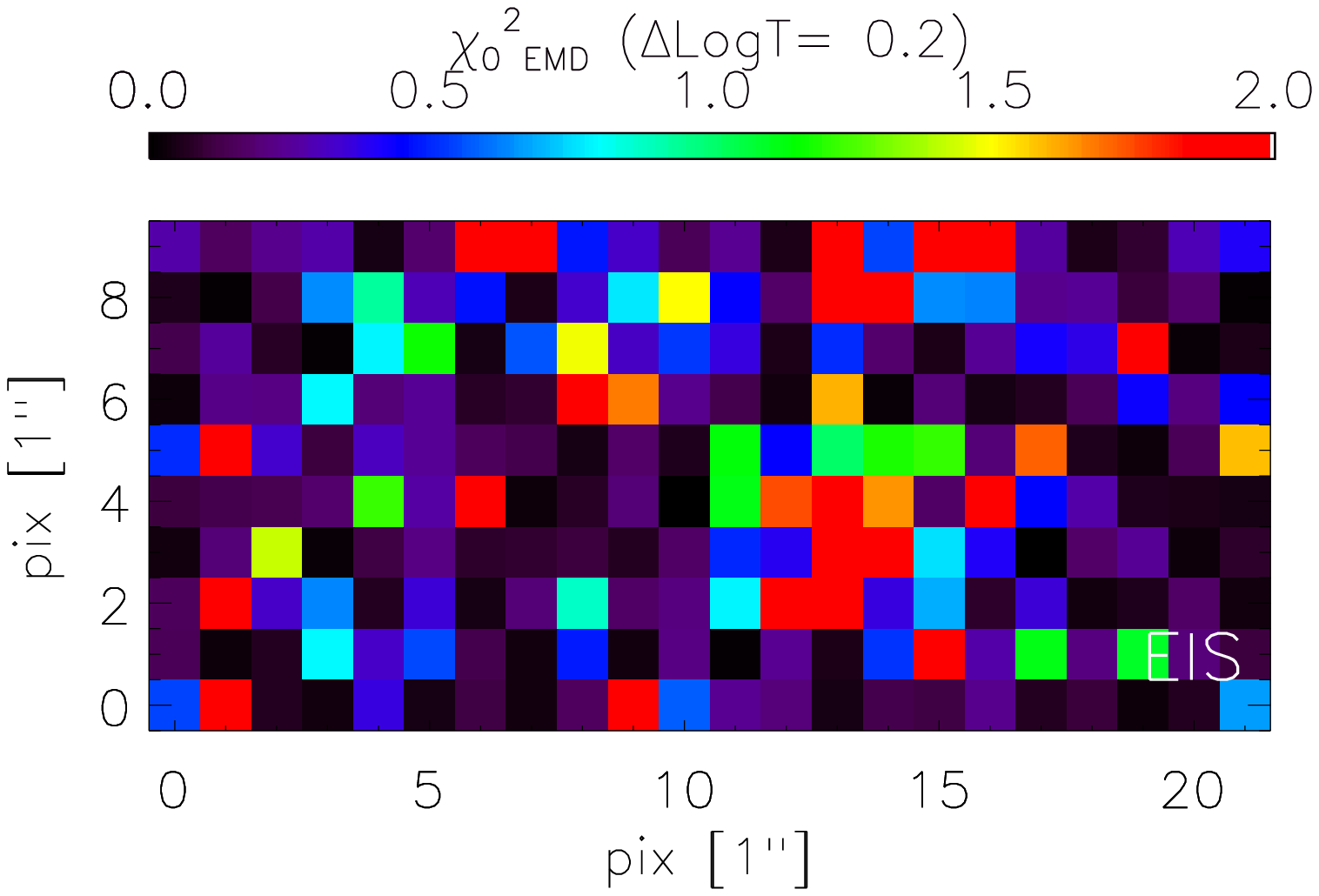}\hspace{-1cm}
  \includegraphics[scale=0.37]{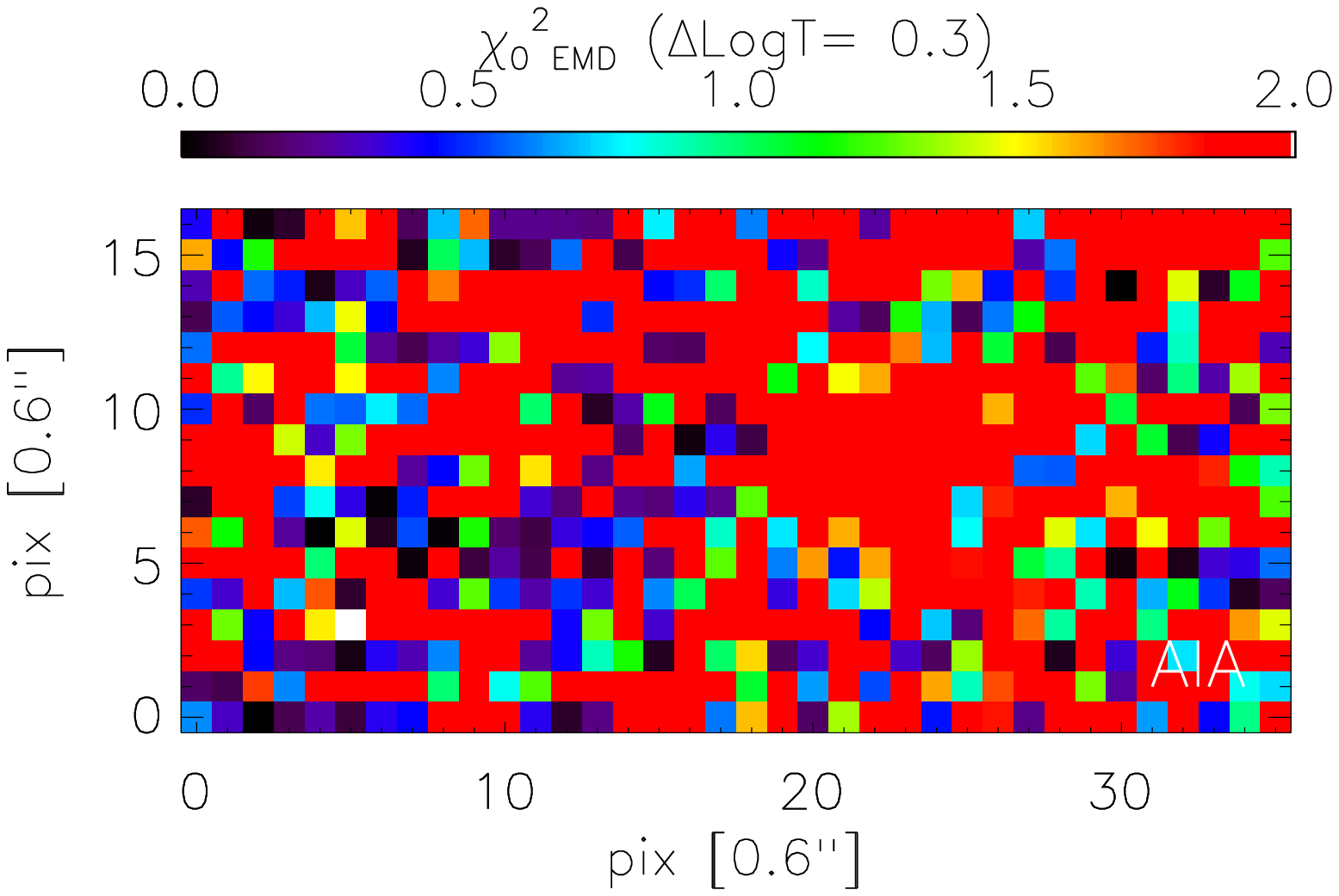}}
 \centerline{ \includegraphics[scale=0.37]{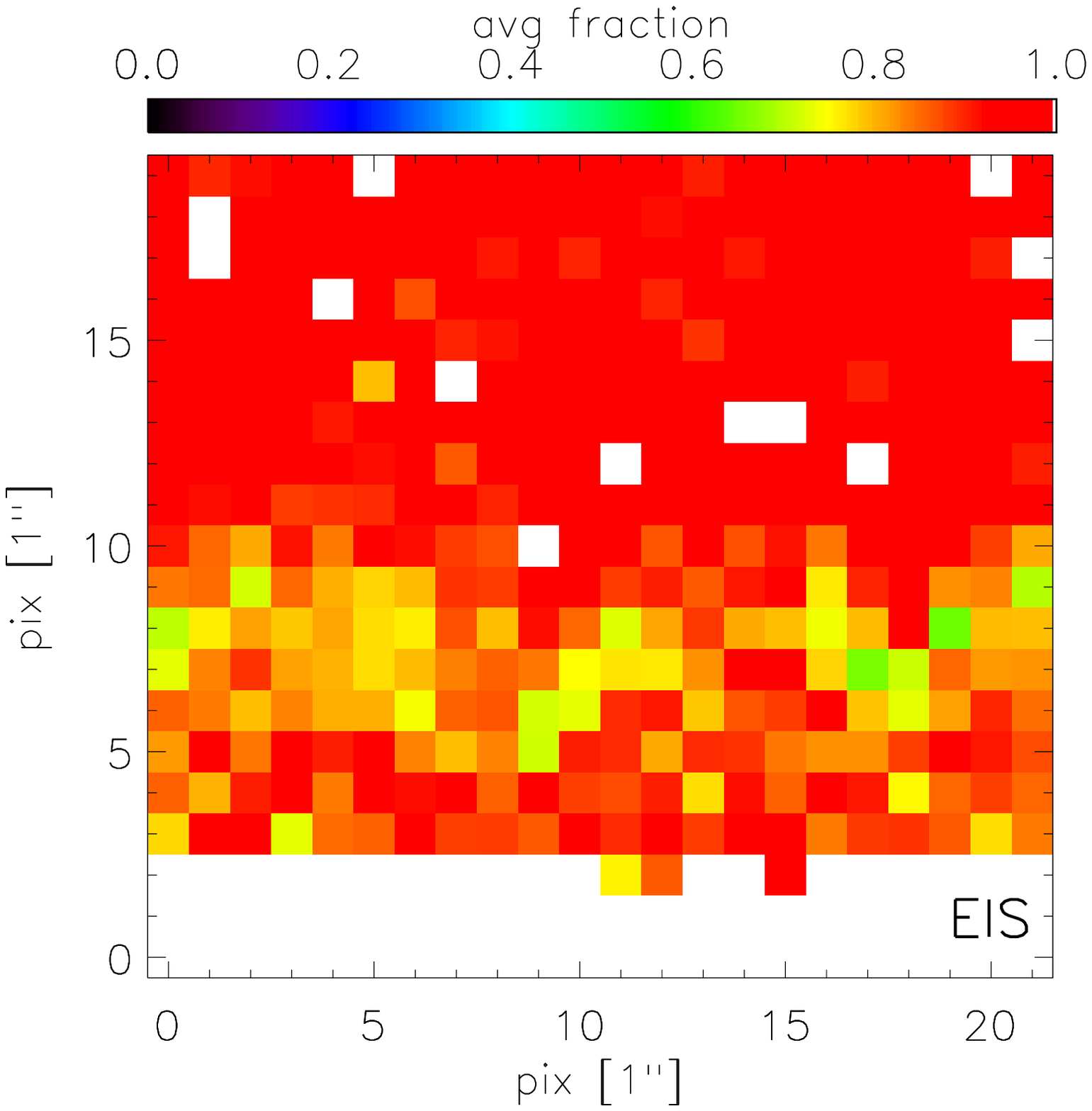}\hspace{-1cm}
  \includegraphics[scale=0.37]{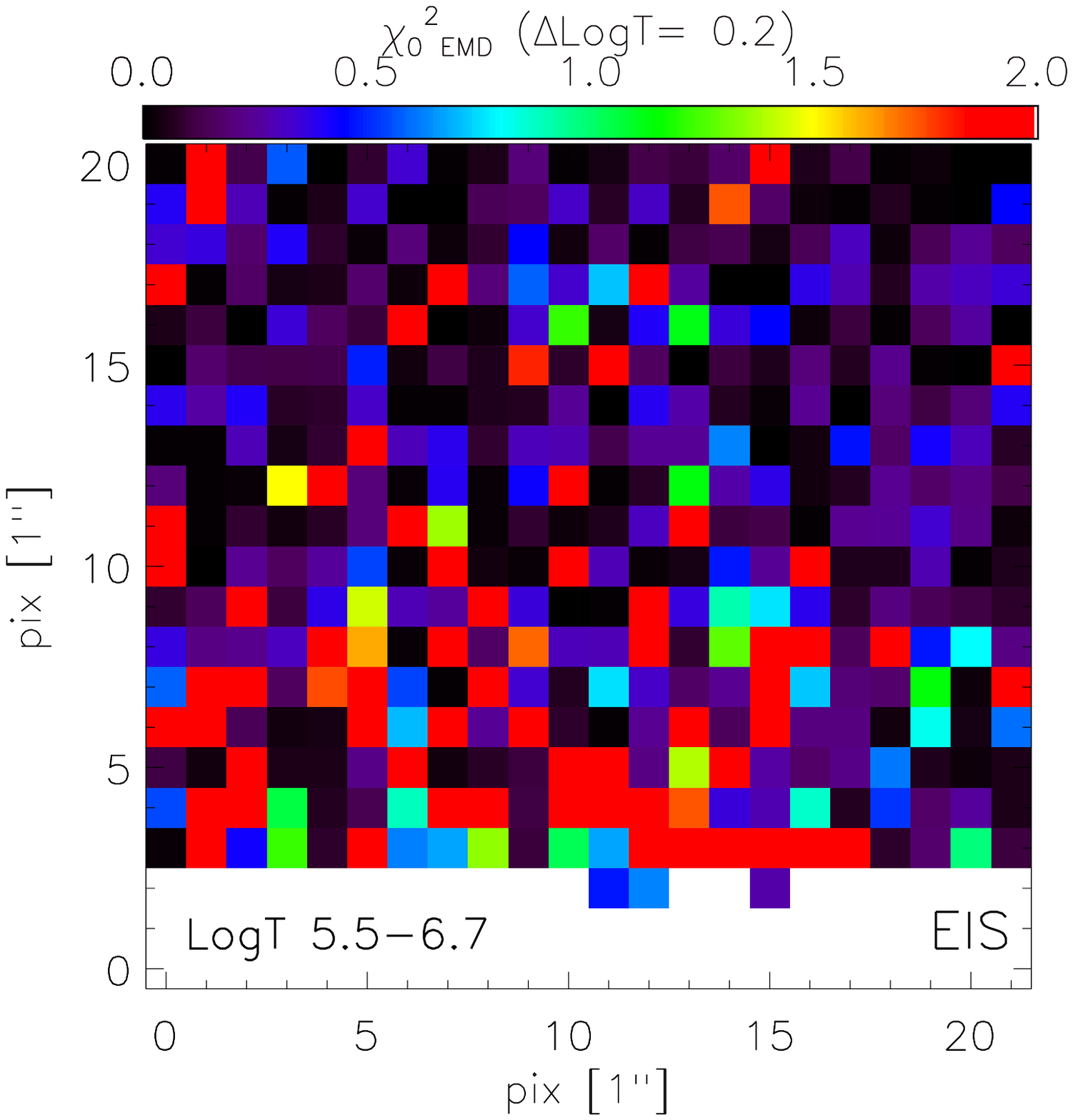}\hspace{-0.8cm}
  \includegraphics[scale=0.35]{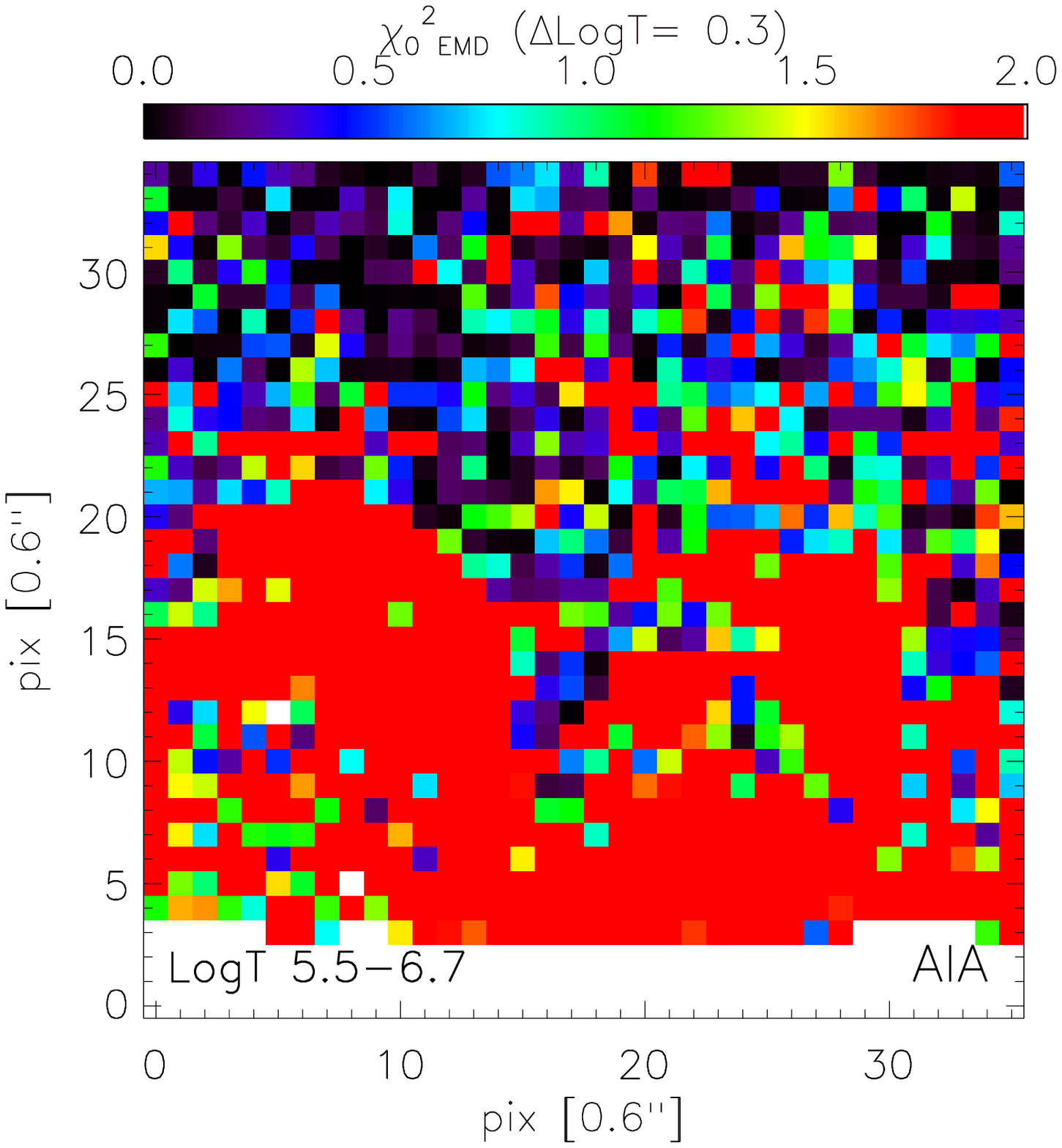}}\vspace{-0.1cm}
\caption{Left: Maps of the ``goodness of fit'' for the EMD for the EIS case 
  both top view (top row) and side view (bottom row) for snapshot C.
  The value in each pixel represents the average over all temperature bins between
  $\log T =5.5$ and 6.7, of the fraction of the solutions falling in the range 
  between the best fit EMD and the true EMD which is used as a measure of the 
  probability that the true EMD is part of the same distribution of the EMD solutions.
  A value close to 1 implies that for most temperature bins the true EMD is far from
  the bulk of the distribution. We use the full temperature resolution 
  ($\Delta \log T [K] =0.05$).
  Middle: Maps of $\chi^2(EMD)$ (see text for definition), parametrizing 
  the ability of the MCMC method to recover the full EMD in each temperature 
  bin from EIS data (including noise). We used $\Delta \log T [K] =0.2$.
  Right:  Maps of $\chi^2(EMD)$ for the EMD derived from AIA data 
  (including noise), and assuming $\Delta \log T [K] =0.3$.
  \label{fig:emtchi2_dyn_wn}}
\end{figure*}

\begin{figure*}[!ht]\vspace{0.5cm}
 \centerline{ \includegraphics[scale=0.37]{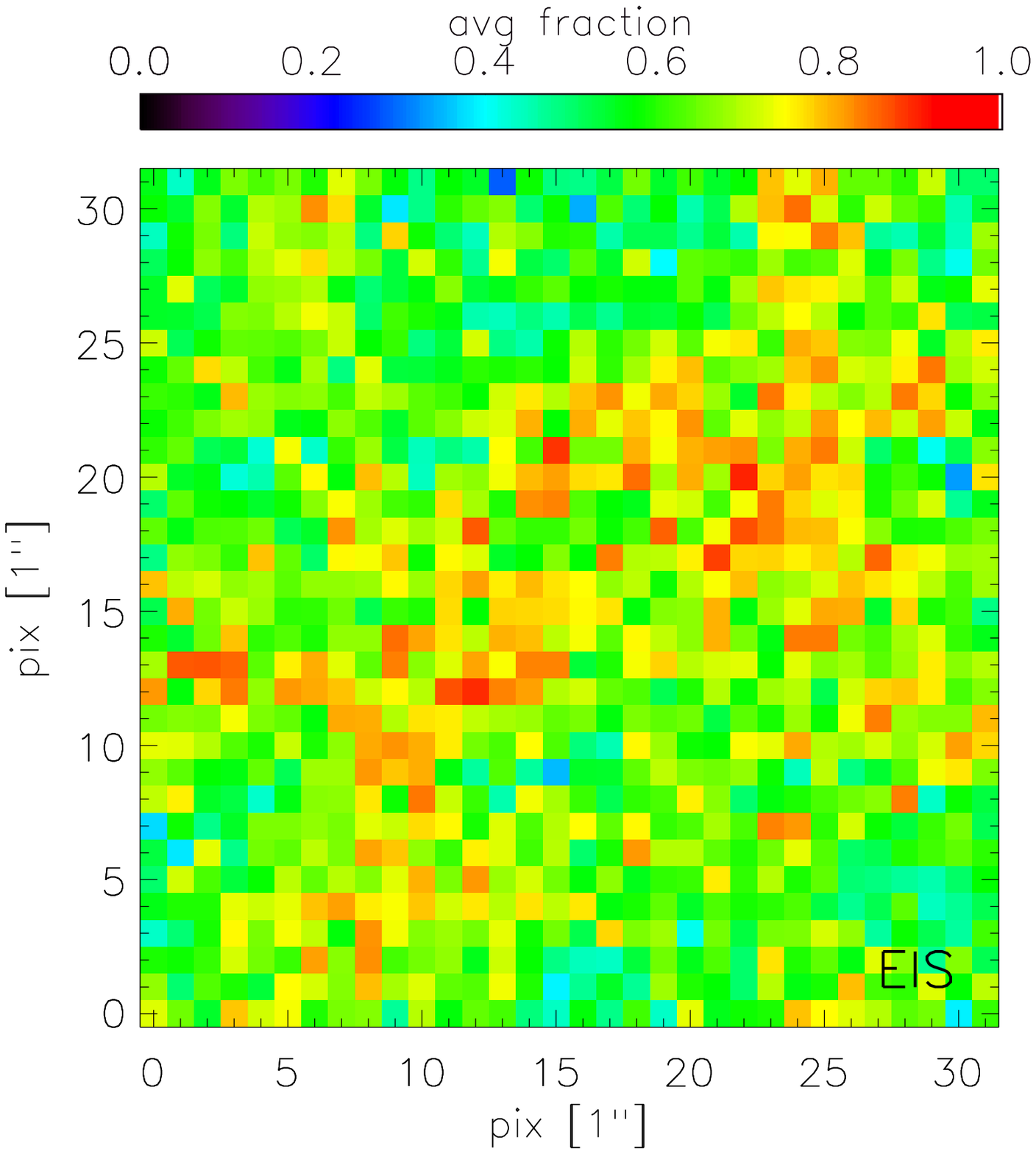}\hspace{-1cm}
  \includegraphics[scale=0.37]{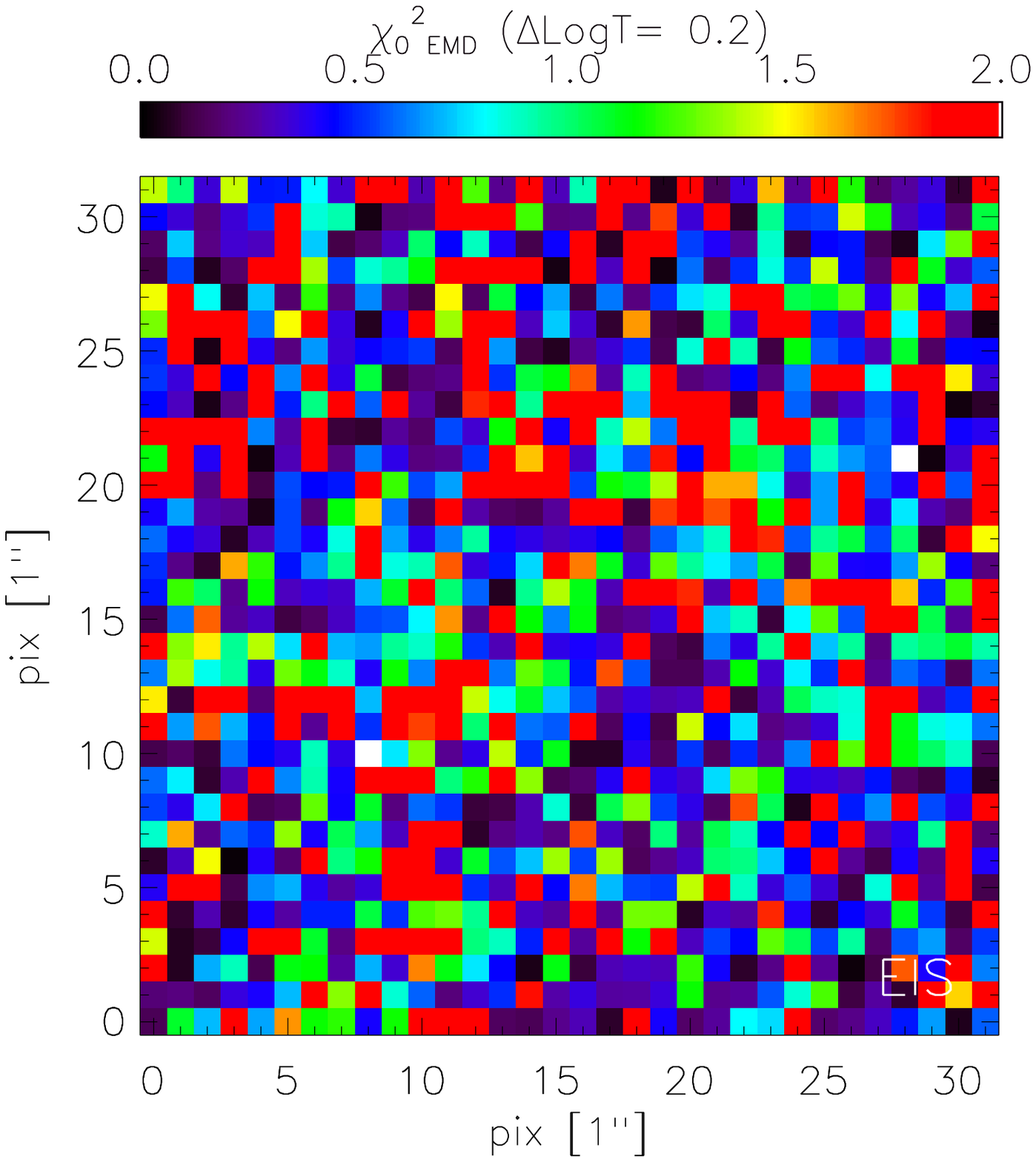}\hspace{-1cm}
  \includegraphics[scale=0.37]{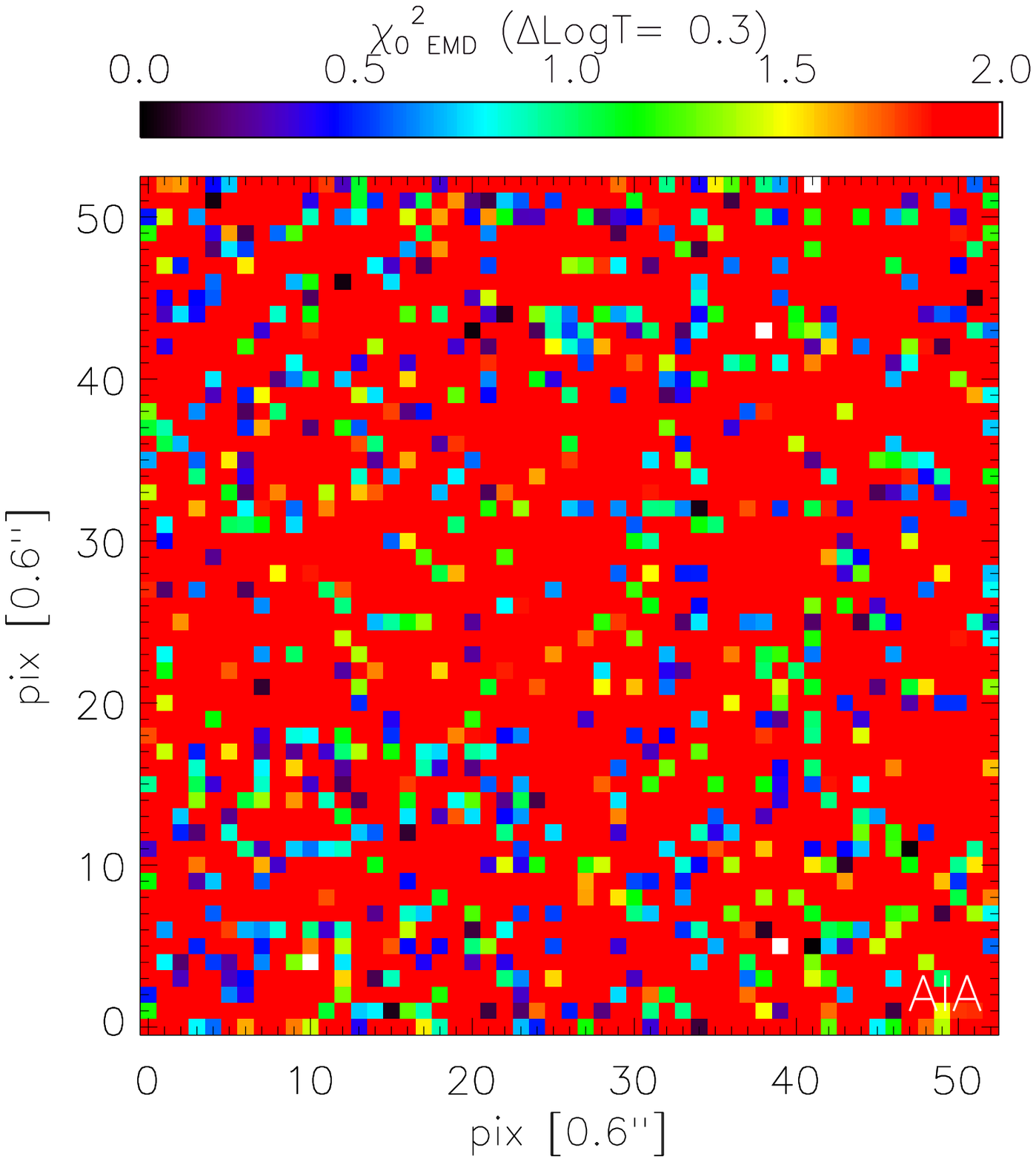}}
 \centerline{ \includegraphics[scale=0.37]{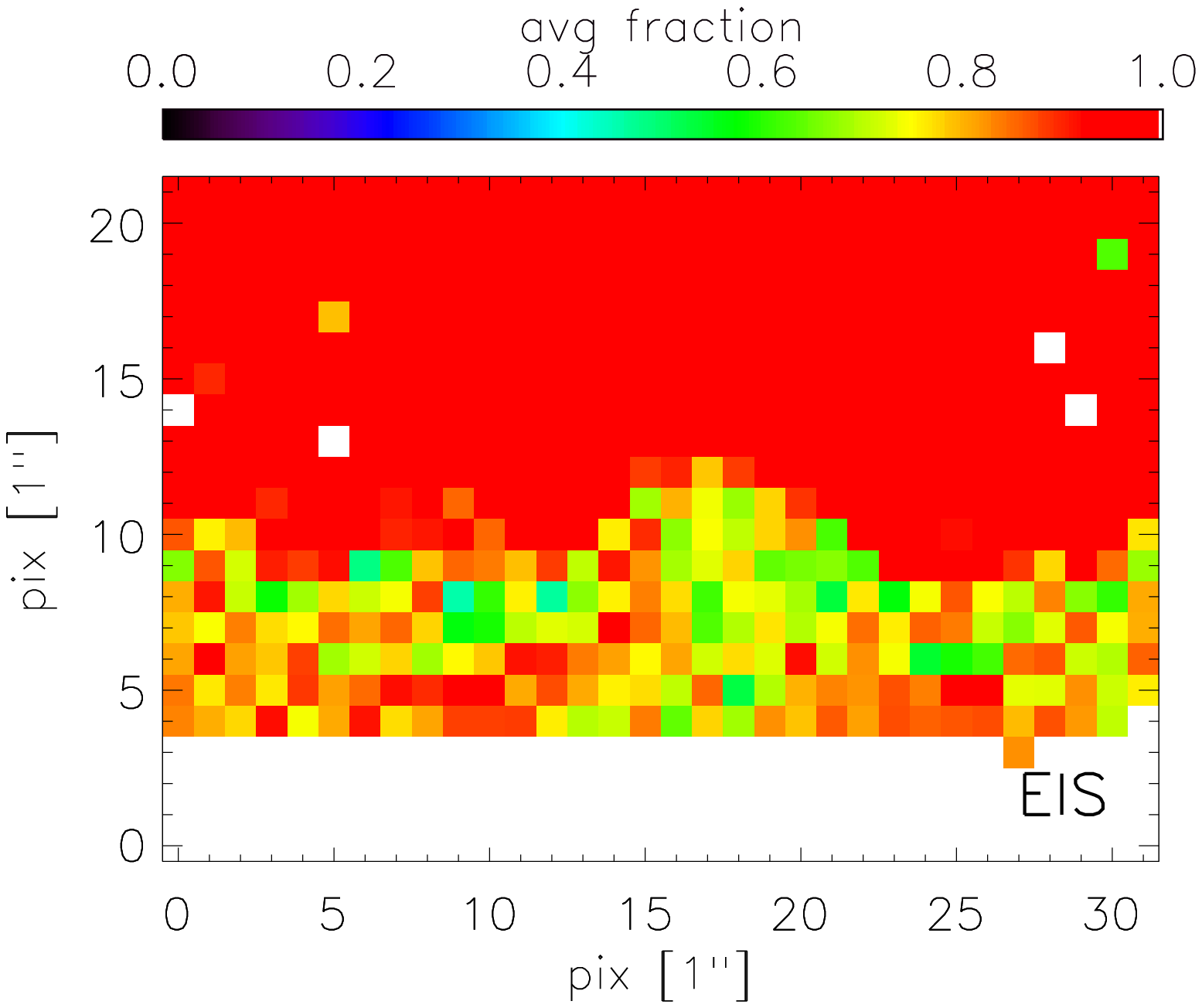}\hspace{-1cm}
  \includegraphics[scale=0.37]{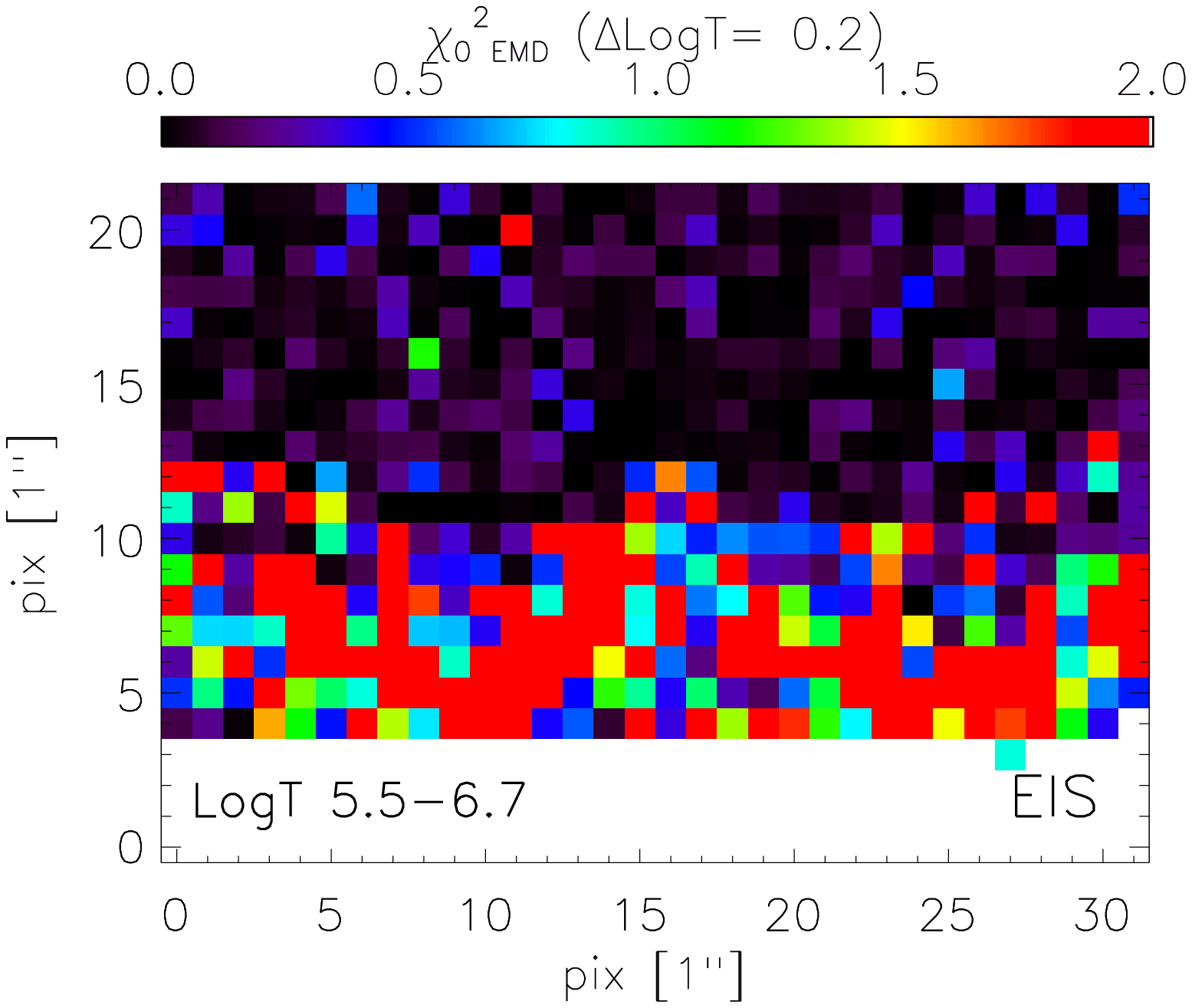}\hspace{-1cm}
  \includegraphics[scale=0.37]{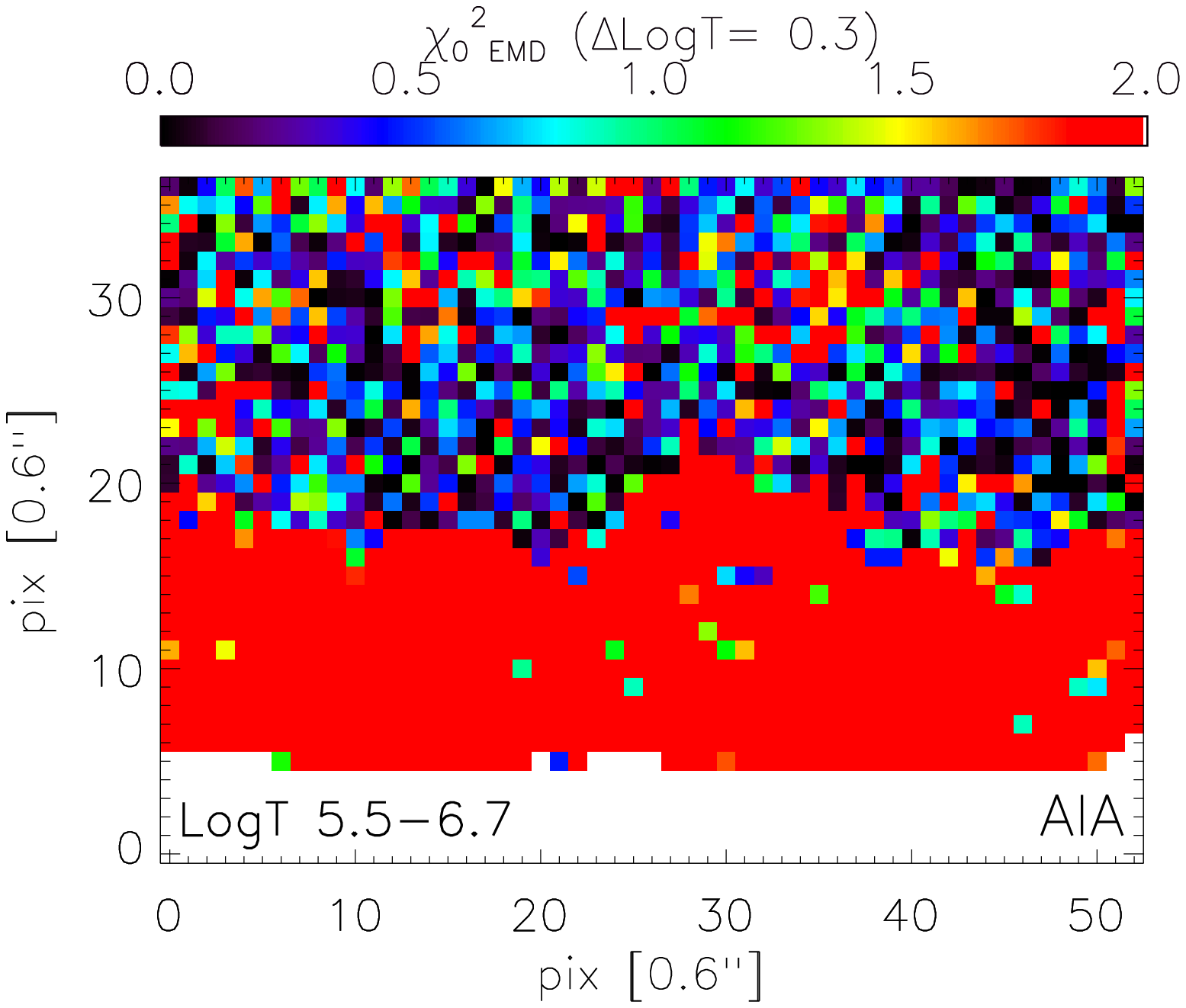}}\vspace{-0.1cm}
\caption{Plots analogous to plots in Figure~\ref{fig:emtchi2_dyn_wn} but for 
  snapshot H.
  \label{fig:emtchi2_cb_wn}}
\end{figure*}

In order to assess the ability of the MCMC method to recover the full EMD 
for all the different cases studied, we use two different ways to parametrize
the ``goodness of the fit''. 
First, we use the full information of the set of EMD solution produced 
by the MCMC routine for each pixel. For a given pixel, and a given 
temperature bin (we use here the full temperature resolution, i.e., 
$\Delta \log T [K] =0.05$), we can calculate the fraction of solutions 
included in the range between the true EMD value and the best fit value. 
This can be interpreted as a probability for the EMD value to be part of the
distribution of Monte Carlo solutions. Since it is not trivial to combine
the results of the different temperature bins, we compute an average
of these fraction over all temperature bins between $\log T =5.5$ 
and 6.7. If the average is close to 1, this implies that for most temperature 
bins the true EMD is far from the bulk of the distribution.
We show the maps of the values obtained from the EIS EMDs in 
this fashion in the left panels of Figure~\ref{fig:emtchi2_dyn_wn} (for
snapshot C) and \ref{fig:emtchi2_cb_wn} (for snapshot H).
The corresponding maps for the AIA EMDs are not shown but
are qualitatively very similar to the EIS cases shown here.
For the top view cases the maps present similar structuring to the 
$\chi^2_0$ maps shown in Figures~\ref{fig:chi2_e_dyn} and 
\ref{fig:chi2_e_cb}, with the worst match of EMD found in the regions
with large superpositions of plasma structures with different 
densities and temperatures.
For the side view, while lower in the atmosphere the values are similar to
the top view case, they become $\sim 1$ everywhere at larger heights.
The reason for this is that at larger heights the true EMD are close 
to isothermal, dropping to zero everywhere outside the peak region.
For all the temperature bins where the true EMD is zero, the derived EMDs
are non-zero causing large discrepancies in most of the temperature bins.

Another way to calculate a measure of how well the derived EMDs match
the true EMDs is to calculate a $\chi^2_0 (EMD)$ for the EMD, which we 
define as $= \Sigma_j [(EMD_{j, MCMC}-EMD_{j, true})/\sigma(EMD)_j]^2/df$, where 
$EMD_{j,true}$  and $EMD_{j, MCMC}$ are respectively the values of the true 
and derived EMD for the temperature bin $j$, $\sigma(EMD)_j$ is the
uncertainty of the EMD in the T bin $j$ defined by the range including 68\%
of the solutions, as described above, and $df$ are the degrees of freedom
(as defined at the beginning of section \ref{s:results}).
We find that when calculating the so defined $\chi^2_0 (EMD)$ using the 
fine temperature grid ($\Delta \log T = 0.05$) the $\chi^2_0 (EMD)$ values
are very large ($\gg 10$) for both the EIS and AIA cases.
Therefore we decrease the temperature resolution until the $\chi^2_0 (EMD)$
values become on average more reasonable. We find that this threshold 
is $\Delta \log T \sim 0.2$ for EIS, and we show the corresponding
maps of $\chi^2_0 (EMD)$ in the middle panels of 
Figure~\ref{fig:emtchi2_dyn_wn} and \ref{fig:emtchi2_cb_wn}.
These maps show the usual pattern of the features with large superposition
of different structures, where the agreement is worse, but for the other regions
the integration over large temperature bins yields a better match of the 
derived and true EMDs, by smoothing out the large variability on small 
temperature scale shown in Figures~29 
and ~\ref{fig:emt_sample}.
We note that for the side view cases, high in the corona, the uncertainties 
of these EMD on the coarser temperature grid appear even overestimated, 
since the $\chi^2_0 (EMD)$ drops to very low values. 
This is consistent with the findings of \cite{Landi12} who explored the 
ability of the MCMC methods to diagnose isothermal EMD. \cite{Landi12}
find indeed that for isothermal plasma a bin width of $\Delta \log T \sim 0.05$
is sufficient to diagnose the temperature distribution from spectral data.
For AIA, even with $\Delta \log T \sim 0.3$ the $\chi^2_0 (EMD)$ 
values (maps are shown in the right panels of Figure~\ref{fig:emtchi2_dyn_wn} 
and \ref{fig:emtchi2_cb_wn}) are large for the top view cases, in 
agreement with the results discussed in above section~\ref{ss:EM_maps}.
The side view cases show that with AIA the temperature binsize 
$\Delta \log T \sim 0.3$ is adequate to diagnose isothermal plasmas,
as shown by the low $\chi^2_0 (EMD)$ values at high $z$ values, but 
is still insufficient to guarantee a good match of the true EMD where
the distributions are significantly multi-thermal.
The EMDs derived AIA are a much less accurate reproduction
of the true EMDs compared to EIS.
This points to a significant problem in relying on AIA data 
exclusively, to diagnose the plasma temperature distribution. 
This can be explained by considering two main factors: (1) the smaller
number of constraints - AIA data provide a considerably more limited 
number of constraints (at best 6) compared to EIS; (2)  imaging data have 
significantly more limited temperature diagnostics than spectral data, 
because of their broader temperature sensitivity (even for the narrow 
AIA passbands) compared to the resolved spectral lines.   
In this respect, though outside the scope of this paper, we note
that narrow band coronal imagers also have the additional 
disadvantage with respect to broad band instruments of being
critically sensitive to the uncertainties in the atomic data
\citep[see, e.g.,][]{Testa12}.
We note that the EMDs derived from AIA present significant 
discrepancies with the true EMDs even in the case that does not 
include noise, or in the test case with very high S/N discussed at 
the beginning of this section when we discussed the fit to the 
observed intensities.
We therefore conclude that the limitations in diagnosing EMDs with 
AIA data are mainly due to the intrinsic characteristics of the (broad) 
AIA temperature responses, and that the noise is not the main cause of 
the discrepancies between true and derived EMDs.

While our results imply that in general the MCMC method does 
not reproduce the true EMDs at the highest temperature resolution, 
it is preferable for several reasons to run the method at high 
resolution ($\Delta \log T \sim 0.05$) and then rebin the solution
a posteriori. 
As discussed above (and shown in Figure~\ref{fig:emtchi2_dyn_wn} 
and \ref{fig:emtchi2_cb_wn}), the high resolution allows one to 
find nearly isothermal solutions. Also, we have run for a few
pixels the MCMC on the EIS synthetic data using different values
of $\Delta \log T$ and found that adopting a too large bin size 
leads to less stable solutions, due to the rebinning of the temperature 
response function and consequent loss of temperature information. 
For instance if the temperature bins become significantly large 
compared to the peaks in the temperature response functions the 
rebinned responses often end up with peaks which are much less 
prominent, and significantly displaced in temperature.

\section{Discussion and conclusions}
\label{s:conclusions}

We have presented the results of a test of the limitations of
plasma temperature diagnostics that are currently available from spectral
and imaging observations of the solar corona. 
Determining the plasma emission measure distribution is fraught with 
difficulties, both because of the intrinsic nature of the mathematical 
inversion problem, which is ill-posed, and because of the limitations of 
the available data.
While imaging instruments provide only moderate temperature diagnostics
(compared with spectrographs that can resolve spectral lines), they
typically provide significantly better spatial coverage and resolution, and 
temporal cadence. 

Here we used advanced 3D radiative MHD simulations of the solar 
atmosphere \citep{Hansteen07,Gudiksen11} as realistic test cases,
from which we produced synthetic observables, and applied
a Monte Carlo Markov chain forward modeling technique to derive
the emission measure distributions (EMD).
We then compared the results of EMD reconstruction from imaging (AIA) 
and spectral (EIS) synthetic data (based on the coronal properties in
the models) with the input ``true'' EMD to establish the limitations 
of the temperature diagnostic power of the available instruments, 
and how these results depend on the characteristics of the underlying 
thermal distributions.  
We also investigated the effect of the photon counting noise, by 
running the analysis with or without randomization accounting for 
Poisson noise.

These 3D simulations provide us with the opportunity to improve upon
previous work by exploring more realistic configurations, with
significant superposition of different structures along the LOS,
allowing a statistical approach to determine the accuracy and
limitations of the plasma diagnostics, for a variety of spatial and
thermal structuring of the plasma.   
The three dimensional nature of the simulations also allows us to
explore a variety of realistic viewing angles, reproducing typical
distributions of structuring ranging from on disk to limb observations. 

We assess the robustness of the EMD reconstruction method by 
using several parameters. We explored the ability of the method
in: (a) reproducing the ``measured'' intensities; (b) determining
the temperature of the peak of the emission measure distribution;
(c) deriving the emission measure values in broad temperature 
bins; (d) reproducing the true emission distribution in its details.

The analysis of the spectral EIS synthetic data show that the measured
intensities are reproduced generally well by the inferred EMD, even
when including the effect of noise, with the exception of regions
with mixing of regions with significantly different densities along 
the LOS. 
The temperature of the peak of the EMD is reproduced reasonably well
for the side view, in larger areas where the EMD is close to isothermal.
For the top view, where the EMD are broad and in particular where there
are large amounts of dense material and large superposition of different
structures (emerging flux region, moss), the temperature of the peak 
EMD appears to be systematically underestimated. 
The maps of emission measure, when integrated in broad temperature 
bins, are similar to the true distributions, in the well constrained range 
($\log (T[K]) \sim [5.6-6.4]$); the low and high temperature ends of 
the considered range are rather noisy and present spurious components. 
However, when considering the direct comparison of the derived EMD 
curves with the true distributions, at the full temperature resolution, 
we find that the detailed properties of the true EMD on the smallest
temperature scale are not accurately reproduced.
The best fit EMD can present peaks which are not displayed by the 
true EMD, or vice versa, narrow peaks might be missed by the 
reconstruction method. The worst cases show how the uncertainties
do not generally account for these discrepancies, and are therefore
likely underestimated by the method.
We explore the temperature resolution at which the discrepancies 
between true and derived EMD become smaller, and we find that for
EIS the $\Delta \log T$ needs to be $\gtrsim 0.2$.
We therefore conclude that the temperature diagnostic provided by 
spectral data with enough observational constrains and good 
temperature coverage, are on average reliable to derive the 
general characteristics of the emission measure distribution, such
as the emission measure in broad temperature bins, but that the
results are not robust enough as far as the fine details of the EMD, 
such as the presence of narrow peaks, are concerned, and that 
features on scales below $\gtrsim 0.2$ in $\log T$ cannot be trusted.

The results obtained from the analysis of the AIA synthetic data
are more unsatisfactory. The synthetic intensities for the side view
LOS are low in some channels, making the results more sensitive 
to the effect of the noise.
The maps of temperature at which the EMDs peak show that, as 
for the EIS case, the temperatures are not very well reproduced, 
and appear systematically underestimated, especially in the top 
view case. 
Also the maps of emission measure integrated in broad temperature 
bins are a poor match for the true distributions, especially for the top
view case.
Finally, the direct comparison of derived and true EMD curves at the 
full temperature resolution indicates discrepancies well beyond the 
errors, implying that the EMD are not at all well constrained by the 
AIA data, often not even in their general characteristics.
Even degrading the temperature resolution to $\Delta \log T = 0.3$
the derived EMD do not match accurately the true EMD, especially
where the temperature distributions are broad.
We therefore conclude that the limited number of constraints and the 
broad temperature sensitivity of the AIA passbands, which often 
present multiple peaks, critically hamper the capability of diagnosing 
the plasma temperature distribution on the basis of AIA imaging 
observations exclusively.
The addition of simultaneous {\em Hinode}/XRT data, which 
provide complementary imaging observations in X-ray
broadbands, might improve considerably the ability to reconstruct
EMDs from AIA data, by providing additional constraints, especially
at the high temperature end ($\log T \sim 6.5-7$), where 
AIA provides limited information. We plan to explore this issue
in a follow-up paper.
In this paper we have not addressed the use of EIS and AIA 
together because, as discussed here above and in other papers
(see e.g., \citealt{Warren11,Testa11}), the spectral data, when 
available, provide much better constraints to the temperature 
distribution of the plasma, compared with imaging data, and 
the imaging data provide only limited additional constraints. 
However, imaging data provide a much better temporal and
spatial coverage of the coronal plasma, so we focused on 
investigating the usefulness of imaging data for thermal 
diagnostics, when spectral data are not available, as in the
large majority of solar observations. 

The approach adopted here, using 3D simulations, has provided 
us with the opportunity to improve upon previous work by 
exploring more realistic configurations, with significant 
superposition of different structures along the LOS, allowing 
a statistical approach to determine the accuracy and limitations 
of the plasma diagnostics, for a variety of spatial and
thermal structuring of the plasma.   
These results provide a stringent and accurate assessment of
the limitations of the temperature diagnostics, indicating the
extent of the sensitivity of EMD measurements to the structuring
along the LOS, and to characteristics inherent to the 
instrumentation. 
However, we emphasize that the exercise we carried out focuses 
on the use of the MCMC reconstruction method, and therefore it 
only assesses the robustness of this particular method, which is 
however largely adopted and trusted for studies of emission
measure distributions of solar and stellar coronal plasmas.
We note that these results actually present an optimistic view, 
because they do not account for several effects that further 
complicate real observations, such as for instance uncertainties 
in both instrument calibration and atomic data, influence of blends
in fitting the spectral data, and distribution of element abundances. 
Finally, the 3D simulations adopted here model limited coronal 
volumes, therefore likely underestimating the effects of superposition
along the LOS.

\begin{acknowledgements}
We thank Vinay Kashyap, Karel Schrijver, Fabio Reale, Ed DeLuca, 
Paul Boerner, Enrico Landi for useful discussions and comments that 
have helped improve the paper. 
PT was supported by contract SP02H1701R from Lockheed-Martin, NASA
contract NNM07AB07C to the Smithsonian Astrophysical Observatory, and
NASA grants NNX10AF29G and NNX11AC20G. 
BDP gratefully acknowledges support by NASA grants NNX08AH45G, 
NNX08BA99G, and NNX11AN98G.
The 3D simulations have been run on clusters from the Notur project, 
and the Pleiades cluster through computing grants 
SMD-07-0434, SMD-08-0743, SMD-09-1128, SMD-09-1336, 
SMD-10-1622, SMD-10-1869, SMD-11-2312, and 
SMD-11-2752 from the High End Computing (HEC) division of NASA. 
We thankfully acknowledge the computer and supercomputer 
resources of the Research Council of Norway through grant 170935/V30 
and through grants of computing time from the Programme for 
Supercomputing. 
This work has benefited from discussions at the International Space 
Science Institute (ISSI) meeting on ``Coronal Heating - Using Observables 
(flows and emission measure) to Settle the Question of Steady vs. 
Impulsive Heating''  February 27- March 1 2012, where many topics 
relevant to this work were discussed with other colleagues.
\end{acknowledgements}

\bibliographystyle{/Users/ptesta/WORK/PAPERS/apj}

\newpage

\appendix
\section{Comparison of synthetic intensities with actual observations
\label{app:comp_obs}}

\begin{figure}[!t]
\centerline{\includegraphics[scale=0.6]{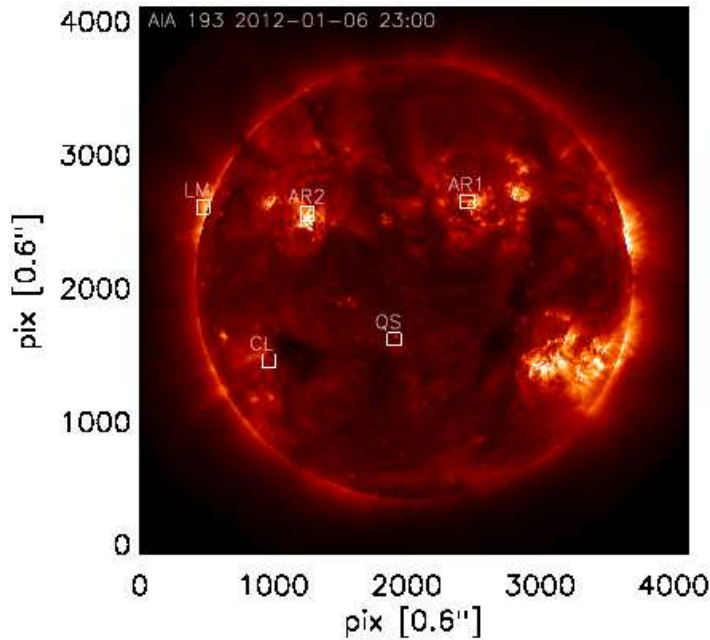}}
\caption{SDO/AIA 193\AA\ channel full disk observation taken on 2012 January 6 
  around 23UT, where we selected a subset of small regions ($100 \times 100$ 
  pixels) of different coronal features: quiet Sun (QS), cool fan loops (CL), limb (LM),
  and active region plasma (AR1, AR2).  In Figure~\ref{fig:his_obs} we show the 
  distributions of the 6 AIA EUV coronal channels intensities for these selected regions,
  to be compared with the distributions of the synthetic intensities derived 
  from the simulations (shown in Figure~\ref{fig:his_sim}).  \label{fig:sel_obs}}
\end{figure}

In this appendix we present a comparison of the synthetic intensities 
derived from the simulations, as described in section \ref{ss:model}, 
with measured intensities from recent SDO/AIA observations.
We considered AIA observations taken on 2012 January 6, around 23UT
(the 193\AA\ full disk image is shown in Figure~\ref{fig:sel_obs}), and 
selected a few small regions (100 pixels $\times$ 100 pixels, 
corresponding to $\sim 60$~arcsec $\times$ 60~arcsec) sampling a 
variety of coronal features, ranging from quiet Sun to active regions, 
and including areas both on-disk and above limb.

\begin{figure*}[!t]
\centerline{\includegraphics[scale=0.4]{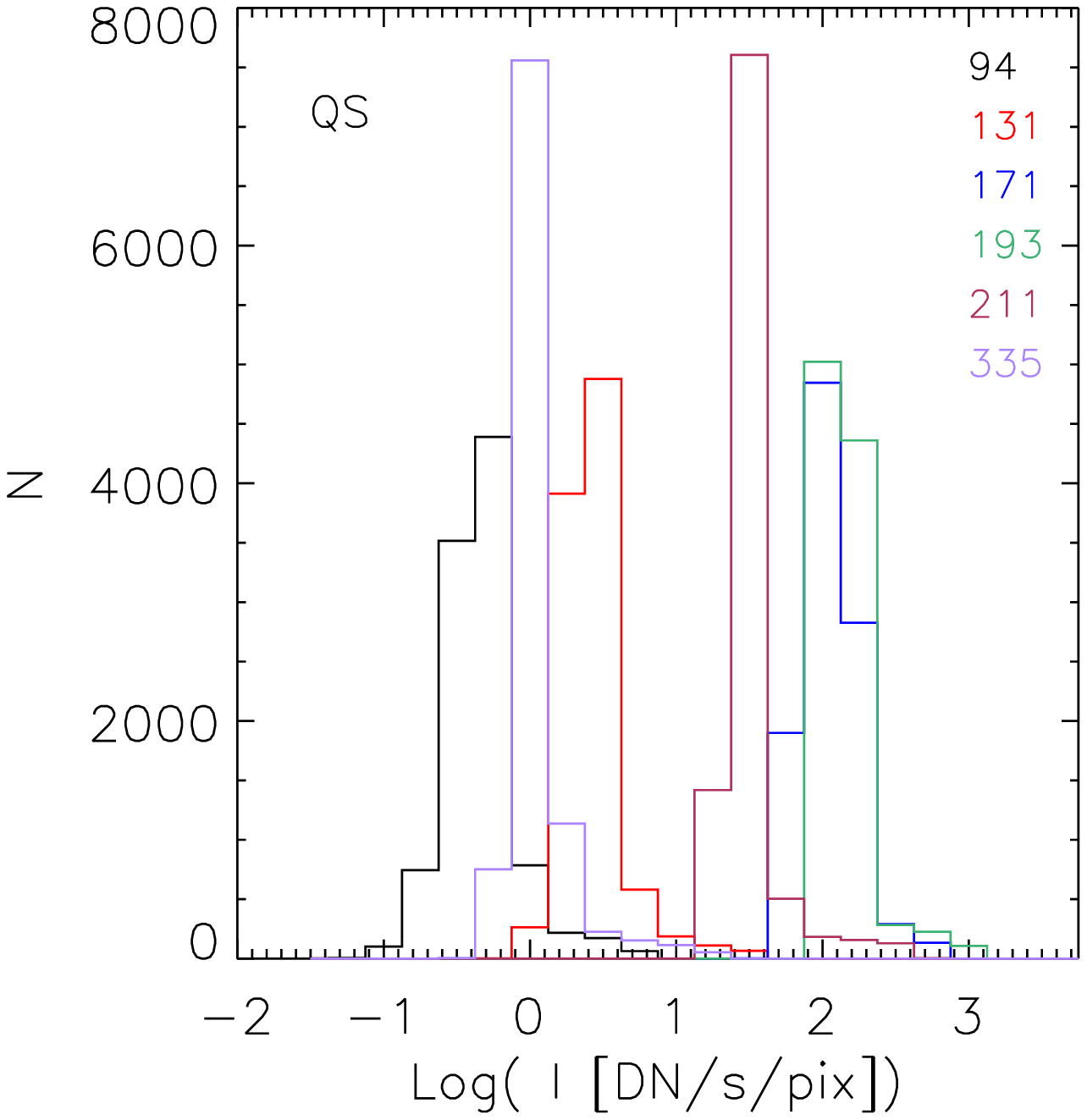}\hspace{-0.5cm}
  \includegraphics[scale=0.4]{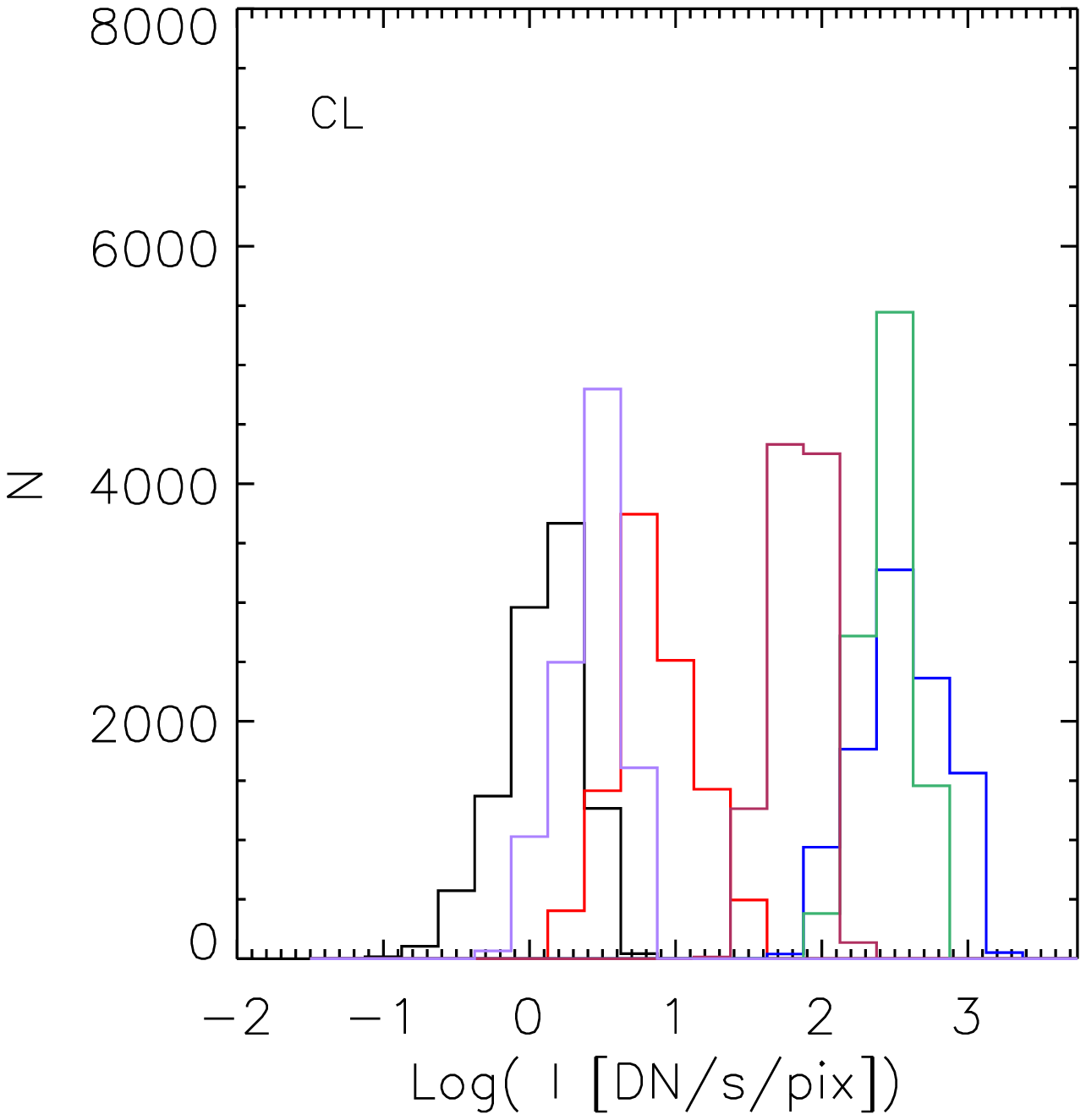}\hspace{-0.5cm}
  \includegraphics[scale=0.4]{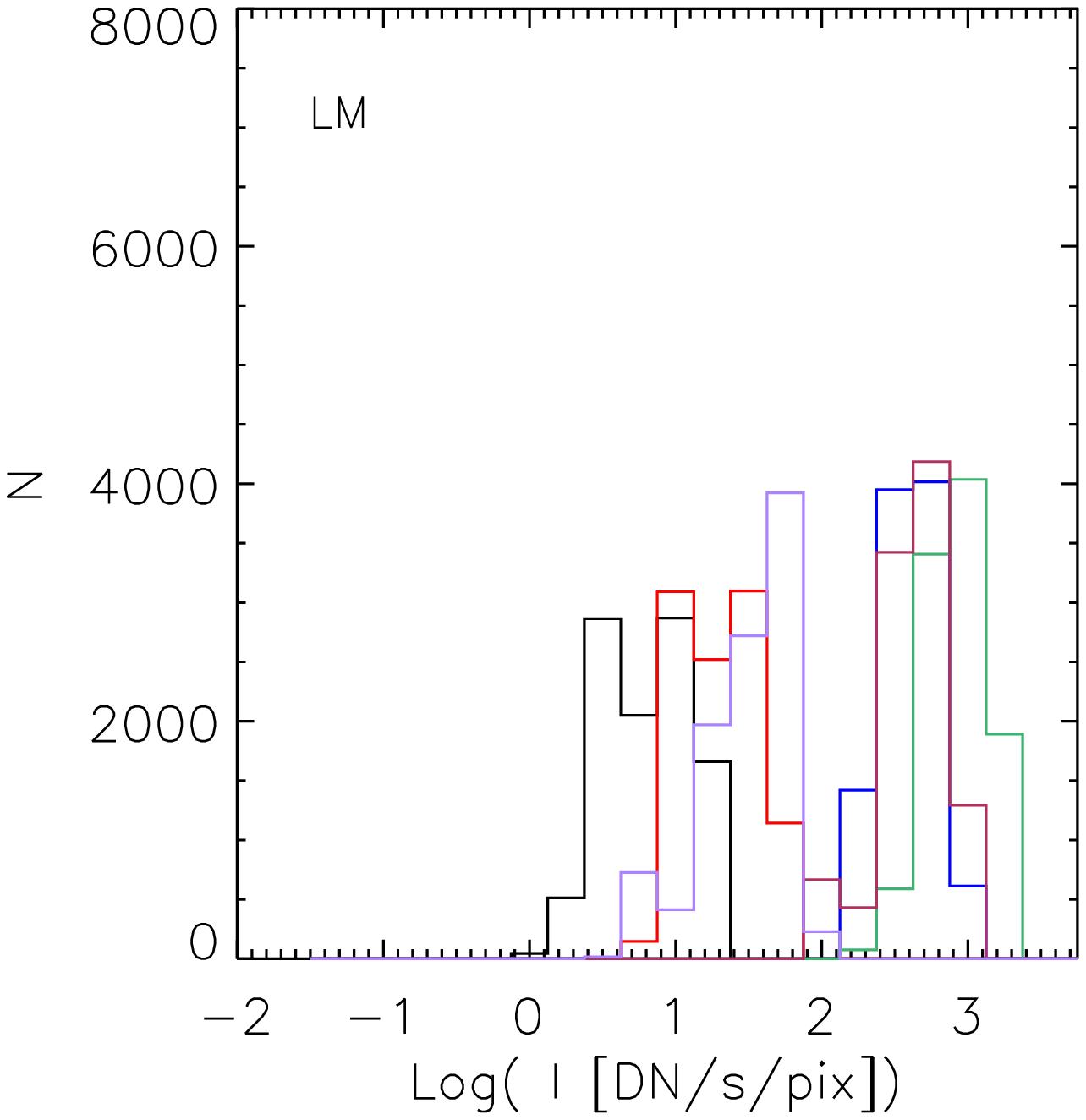}\vspace{-0.2cm}}
\centerline{\includegraphics[scale=0.4]{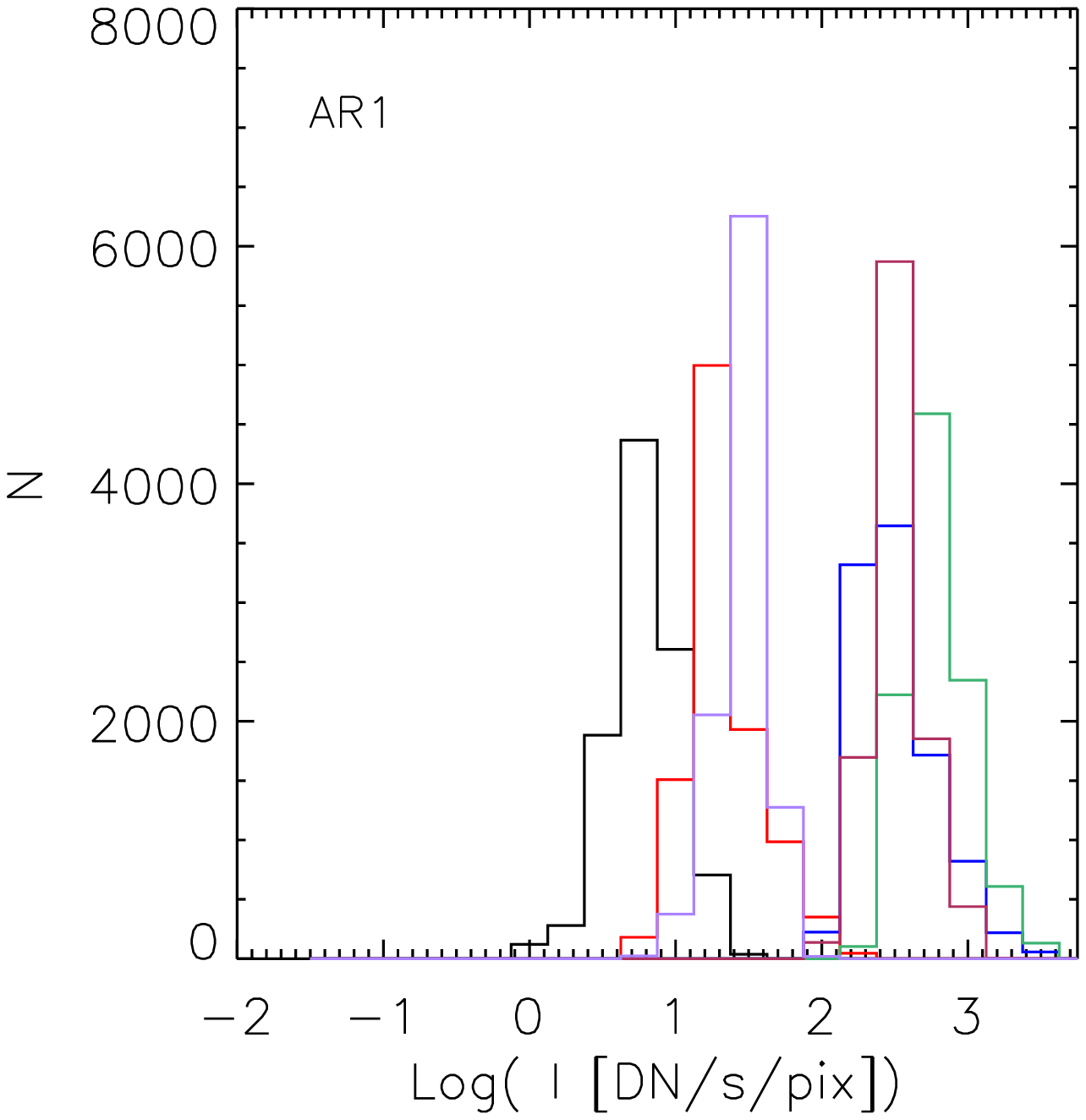}\hspace{-0.5cm}
  \includegraphics[scale=0.4]{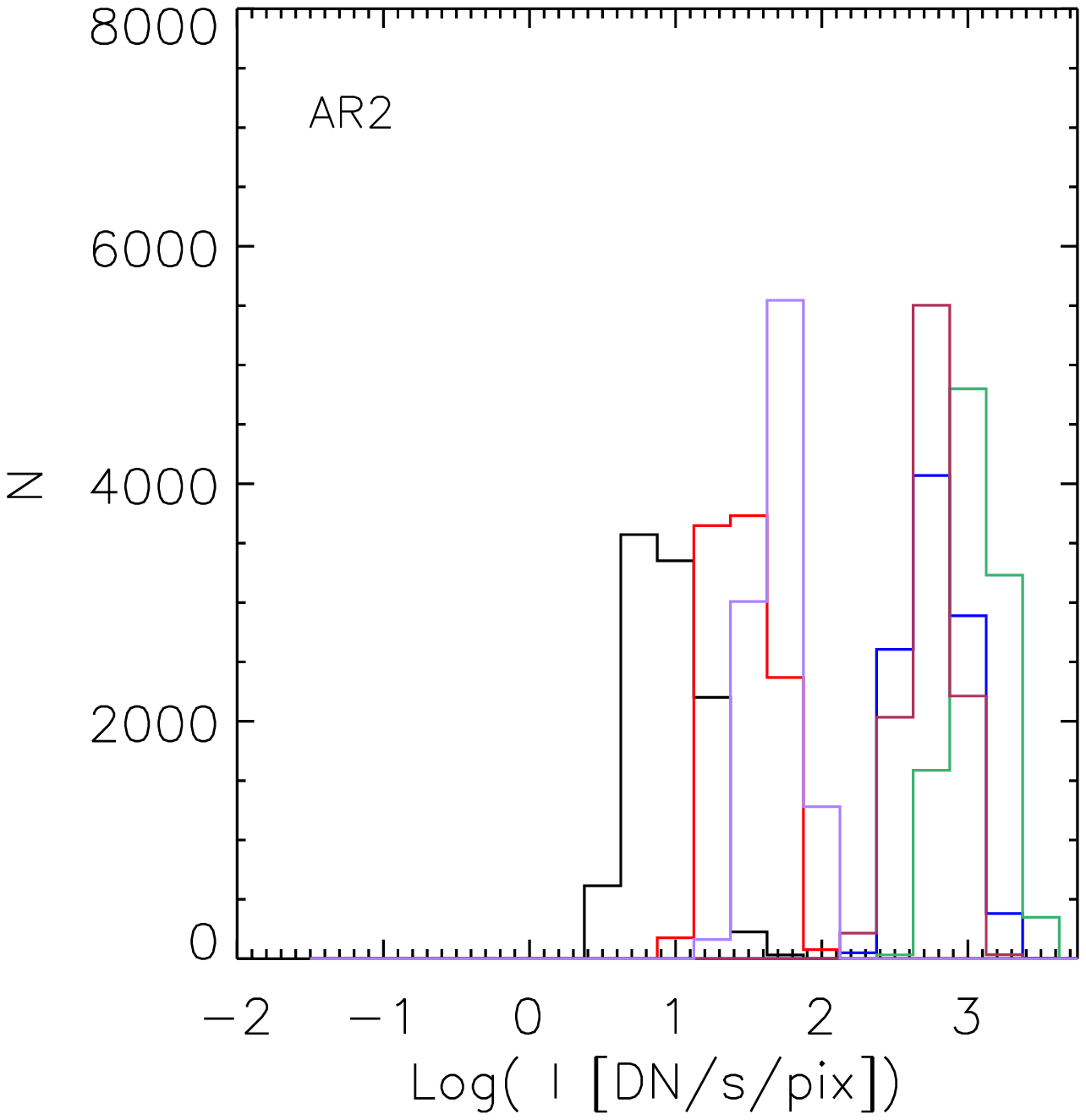}}
\caption{Histograms showing the distributions of the observed 
  intensities (in DN~s$^{-1}$~pix$^{-1}$) in the AIA channels (shown in 
  different colors, as labeled in the top left panel, QS), for the regions 
  indicated in Figure~\ref{fig:sel_obs}: quiet Sun (QS), cool fan loops (CL), 
  limb (LM), and active region plasma (AR1, AR2).
  \label{fig:his_obs}}
\end{figure*}

In Figure~\ref{fig:his_obs} we show the distribution of the observed intensities
(in units of DN~s$^{-1}$~pix$^{-1}$) in the six selected AIA EUV bands, for the 
different regions shown and labeled in Figure~\ref{fig:sel_obs}. These plots 
clearly show that: (a) average absolute intensities change by more than an
order of magnitude from region to region; (b) the distribution in each channel,
in each of the regions is rather narrow; (c) the distributions of intensities in the
94\AA, 131\AA, and 335\AA\ channels typically peak at values about two orders
of magnitude lower than the intensities in the other three bright channels; (d)
the intensities in the 211\AA, and 335\AA\ channels typically have relative
lower values in cooler regions (quiet Sun and cool fan loops; regions QS and CL).

\begin{figure*}[!t]
\centerline{\includegraphics[scale=0.45]{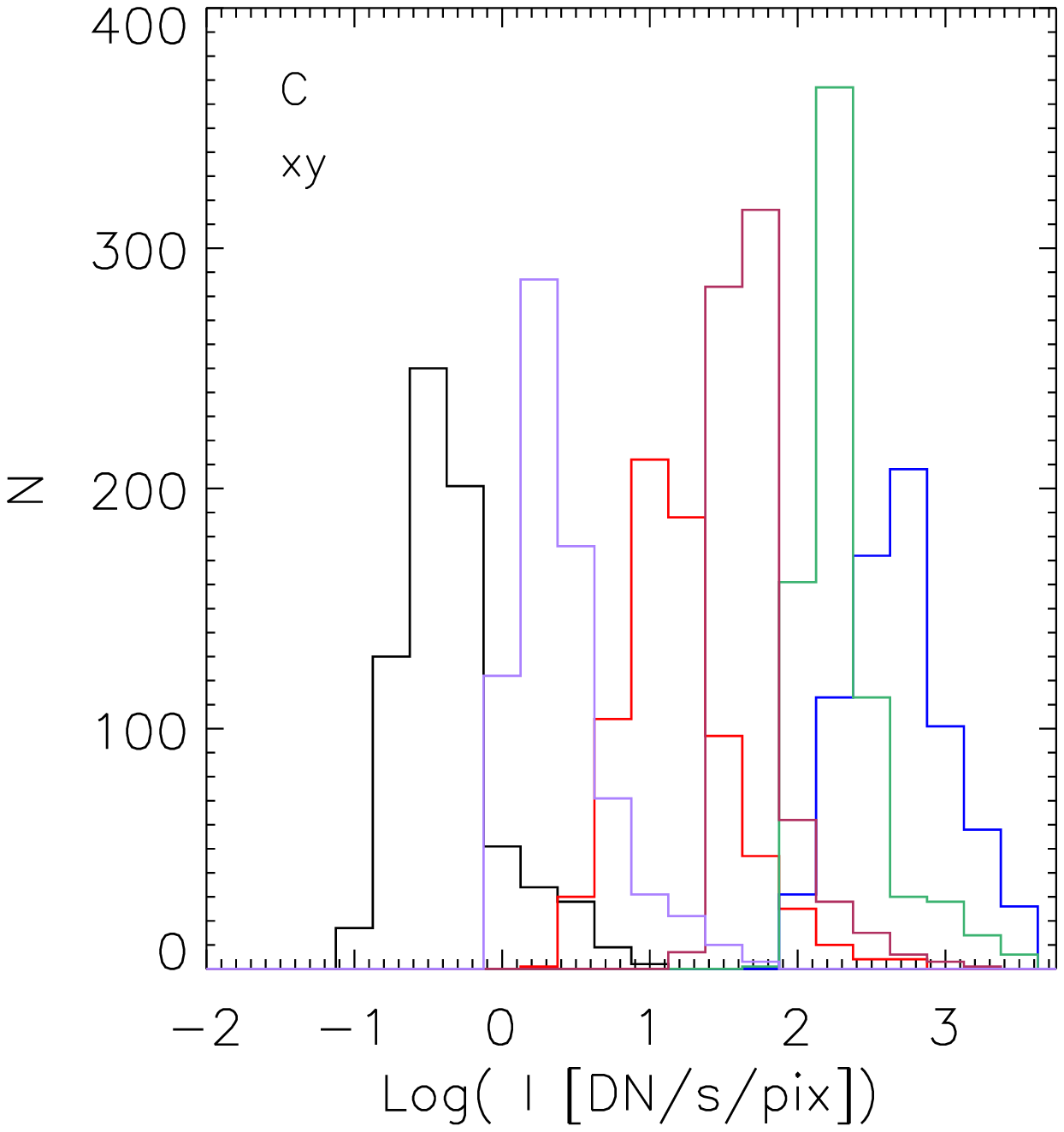}\hspace{-0.5cm}
  \includegraphics[scale=0.45]{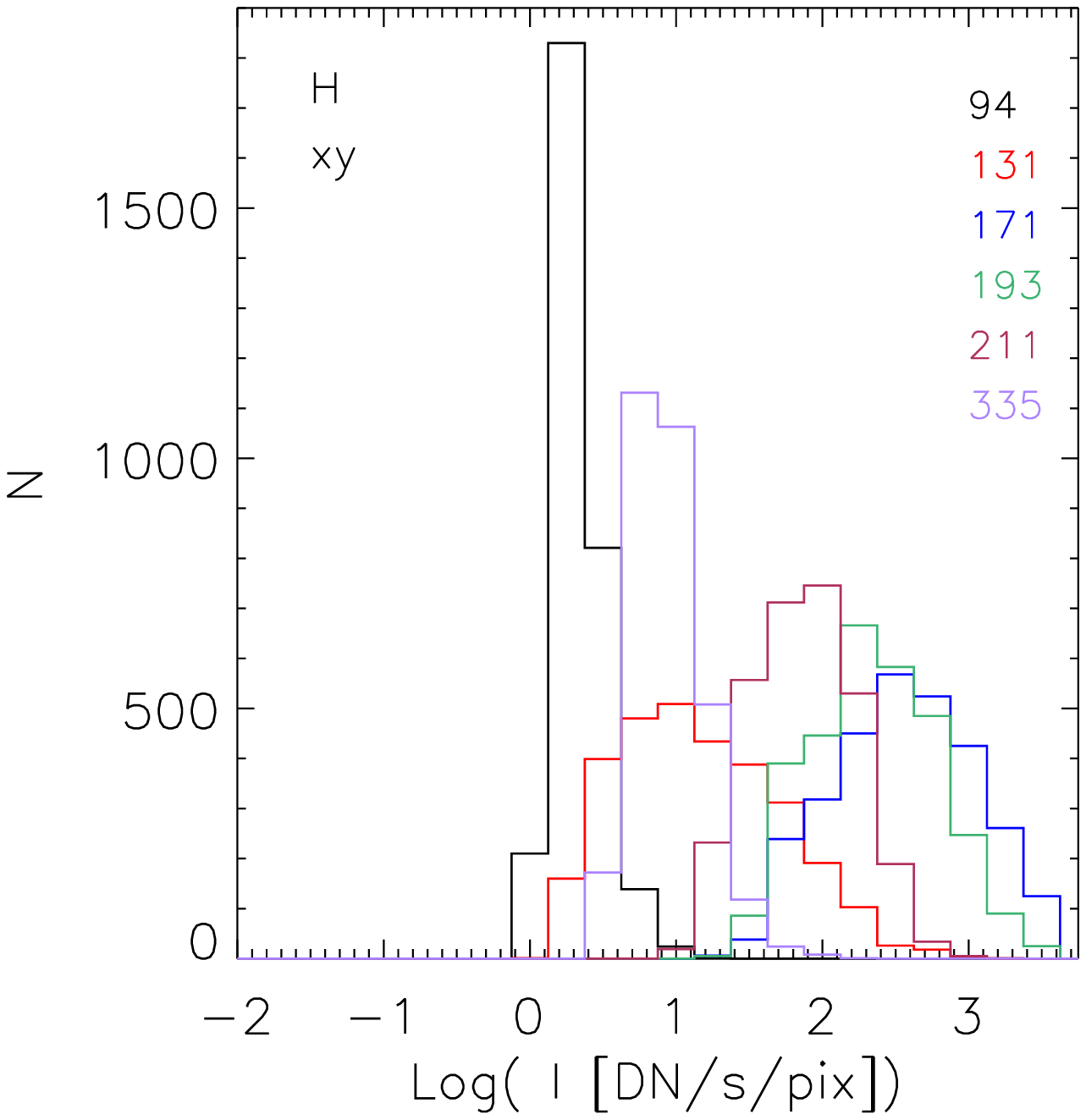}\vspace{-0.4cm}}
\centerline{\includegraphics[scale=0.45]{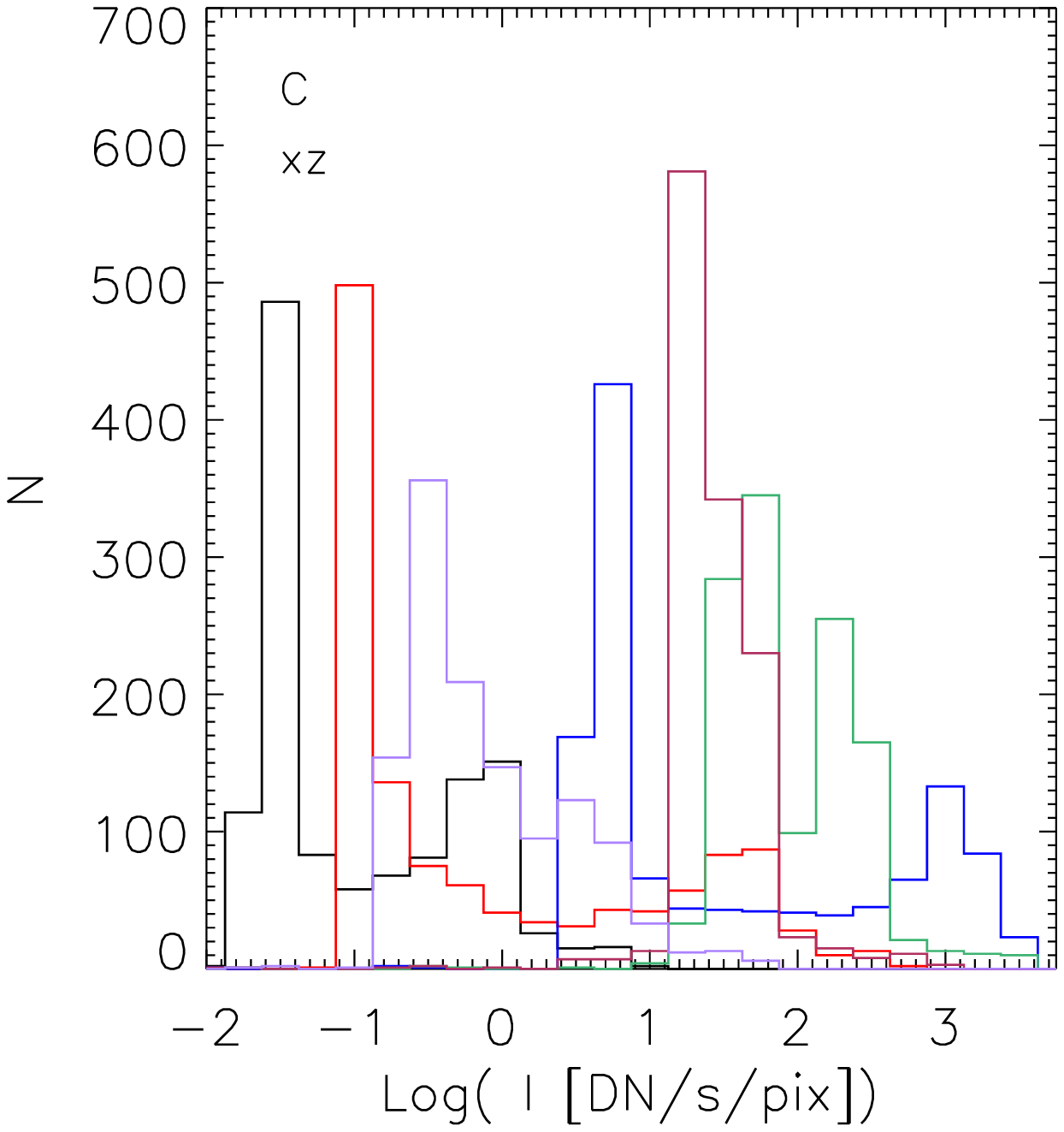}\hspace{-0.5cm}
  \includegraphics[scale=0.45]{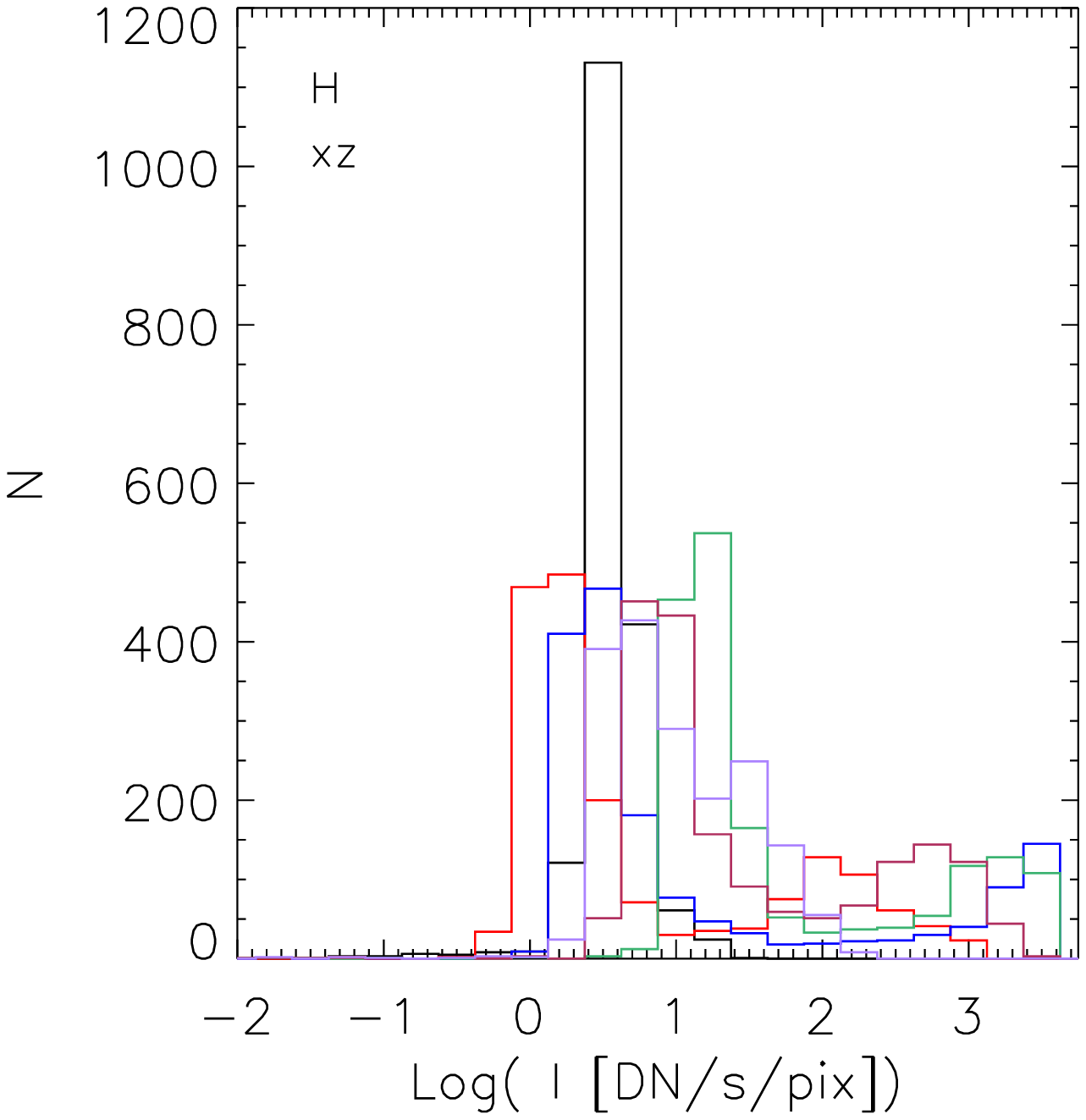}}
\caption{Histograms showing the distributions of the AIA intensities synthesized 
  from the two snapshots (left panels: smaller cooler snapshot, C; right panels: 
  larger hotter snapshot, H), and for the two different views (``xy'' in the top panels
  and ``xz'' in the bottom panels).  \label{fig:his_sim}}
\end{figure*}

In Figure~\ref{fig:his_sim} we show the distribution of the synthetic 
intensities in the AIA passbands, for the two lines of sight for each 
of the two snapshots.
The smaller, cooler snapshot (C) is characterized by intensities similar 
to the values observed in cool fan loops (region CL) and quiet Sun 
(region QS), whereas the larger and hotter snapshot (H) produces 
intensities more similar to the observed values of active region 
plasma (regions AR1 and AR2). 
We note that, as expected, the top view (``xy'') provides a more 
realistic comparison for real coronal observations, while for the side 
view (``xz'') the distributions have tails at low intensity values, which 
are not observed in the AIA selected data. 
This is because the small box size does not accurately capture the 
enormous line-of-sight at the limb in the optically thin corona.  
However, for the side view of snapshot C the inversion of the peaks 
of the 335\AA\ and 131\AA\ channels with respect to the top view (i.e., 
for the side view the 131\AA\ distribution peaks at lower intensity values 
than the 335\AA) reproduces what is observed in actual data at the limb 
(region LM) compared with quiet on disk regions (QS, CL).
We also note that the distributions for snapshot H appear generally 
broader than the observed distributions. For snapshot C the distribution 
of each passband peaks around a different value, whereas the observed 
distributions typically show a common peak for the weaker channels 
and a separate common peak for the stronger channels.
Finally, the intensities in the 94\AA\ channel are systematically lower 
than the observed values, but this is expected on the basis of recent 
studies arguing for incompleteness of atomic databases, especially in 
the 94\AA\ passband \citep[e.g.,][]{Testa12}. 
By and large, these qualitative comparisons indicate that the simulations produce
emission values of similar order of magnitude of real observations, and therefore
represent a reasonable test case for the coronal diagnostics.

\section{Emission measure peak temperature \label{app:tmax}}

See complete preprint version at http://folk.uio.no/bdp/papers/3dEMD\_ptesta.pdf

\end{document}

%% file: tab_lines.tex
\begin{deluxetable}{rlcc} 
\footnotesize
\tablecolumns{4} 
\tablewidth{0pc} 
\tablecaption{EIS lines synthesized from 3D models and used for 
  reconstructing the emission measure distribution.
	\label{tab:lines}} 
\tablehead{ 
\colhead{\ll [\AA] \tablenotemark{a}}  & \colhead{Ion} & 
\colhead{$\log (T_{\rm max} [{\rm K}])$}   & \colhead{Notes \tablenotemark{b}} 
}
\startdata 
  
268.991  &  \mgvi    &  5.65  & sb   \\ 
185.213  &  \feviii    &  5.70  &     \\ 
278.404  &  \mgvii   &  5.80  &      \\
275.361  &  \sivii     &  5.80  &      \\ 
188.497  &  \feix      &  5.90  &     \\ 
184.537  &  \fex       &  6.05  &     \\ 
188.216  &  \fexi      &  6.15  &     \\ 
195.119  &  \fexii     &  6.20  & sb  \\ 
274.204  &  \fexiv    &  6.30  &     \\ 
284.163  &  \fexv     &  6.35  &     \\ 
262.976  &  \fexvi    &  6.45  &     \\ 
208.604  &  \caxvi   &  6.70  &     \\ 
192.853  &  \caxvii  &  6.75  & bl  \\ 
254.347  &  \fexvii   &  6.75  &   \\ 

\enddata 
\tablenotetext{a}{The wavelengths are from CHIANTI (in case of 
self-blend we list the wavelength of the strongest line).}
\tablenotetext{b}{The label ``sb'' indicates that the spectral feature is 
a self-blend of lines from the same ion, all included in synthesizing the data.
The label ``bl'' for the \caxvii\ line indicates that at the EIS spectral resolution 
this line is blended with \fexi\ and \ov\ lines \citep{Ko09}. For this
paper however we only synthetize the \caxvii\ intensity, i.e., we do
not calculate the blending lines.}
\end{deluxetable}

%% file: 3dEMD_ptesta.bbl
\begin{thebibliography}{62}
\expandafter\ifx\csname natexlab\endcsname\relax\def\natexlab#1{#1}\fi

\bibitem[{{Aschwanden} \& {Boerner}(2011)}]{Aschwanden11}
{Aschwanden}, M.~J., \& {Boerner}, P. 2011, \apj, 732, 81

\bibitem[{{Aschwanden} {et~al.}(2011){Aschwanden}, {Boerner}, {Schrijver}, \&
  {Malanushenko}}]{Aschwanden11b}
{Aschwanden}, M.~J., {Boerner}, P., {Schrijver}, C.~J., \& {Malanushenko}, A.
  2011, \solphys, 384

\bibitem[{{Aschwanden} \& {Nightingale}(2005)}]{Aschwanden05}
{Aschwanden}, M.~J., \& {Nightingale}, R.~W. 2005, \apj, 633, 499

\bibitem[{{Aschwanden} {et~al.}(2000){Aschwanden}, {Nightingale}, \&
  {Alexander}}]{Aschwanden00}
{Aschwanden}, M.~J., {Nightingale}, R.~W., \& {Alexander}, D. 2000, \apj, 541,
  1059

\bibitem[{{Boerner} {et~al.}(2012){Boerner}, {Edwards}, {Lemen}, {Rausch},
  {Schrijver}, {Shine}, {Shing}, {Stern}, {Tarbell}, {Title}, {Wolfson},
  {Soufli}, {Spiller}, {Gullikson}, {McKenzie}, {Windt}, {Golub}, {Podgorski},
  {Testa}, \& {Weber}}]{Boerner12}
 {Boerner}, P., {Edwards}, C., {Lemen}, J., et al.\ 2012, \solphys,
  275, 41

\bibitem[{{Brooks} {et~al.}(2009){Brooks}, {Warren}, {Williams}, \&
  {Watanabe}}]{Brooks09}
{Brooks}, D.~H., {Warren}, H.~P., {Williams}, D.~R., \& {Watanabe}, T. 2009,
  \apj, 705, 1522

\bibitem[{{Brooks} {et~al.}(2011){Brooks}, {Warren}, \& {Young}}]{Brooks11}
{Brooks}, D.~H., {Warren}, H.~P., \& {Young}, P.~R. 2011, \apj, 730, 85

\bibitem[{{Brosius} {et~al.}(1996){Brosius}, {Davila}, {Thomas}, \&
  {Monsignori-Fossi}}]{Brosius96}
{Brosius}, J.~W., {Davila}, J.~M., {Thomas}, R.~J., \& {Monsignori-Fossi},
  B.~C. 1996, \apjs, 106, 143

\bibitem[{{Cargill} \& {Klimchuk}(2004)}]{Cargill04}
{Cargill}, P.~J., \& {Klimchuk}, J.~A. 2004, \apj, 605, 911

\bibitem[{{Craig} \& {Brown}(1976)}]{Craig76}
{Craig}, I.~J.~D., \& {Brown}, J.~C. 1976, \aap, 49, 239

\bibitem[{{Culhane} {et~al.}(2007){Culhane}, {Harra}, {James}, {Al-Janabi},
  {Bradley}, {Chaudry}, {Rees}, {Tandy}, {Thomas}, {Whillock}, {Winter},
  {Doschek}, {Korendyke}, {Brown}, {Myers}, {Mariska}, {Seely}, {Lang}, {Kent},
  {Shaughnessy}, {Young}, {Simnett}, {Castelli}, {Mahmoud}, {Mapson-Menard},
  {Probyn}, {Thomas}, {Davila}, {Dere}, {Windt}, {Shea}, {Hagood}, {Moye},
  {Hara}, {Watanabe}, {Matsuzaki}, {Kosugi}, {Hansteen}, \&
  {Wikstol}}]{Culhane07}
{Culhane}, J.~L., {Harra}, L.~K., {James}, A.~M., et al. 2007, \solphys, 243, 19

\bibitem[{{Del Zanna}(2003)}]{Delzanna03}
{Del Zanna}, G. 2003, \aap, 406, L5

\bibitem[{{Del Zanna} \& {Mason}(2003)}]{DZM03}
{Del Zanna}, G., \& {Mason}, H.~E. 2003, \aap, 406, 1089

\bibitem[{{Dere} {et~al.}(1997){Dere}, {Landi}, {Mason}, {Monsignori Fossi}, \&
  {Young}}]{chianti}
{Dere}, K.~P., {Landi}, E., {Mason}, H.~E., {Monsignori Fossi}, B.~C., \&
  {Young}, P.~R. 1997, \aaps, 125, 149

\bibitem[{{Dere} {et~al.}(2009){Dere}, {Landi}, {Young}, {Del Zanna},
  {Landini}, \& {Mason}}]{chianti6}
{Dere}, K.~P., {Landi}, E., {Young}, P.~R., {Del Zanna}, G., {Landini}, M., \&
  {Mason}, H.~E. 2009, \aap, 498, 915

\bibitem[{{Feldman}(1992)}]{Feldman92}
{Feldman}, U. 1992, \physscr, 46, 202

\bibitem[{{Gudiksen} {et~al.}(2011){Gudiksen}, {Carlsson}, {Hansteen}, {Hayek},
  {Leenaarts}, \& {Mart{\'{\i}}nez-Sykora}}]{Gudiksen11}
{Gudiksen}, B.~V., {Carlsson}, M., {Hansteen}, V.~H., {Hayek}, W., {Leenaarts},
  J., \& {Mart{\'{\i}}nez-Sykora}, J. 2011, \aap, 531, A154

\bibitem[{{Hannah} \& {Kontar}(2012)}]{Hannah12}
{Hannah}, I.~G., \& {Kontar}, E.~P. 2012, \aap, 539, A146

\bibitem[{{Hansteen} {et~al.}(2007){Hansteen}, {Carlsson}, \&
  {Gudiksen}}]{Hansteen07}
{Hansteen}, V.~H., {Carlsson}, M., \& {Gudiksen}, B. 2007, in Astronomical
  Society of the Pacific Conference Series, Vol. 368, The Physics of
  Chromospheric Plasmas, ed. {P.~Heinzel, I.~Dorotovi{\v c}, \& R.~J.~Rutten},
  107

\bibitem[{{Judge}(2010)}]{Judge10}
{Judge}, P.~G. 2010, \apj, 708, 1238

\bibitem[{{Judge} {et~al.}(1997){Judge}, {Hubeny}, \& {Brown}}]{Judge97}
{Judge}, P.~G., {Hubeny}, V., \& {Brown}, J.~C. 1997, \apj, 475, 275

\bibitem[{{Kashyap} \& {Drake}(1998)}]{Kashyap98}
{Kashyap}, V., \& {Drake}, J.~J. 1998, \apj, 503, 450

\bibitem[{{Kashyap} \& {Drake}(2000)}]{PoA}
---. 2000, Bulletin of the Astronomical Society of India, 28, 475

\bibitem[{{Klimchuk}(2006)}]{Klimchuk06}
{Klimchuk}, J.~A. 2006, \solphys, 234, 41

\bibitem[{{Klimchuk} \& {Cargill}(2001)}]{Klimchuk01}
{Klimchuk}, J.~A., \& {Cargill}, P.~J. 2001, \apj, 553, 440

\bibitem[{{Ko} {et~al.}(2009){Ko}, {Doschek}, {Warren}, \& {Young}}]{Ko09}
{Ko}, Y., {Doschek}, G.~A., {Warren}, H.~P., \& {Young}, P.~R. 2009, \apj, 697,
  1956

\bibitem[{{Landi} {et~al.}(2006){Landi}, {Del Zanna}, {Young}, {Dere}, {Mason},
  \& {Landini}}]{Landi06}
{Landi}, E., {Del Zanna}, G., {Young}, P.~R., {Dere}, K.~P., {Mason}, H.~E., \&
  {Landini}, M. 2006, \apjs, 162, 261

\bibitem[{{Landi} \& {Feldman}(2008)}]{Landi08}
{Landi}, E., \& {Feldman}, U. 2008, \apj, 672, 674

\bibitem[{{Landi} {et~al.}(2002){Landi}, {Feldman}, \& {Dere}}]{Landi02}
{Landi}, E., {Feldman}, U., \& {Dere}, K.~P. 2002, \apjs, 139, 281

\bibitem[{{Landi} \& {Klimchuk}(2010)}]{LandiKlimchuk10}
{Landi}, E., \& {Klimchuk}, J.~A. 2010, \apj, 723, 320

\bibitem[{{Landi} \& {Landini}(1998)}]{Landi98}
{Landi}, E., \& {Landini}, M. 1998, \aap, 340, 265

\bibitem[{{Landi} {et~al.}(2009){Landi}, {Miralles}, {Curdt}, \&
  {Hara}}]{Landi09}
{Landi}, E., {Miralles}, M.~P., {Curdt}, W., \& {Hara}, H. 2009, \apj, 695, 221

\bibitem[{{Landi} {et~al.}(2012){Landi}, {Reale}, \& {Testa}}]{Landi12}
{Landi}, E., {Reale}, F., \& {Testa}, P. 2012, \aap, 538, A111

\bibitem[{{Lemen} {et~al.}(2012){Lemen}, {Title}, {Akin}, {Boerner}, {Chou},
  {Drake}, {Duncan}, {Edwards}, {Friedlaender}, {Heyman}, {Hurlburt}, {Katz},
  {Kushner}, {Levay}, {Lindgren}, {Mathur}, {McFeaters}, {Mitchell}, {Rehse},
  {Schrijver}, {Springer}, {Stern}, {Tarbell}, {Wuelser}, {Wolfson}, {Yanari},
  {Bookbinder}, {Cheimets}, {Caldwell}, {Deluca}, {Gates}, {Golub}, {Park},
  {Podgorski}, {Bush}, {Scherrer}, {Gummin}, {Smith}, {Auker}, {Jerram},
  {Pool}, {Soufli}, {Windt}, {Beardsley}, {Clapp}, {Lang}, \&
  {Waltham}}]{Lemen12}
{Lemen}, J.~R., {Title}, A.~M., {Akin}, D.~J., et al. 2012, \solphys, 275, 17

\bibitem[{{Mart{\'{\i}}nez-Sykora} {et~al.}(2011){Mart{\'{\i}}nez-Sykora}, {De
  Pontieu}, {Testa}, \& {Hansteen}}]{MartinezSykora11}
{Mart{\'{\i}}nez-Sykora}, J., {De Pontieu}, B., {Testa}, P., \& {Hansteen}, V.
  2011, \apj, 743, 23

\bibitem[{{McIntosh}(2000)}]{McIntosh00}
{McIntosh}, S.~W. 2000, \apj, 533, 1043

\bibitem[{{Patsourakos} \& {Klimchuk}(2006)}]{Patsourakos06}
{Patsourakos}, S., \& {Klimchuk}, J.~A. 2006, \apj, 647, 1452

\bibitem[{{Patsourakos} \& {Klimchuk}(2009)}]{Patsourakos09}
---. 2009, \apj, 696, 760

\bibitem[{{Phillips} {et~al.}(2008){Phillips}, {Feldman}, \&
  {Landi}}]{Phillips08}
{Phillips}, K.~J.~H., {Feldman}, U., \& {Landi}, E. 2008, {Ultraviolet and
  X-ray Spectroscopy of the Solar Atmosphere} (Cambridge University Press)

\bibitem[{{Reale}(2010)}]{Reale10}
{Reale}, F. 2010, Living Reviews in Solar Physics, 7, 5

\bibitem[{{Reale} {et~al.}(2011){Reale}, {Guarrasi}, {Testa}, {DeLuca},
  {Peres}, \& {Golub}}]{Reale11}
{Reale}, F., {Guarrasi}, M., {Testa}, P., {DeLuca}, E.~E., {Peres}, G., \&
  {Golub}, L. 2011, \apjl, 736, L16

\bibitem[{{Reale} {et~al.}(2009{\natexlab{a}}){Reale}, {McTiernan}, \&
  {Testa}}]{Reale09b}
{Reale}, F., {McTiernan}, J.~M., \& {Testa}, P. 2009{\natexlab{a}}, \apjl, 704,
  L58

\bibitem[{{Reale} {et~al.}(2007){Reale}, {Parenti}, {Reeves}, {Weber}, {Bobra},
  {Barbera}, {Kano}, {Narukage}, {Shimojo}, {Sakao}, {Peres}, \&
  {Golub}}]{Reale07}
{Reale}, F., {Parenti}, S., {Reeves}, K.~K., et al. 2007, Science, 318, 1582

\bibitem[{{Reale} {et~al.}(2009{\natexlab{b}}){Reale}, {Testa}, {Klimchuk}, \&
  {Parenti}}]{Reale09}
{Reale}, F., {Testa}, P., {Klimchuk}, J.~A., \& {Parenti}, S.
  2009{\natexlab{b}}, \apj, 698, 756

\bibitem[{{Schmelz} {et~al.}(2011){Schmelz}, {Jenkins}, {Worley}, {Anderson},
  {Pathak}, \& {Kimble}}]{Schmelz11}
{Schmelz}, J.~T., {Jenkins}, B.~S., {Worley}, B.~T., {Anderson}, D.~J.,
  {Pathak}, S., \& {Kimble}, J.~A. 2011, \apj, 731, 49

\bibitem[{{Schmelz} {et~al.}(2010){Schmelz}, {Kimble}, {Jenkins}, {Worley},
  {Anderson}, {Pathak}, \& {Saar}}]{Schmelz10}
{Schmelz}, J.~T., {Kimble}, J.~A., {Jenkins}, B.~S., {Worley}, B.~T.,
  {Anderson}, D.~J., {Pathak}, S., \& {Saar}, S.~H. 2010, \apjl, 725, L34

\bibitem[{{Schmelz} {et~al.}(2005){Schmelz}, {Nasraoui}, {Roames}, {Lippner},
  \& {Garst}}]{Schmelz05}
{Schmelz}, J.~T., {Nasraoui}, K., {Roames}, J.~K., {Lippner}, L.~A., \&
  {Garst}, J.~W. 2005, \apjl, 634, L197

\bibitem[{{Schmelz} {et~al.}(2009{\natexlab{a}}){Schmelz}, {Saar}, {DeLuca},
  {Golub}, {Kashyap}, {Weber}, \& {Klimchuk}}]{Schmelz09a}
{Schmelz}, J.~T., {Saar}, S.~H., {DeLuca}, E.~E., {Golub}, L., {Kashyap},
  V.~L., {Weber}, M.~A., \& {Klimchuk}, J.~A. 2009{\natexlab{a}}, \apjl, 693,
  L131

\bibitem[{{Schmelz} {et~al.}(2009{\natexlab{b}}){Schmelz}, {Saar}, {Weber},
  {Deluca}, \& {Golub}}]{Schmelz09c}
{Schmelz}, J.~T., {Saar}, S.~H., {Weber}, M.~A., {Deluca}, E.~E., \& {Golub},
  L. 2009{\natexlab{b}}, in Astronomical Society of the Pacific Conference
  Series, Vol. 415, The Second Hinode Science Meeting: Beyond Discovery-Toward
  Understanding, ed. B.~{Lites}, M.~{Cheung}, T.~{Magara}, J.~{Mariska}, \&
  K.~{Reeves}, 299

\bibitem[{{Schmelz} {et~al.}(2001){Schmelz}, {Scopes}, {Cirtain}, {Winter}, \&
  {Allen}}]{Schmelz01}
{Schmelz}, J.~T., {Scopes}, R.~T., {Cirtain}, J.~W., {Winter}, H.~D., \&
  {Allen}, J.~D. 2001, \apj, 556, 896

\bibitem[{{Shestov} {et~al.}(2010){Shestov}, {Kuzin}, {Urnov}, {Ul'Yanov}, \&
  {Bogachev}}]{Shestov10}
{Shestov}, S.~V., {Kuzin}, S.~V., {Urnov}, A.~M., {Ul'Yanov}, A.~S., \&
  {Bogachev}, S.~A. 2010, Astronomy Letters, 36, 44

\bibitem[{{Sylwester} {et~al.}(2010){Sylwester}, {Sylwester}, \&
  {Phillips}}]{Sylwester10}
{Sylwester}, B., {Sylwester}, J., \& {Phillips}, K.~J.~H. 2010, \aap, 514, A82+

\bibitem[{{Testa} {et~al.}(2012){Testa}, {Drake}, \& {Landi}}]{Testa12}
{Testa}, P., {Drake}, J.~J., \& {Landi}, E. 2012, \apj, 745, 111

\bibitem[{{Testa} {et~al.}(2005){Testa}, {Peres}, \& {Reale}}]{Testa05}
{Testa}, P., {Peres}, G., \& {Reale}, F. 2005, \apj, 622, 695

\bibitem[{{Testa} {et~al.}(2002){Testa}, {Peres}, {Reale}, \&
  {Orlando}}]{Testa02}
{Testa}, P., {Peres}, G., {Reale}, F., \& {Orlando}, S. 2002, \apj, 580, 1159

\bibitem[{{Testa} {et~al.}(2011){Testa}, {Reale}, {Landi}, {DeLuca}, \&
  {Kashyap}}]{Testa11}
{Testa}, P., {Reale}, F., {Landi}, E., {DeLuca}, E.~E., \& {Kashyap}, V. 2011,
  \apj, 728, 30

\bibitem[{{Testa} \& {Reale}(2012)}]{Testa12b}
{Testa}, P., \& {Reale}, F. 2012, \apjl, 750, L10

\bibitem[{{Tripathi} {et~al.}(2011){Tripathi}, {Klimchuk}, \&
  {Mason}}]{Tripathi11}
{Tripathi}, D., {Klimchuk}, J.~A., \& {Mason}, H.~E. 2011, \apj, 740, 111

\bibitem[{{Warren} \& {Brooks}(2009)}]{Warren09}
{Warren}, H.~P., \& {Brooks}, D.~H. 2009, \apj, 700, 762

\bibitem[{{Warren} {et~al.}(2011){Warren}, {Brooks}, \&
  {Winebarger}}]{Warren11}
{Warren}, H.~P., {Brooks}, D.~H., \& {Winebarger}, A.~R. 2011, \apj, 734, 90

\bibitem[{{Warren} {et~al.}(2008){Warren}, {Ugarte-Urra}, {Doschek}, {Brooks},
  \& {Williams}}]{Warren08loops}
{Warren}, H.~P., {Ugarte-Urra}, I., {Doschek}, G.~A., {Brooks}, D.~H., \&
  {Williams}, D.~R. 2008, \apjl, 686, L131

\bibitem[{{Watanabe} {et~al.}(2007){Watanabe}, {Hara}, {Culhane}, {Harra},
  {Doschek}, {Mariska}, \& {Young}}]{Watanabe07}
{Watanabe}, T., {Hara}, H., {Culhane}, L., {Harra}, L.~K., {Doschek}, G.~A.,
  {Mariska}, J.~T., \& {Young}, P.~R. 2007, \pasj, 59, 669

\bibitem[{{Young} {et~al.}(2009){Young}, {Watanabe}, {Hara}, \&
  {Mariska}}]{Young09}
{Young}, P.~R., {Watanabe}, T., {Hara}, H., \& {Mariska}, J.~T. 2009, \aap,
  495, 587

\end{thebibliography}
